
\font\tenmib=cmmib10
\font\sevenmib=cmmib10 scaled 800
\font\titolo=cmbx12
\font\titolone=cmbx10 scaled\magstep 2

\font\cs=cmcsc10
\font\sc=cmcsc10
\font\css=cmcsc8

\font\ninerm=cmr9
\font\ottorm=cmr8
\textfont5=\tenmib
\scriptfont5=\sevenmib
\scriptscriptfont5=\fivei

\font\euftw=eufm9 scaled\magstep1
\font\euftww=eufm7 scaled\magstep1
\font\euftwww=eufm5 scaled\magstep1
\font\msytw=msbm9 scaled\magstep1

\font\msytwww=msbm5 scaled\magstep1
\font\indbf=cmbx10 scaled\magstep2

\def\st{\scriptstyle}

\font\ottorm=cmr8\font\ottoi=cmmi8\font\ottosy=cmsy8%
\font\ottobf=cmbx8\font\ottott=cmtt8%
\font\ottocss=cmcsc8%
\font\ottosl=cmsl8\font\ottoit=cmti8%
\font\sixrm=cmr6\font\sixbf=cmbx6\font\sixi=cmmi6\font\sixsy=cmsy6%
\font\fiverm=cmr5\font\fivesy=cmsy5
\font\fivei=cmmi5
\font\fivebf=cmbx5%

\def\ottopunti{\def\rm{\fam0\ottorm}%
\textfont0=\ottorm\scriptfont0=\sixrm\scriptscriptfont0=\fiverm%
\textfont1=\ottoi\scriptfont1=\sixi\scriptscriptfont1=\fivei%
\textfont2=\ottosy\scriptfont2=\sixsy\scriptscriptfont2=\fivesy%
\textfont3=\tenex\scriptfont3=\tenex\scriptscriptfont3=\tenex%
\textfont4=\ottocss\scriptfont4=\sc\scriptscriptfont4=\sc%
\textfont5=\tenmib\scriptfont5=\sevenmib\scriptscriptfont5=\fivei
\textfont\itfam=\ottoit\def\it{\fam\itfam\ottoit}%
\textfont\slfam=\ottosl\def\sl{\fam\slfam\ottosl}%
\textfont\ttfam=\ottott\def\tt{\fam\ttfam\ottott}%
\textfont\bffam=\ottobf\scriptfont\bffam=\sixbf%
\scriptscriptfont\bffam=\fivebf\def\bf{\fam\bffam\ottobf}%
\setbox\strutbox=\hbox{\vrule height7pt depth2pt width0pt}%
\normalbaselineskip=9pt\let\sc=\sixrm\normalbaselines\rm}
\let\nota=\ottopunti%

\mathchardef\BDpr = "0540  
\mathchardef\Bg   = "050D  

%
%
%
%
%
%
%

\global\newcount\numsec\global\newcount\numapp
\global\newcount\numfor\global\newcount\numfig
\global\newcount\numsub
\numsec=0\numapp=0\numfig=0
\def\veroparagrafo{\number\numsec}\def\veraformula{\number\numfor}
\def\veraappendice{\number\numapp}\def\verasub{\number\numsub}
\def\verafigura{\number\numfig}
\def\section(#1,#2){\advance\numsec by 1\numfor=1\numsub=1\numfig=1%
\SIA p,#1,{\veroparagrafo} %
\write15{\string\Fp (#1){\secc(#1)}}%
\write16{ sec. #1 ==> \secc(#1)  }%
\hbox to \hsize{\titolo\hfill \number\numsec. #2\hfill%
\expandafter{\alato(sec. #1)}}\*}

\def\appendix(#1,#2){\advance\numapp by 1\numfor=1\numsub=1\numfig=1%
\SIA p,#1,{A\veraappendice} %
\write15{\string\Fp (#1){\secc(#1)}}%
\write16{ app. #1 ==> \secc(#1)  }%
\hbox to \hsize{\titolo\hfill Appendix A\number\numapp. #2\hfill%
\expandafter{\alato(app. #1)}}\*}

\def\senondefinito#1{\expandafter\ifx\csname#1\endcsname\relax}

\def\SIA #1,#2,#3 {\senondefinito{#1#2}%
\expandafter\xdef\csname #1#2\endcsname{#3}\else
\write16{???? ma #1#2 e' gia' stato definito !!!!} \fi}

\def \Fe(#1)#2{\SIA fe,#1,#2 }
\def \Fp(#1)#2{\SIA fp,#1,#2 }
\def \Fg(#1)#2{\SIA fg,#1,#2 }

\def\etichetta(#1){(\veroparagrafo.\veraformula)%
\SIA e,#1,(\veroparagrafo.\veraformula) %
\global\advance\numfor by 1%
\write15{\string\Fe (#1){\equ(#1)}}%
\write16{ EQ #1 ==> \equ(#1)  }}

\def\etichettaa(#1){(A\veraappendice.\veraformula)%
\SIA e,#1,(A\veraappendice.\veraformula) %
\global\advance\numfor by 1%
\write15{\string\Fe (#1){\equ(#1)}}%
\write16{ EQ #1 ==> \equ(#1) }}

\def\getichetta(#1){\veroparagrafo.\verafigura%
\SIA g,#1,{\veroparagrafo.\verafigura} %
\global\advance\numfig by 1%
\write15{\string\Fg (#1){\graf(#1)}}%
\write16{ Fig. #1 ==> \graf(#1) }}

\def\etichettap(#1){\veroparagrafo.\verasub%
\SIA p,#1,{\veroparagrafo.\verasub} %
\global\advance\numsub by 1%
\write15{\string\Fp (#1){\secc(#1)}}%
\write16{ par #1 ==> \secc(#1)  }}

\def\etichettapa(#1){A\veraappendice.\verasub%
\SIA p,#1,{A\veraappendice.\verasub} %
\global\advance\numsub by 1%
\write15{\string\Fp (#1){\secc(#1)}}%
\write16{ par #1 ==> \secc(#1)  }}

\def\Eq(#1){\eqno{\etichetta(#1)\alato(#1)}}
\def\eq(#1){\etichetta(#1)\alato(#1)}
\def\Eqa(#1){\eqno{\etichettaa(#1)\alato(#1)}}
\def\eqa(#1){\etichettaa(#1)\alato(#1)}
\def\eqg(#1){\getichetta(#1)\alato(fig. #1)}
\def\sub(#1){\0\palato(p. #1){\bf \etichettap(#1).}}
\def\asub(#1){\0\palato(p. #1){\bf \etichettapa(#1).}}

\def\equv(#1){\senondefinito{fe#1}$\clubsuit$#1%
\write16{eq. #1 non e' (ancora) definita}%
\else\csname fe#1\endcsname\fi}
\def\grafv(#1){\senondefinito{fg#1}$\clubsuit$#1%
\write16{fig. #1 non e' (ancora) definito}%
\else\csname fg#1\endcsname\fi}
\def\secv(#1){\senondefinito{fp#1}$\clubsuit$#1%
\write16{par. #1 non e' (ancora) definito}%
\else\csname fp#1\endcsname\fi}

\def\equ(#1){\senondefinito{e#1}\equv(#1)\else\csname e#1\endcsname\fi}
\def\graf(#1){\senondefinito{g#1}\grafv(#1)\else\csname g#1\endcsname\fi}
\def\figura(#1){{\css Figura} \getichetta(#1)}
\def\secc(#1){\senondefinito{p#1}\secv(#1)\else\csname p#1\endcsname\fi}
\def\sec(#1){{\secc(#1)}}
\def\refe(#1){{[\secc(#1)]}}

\def\BOZZA{\bz=1
\def\alato(##1){\rlap{\kern-\hsize\kern-1.2truecm{$\scriptstyle##1$}}}
\def\palato(##1){\rlap{\kern-1.2truecm{$\scriptstyle##1$}}}
}

\def\alato(#1){}
\def\galato(#1){}
\def\palato(#1){}


{\count255=\time\divide\count255 by 60 \xdef\hourmin{\number\count255}
        \multiply\count255 by-60\advance\count255 by\time
   \xdef\hourmin{\hourmin:\ifnum\count255<10 0\fi\the\count255}}

\def\oramin{\hourmin }

\def\data{\number\day/\ifcase\month\or gennaio \or febbraio \or marzo \or
aprile \or maggio \or giugno \or luglio \or agosto \or settembre
\or ottobre \or novembre \or dicembre \fi/\number\year;\ \oramin}

\newdimen\xshift \newdimen\xwidth \newdimen\yshift \newdimen\ywidth

\def\ins#1#2#3{\vbox to0pt{\kern-#2\hbox{\kern#1 #3}\vss}\nointerlineskip}

\def\eqfig#1#2#3#4#5{
\par\xwidth=#1 \xshift=\hsize \advance\xshift
by-\xwidth \divide\xshift by 2
\yshift=#2 \divide\yshift by 2
\line{\hglue\xshift \vbox to #2{\vfil
#3 \includegraphics{#4.ps}
}\hfill\raise\yshift\hbox{#5}}}

\def\8{\write12}  

\def\figini#1{
\catcode`\%=12\catcode`\{=12\catcode`\}=12
\catcode`\<=1\catcode`\>=2
\openout12=#1.ps}

\def\figfin{
\closeout12
\catcode`\%=14\catcode`\{=1%
\catcode`\}=2\catcode`\<=12\catcode`\>=12}


\let\a=\alpha \let\b=\beta  \let\g=\gamma  \let\d=\delta \let\e=\varepsilon
  \let\h=\eta   \let\th=\theta \let\k=\kappa \let\l=\lambda
\let\m=\mu    \let\n=\nu    \let\x=\xi     \let\p=\pi    \let\r=\rho
\let\s=\sigma \let\t=\tau   \let\f=\varphi \let\c=\chi
\let\ps=\psi   \let\o=\omega
\let\G=\Gamma \let\D=\Delta  \let\Th=\Theta\let\L=\Lambda 
         \let\Ps=\Psi
\let\O=\Omega 

\def\\{\hfill\break} \let\==\equiv

\let\io=\infty \def\Dpr{\V\dpr\,}

\let\0=\noindent

\def\ie{{i.e. }}
\let\dpr=\partial

\def\tende#1{\,\vtop{\ialign{##\crcr\rightarrowfill\crcr
 \noalign{\kern-1pt\nointerlineskip} \hskip3.pt${\scriptstyle
 #1}$\hskip3.pt\crcr}}\,}
\def\circage{\lower2pt\hbox{$\,\buildrel > \over {\scriptstyle \sim}\,$}}
\def\otto{\,{\kern-1.truept\leftarrow\kern-5.truept\to\kern-1.truept}\,}
\def\fra#1#2{{#1\over#2}}

\def\EE{{\cal E}}\def\MM{{\cal M}} \def\VV{{\cal V}}
\def\FF{{\cal F}} \def\HHH{{\cal H}}
 
\def\RR{{\cal R}}\def\LL{{\cal L}}  
\def\DD{{\cal D}}\def\AAA{{\cal A}} \def\SS{{\cal S}}

\def\T#1{{#1_{\kern-3pt\lower7pt\hbox{$\widetilde{}$}}\kern3pt}}
\def\VVV#1{{\underline #1}_{\kern-3pt
\lower7pt\hbox{$\widetilde{}$}}\kern3pt\,}
\def\W#1{#1_{\kern-3pt\lower7.5pt\hbox{$\widetilde{}$}}\kern2pt\,}

\def\lis{\overline}\def\tto{\Rightarrow}
 
  \def\sign{{\rm sign}\,}
\def\indica{\leaders \hbox to 0.5cm{\hss.\hss}\hfill}
\def\guida{\leaders\hbox to 1em{\hss.\hss}\hfill}

  \def\AA{{\bf A}}

\def\aaa{{\bf a}}\def\hhh{{\bf h}}

\def\ul{\underline}

\def\defin{{\buildrel def\over=}}

\def\Dpr{\BDpr\,}


\mathchardef\aa   = "050B
\mathchardef\bb   = "050C
\mathchardef\ggg  = "050D
\mathchardef\xxx  = "0518
\mathchardef\zzzzz= "0510
\mathchardef\oo   = "0521
\mathchardef\lll  = "0515
\mathchardef\mm   = "0516
\mathchardef\Dp   = "0540
\mathchardef\H    = "0548
\mathchardef\FFF  = "0546
\mathchardef\ppp  = "0570
\mathchardef\nn   = "0517
\mathchardef\pps  = "0520
\mathchardef\fff  = "0527
\mathchardef\FFF  = "0508
\mathchardef\nnnnn= "056E

\def\to{\rightarrow}

\let\ciao=\bye
\def\qed{\hfill\raise1pt\hbox{\vrule height5pt width5pt depth0pt}}

\def\Val{{\rm Val}}
\def\indic{\hbox{\raise-2pt \hbox{\indbf 1}}}

\def\RRR{\hbox{\msytw R}} 
 
 \def\ccc{\hbox{\msytwww C}}
 
 \def\ZZZ{\hbox{\msytw Z}}
 \def\zzz{\hbox{\msytwww Z}}
\def\TTT{\hbox{\msytw T}}

\def\vvv{\hbox{\euftw v}}    \def\vvvv{\hbox{\euftww v}}
\def\vvvvv{\hbox{\euftwww v}}\def\www{\hbox{\euftw w}}
\def\wwww{\hbox{\euftww w}}

\def\ul#1{{\underline#1}}
\def\Sqrt#1{{\sqrt{#1}}}
\def\V0{{\bf 0}}
\def\defi{\,{\buildrel def\over=}\,}

\def\rhs{{\it r.h.s.}\ }


\newcount\mgnf  
\mgnf=0 

\ifnum\mgnf=0
\def\openone{\leavevmode\hbox{\ninerm 1\kern-3.3pt\tenrm1}}%
\def\*{\vglue0.3truecm}\fi
\ifnum\mgnf=1
\def\openone{\leavevmode\hbox{\ninerm 1\kern-3.63pt\tenrm1}}%
\def\*{\vglue0.5truecm}\fi


\newcount\tipobib\newcount\bz\bz=0\newcount\aux\aux=1
\newdimen\bibskip\newdimen\maxit\maxit=0pt


\tipobib=2
\def\9#1{\ifnum\aux=1#1\else\relax\fi}

\newwrite\bib
\immediate\openout\bib=\jobname.bib
\global\newcount\bibex
\bibex=0
\def\verabib{\number\bibex}

\ifnum\tipobib=0
\def\cita#1{\expandafter\ifx\csname c#1\endcsname\relax
\hbox{$\clubsuit$}#1\write16{Manca #1 !}%
\else\csname c#1\endcsname\fi}
\def\rife#1#2#3{\immediate\write\bib{\string\raf{#2}{#3}{#1}}
\immediate\write15{\string\C(#1){[#2]}}
\setbox199=\hbox{#2}\ifnum\maxit < \wd199 \maxit=\wd199\fi}
\fi
\ifnum\tipobib=1
\def\cita#1{%
\expandafter\ifx\csname d#1\endcsname\relax%
\expandafter\ifx\csname c#1\endcsname\relax%
\hbox{$\clubsuit$}#1\write16{Manca #1 !}%
\else\probib(ref. numero )(#1)%
\csname c#1\endcsname%
\fi\else\csname d#1\endcsname\fi}%
\def\rife#1#2#3{\immediate\write15{\string\Cp(#1){%
\string\immediate\string\write\string\bib{\string\string\string\raf%
{\string\verabib}{#3}{#1}}%
\string\Cn(#1){[\string\verabib]}%
\string\CCc(#1)%
}}}%
\fi
\ifnum\tipobib=2%
\def\cita#1{\expandafter\ifx\csname c#1\endcsname\relax
\hbox{$\clubsuit$}#1\write16{Manca #1 !}%
\else\csname c#1\endcsname\fi}
\def\rife#1#2#3{\immediate\write\bib{\string\raf{#1}{#3}{#2}}
\immediate\write15{\string\C(#1){[#1]}}
\setbox199=\hbox{#2}\ifnum\maxit < \wd199 \maxit=\wd199\fi}
\fi

\def\Cn(#1)#2{\expandafter\xdef\csname d#1\endcsname{#2}}
\def\CCc(#1){\csname d#1\endcsname}
\def\probib(#1)(#2){\global\advance\bibex+1%
\9{\immediate\write16{#1\verabib => #2}}%
}

\def\C(#1)#2{\SIA c,#1,{#2}}
\def\Cp(#1)#2{\SIAnx c,#1,{#2}}

\def\SIAnx #1,#2,#3 {\senondefinito{#1#2}%
\expandafter\def\csname#1#2\endcsname{#3}\else%
\write16{???? ma #1,#2 e' gia' stato definito !!!!}\fi}

\bibskip=10truept
\def\hboxto{\hbox to}

\catcode`\{=12\catcode`\}=12
\catcode`\<=1\catcode`\>=2
\immediate\write\bib<
        \string\halign{\string\hboxto \string\maxit%
        {\string #\string\hfill}&%
        \string\vtop{\string\parindent=0pt\string\advance\string\hsize%
        by -.5truecm%
        \string#\string\vskip \bibskip
        }\string\cr%
>
\catcode`\{=1\catcode`\}=2
\catcode`\<=12\catcode`\>=12

\def\raf#1#2#3{\ifnum \bz=0 [#1]&#2 \cr\else
\llap{${}_{\rm #3}$}[#1]&#2\cr\fi}

\newread\bibin

\catcode`\{=12\catcode`\}=12
\catcode`\<=1\catcode`\>=2
\def\chiudibib<
\catcode`\{=12\catcode`\}=12
\catcode`\<=1\catcode`\>=2
\immediate\write\bib<}>
\catcode`\{=1\catcode`\}=2
\catcode`\<=12\catcode`\>=12
>
\catcode`\{=1\catcode`\}=2
\catcode`\<=12\catcode`\>=12

\def\makebiblio{
\ifnum\tipobib=0
\advance \maxit by 10pt
\else
\maxit=1.truecm
\fi
\chiudibib
\immediate \closeout\bib
\openin\bibin=\jobname.bib
\ifeof\bibin\relax\else
\raggedbottom
\input \jobname.bib
\fi
}

\openin13=#1.aux \ifeof13 \relax \else
\input #1.aux \closein13\fi
\openin14=\jobname.aux \ifeof14 \relax \else
\input \jobname.aux \closein14 \fi
\immediate\openout15=\jobname.aux

\def\biblio{\*\*\centerline{\titolo References}\*\nobreak\makebiblio}


\ifnum\mgnf=0
   \magnification=\magstep0
   \hsize=15truecm\vsize=19.6truecm\voffset2.truecm\hoffset.5truecm
   \parindent=0.3cm\baselineskip=0.45cm\fi
\ifnum\mgnf=1
   \magnification=\magstep1\hoffset=0.truecm
   \hsize=15truecm\vsize=19.6truecm
   \baselineskip=18truept plus0.1pt minus0.1pt \parindent=0.9truecm
   \lineskip=0.5truecm\lineskiplimit=0.1pt      \parskip=0.1pt plus1pt\fi


\mgnf=0   
\openin14=\jobname.aux \ifeof14 \relax \else
\input \jobname.aux \closein14 \fi
\openout15=\jobname.aux

\footline={\rlap{\hbox{\copy200}}\tenrm\hss \number\pageno\hss}
\def\fiat{}


\font\tenmib=cmmib10 
\font\sevenmib=cmmib7\font\fivemib=cmmib5 
\font\ottoit=cmti8
\font\fivei=cmmi5\font\sixi=cmmi6\font\ottoi=cmmi8
\font\ottorm=cmr8\font\fiverm=cmr5\font\sixrm=cmr6
\font\ottosy=cmsy8\font\sixsy=cmsy6\font\fivesy=cmsy5
\font\ottobf=cmbx8\font\sixbf=cmbx6\font\fivebf=cmbx5%
\font\ottott=cmtt8%
\font\ottocss=cmcsc8%
\font\ottosl=cmsl8%

\textfont5=\tenmib\scriptfont5=\sevenmib\scriptscriptfont5=\fivemib
\mathchardef\Ba   = "050B  
\mathchardef\Bb   = "050C  
\mathchardef\Bg   = "050D  
\mathchardef\Bd   = "050E  
\mathchardef\Be   = "0522  
\mathchardef\Bee  = "050F  
\mathchardef\Bz   = "0510  
\mathchardef\Bh   = "0511  
\mathchardef\Bthh = "0512  
\mathchardef\Bth  = "0523  
\mathchardef\Bi   = "0513  
\mathchardef\Bk   = "0514  
\mathchardef\Bl   = "0515  
\mathchardef\Bm   = "0516  
\mathchardef\Bn   = "0517  
\mathchardef\Bx   = "0518  
\mathchardef\Bom  = "0530  
\mathchardef\Bp   = "0519  
\mathchardef\Br   = "0525  
\mathchardef\Bro  = "051A  
\mathchardef\Bs   = "051B  
\mathchardef\Bsi  = "0526  
\mathchardef\Bt   = "051C  
\mathchardef\Bu   = "051D  
\mathchardef\Bf   = "0527  
\mathchardef\Bff  = "051E  
\mathchardef\Bch  = "051F  
\mathchardef\Bps  = "0520  
\mathchardef\Bo   = "0521  
\mathchardef\Bome = "0524  
\mathchardef\BG   = "0500  
\mathchardef\BD   = "0501  
\mathchardef\BTh  = "0502  
\mathchardef\BL   = "0503  
\mathchardef\BX   = "0504  
\mathchardef\BP   = "0505  
\mathchardef\BS   = "0506  
\mathchardef\BU   = "0507  
\mathchardef\BF   = "0508  
\mathchardef\BPs  = "0509  
\mathchardef\BO   = "050A  
\mathchardef\BDpr = "0540  
\mathchardef\Bstl = "053F  

%
\fiat


\overfullrule=0pt
\headline{{\tenrm \number\pageno:\ {\it Fractional Lindstedt series}\hss
}}


\def\V#1{{\bf#1}}
\let\aa=\Ba\let\fff=\Bf\let\defin=\defi\def\HHH{{\cal H}}
\def\AAA{{\cal A}}\let\oo=\Bo\let\nn=\Bn
\let\aaa=\Ba\let\pps=\Bps\def\hhh={\V h}
\let\bb=\Bb

\def\RRR{\hbox{\msytw R}} 
 
 \def\ccc{\hbox{\msytwww C}}
 
 \def\ZZZ{\hbox{\msytw Z}}
 \def\zzz{\hbox{\msytwww Z}}
\def\TTT{\hbox{\msytw T}}

\let\ul=\underline
\def\cfr{{cf. }}\let\ig=\int
\def\Tr{{\rm Tr}}
\fiat

\centerline{\titolone Fractional Lindstedt series}

\vskip1.truecm 
\centerline{\titolo
G. Gallavotti${}^{1}$,
G. Gentile$^{2}$ {\rm and}
A. Giuliani${}^{3}$}
\vskip.2truecm
\centerline{{}$^{1}$ Dipartimento di Fisica,
Universit\`a di Roma Uno, I-00185}
\vskip.1truecm
\centerline{{}$^{2}$ Dipartimento di Matematica,
Universit\`a di Roma Tre, Roma, I-00146}
\vskip.1truecm
\centerline{{}$^{3}$ Dipartimento di Matematica,
Universit\`a di Roma Due, Roma, I-00133}
\vskip1.truecm

\line{\vtop{ \line{\hskip1truecm\vbox{\advance \hsize by -2.1 truecm
        \0{\cs Abstract.}  {\it The parametric equations of the
          surfaces on which highly resonant quasi-periodic motions
          develop (lower-dimensional tori) cannot be analytically
          continued, in general, in the perturbation parameter $\e$,
          i.e. they are not analytic functions of $\e$.
          However rather generally quasi-periodic motions
          whose frequencies satisfy only one rational relation
          (``resonances of order $1$'') admit formal
          perturbation expansions in terms of a fractional power of
          $\e$ depending on the degeneration of the
          resonance. We find conditions for this to happen,
          and in such a case we prove that the formal expansion is
          convergent after suitable resummation.}} \hfill} }}

\*\*\*\*\*\*
\section(1,Introduction)

\0Resonances play an important role in the theory of dynamical
systems. A possible application is provided by problems of
celestial mechanics, such as the phenomenon of resonance locking
between rotation and orbital periods of the satellites \cita{GP}. 
In fact, the presence of friction can select resonant motions
which remain stable when friction (on astronomical time scales) 
becomes negligible.
In such a case maximal KAM tori can be really observed
only approximately and on very short time scale, whereas,
on very large time scales one expects that only
periodic motions survive. On intermediate time scale one can imagine
that quasi-periodic motions, involving a number of frequencies less
than the total number of degrees of freedom (and decreasing with
time), describe most of
the observed dynamics. This makes interesting and important
to study quasi-periodic motions occurring on lower-dimensional tori
for nearly-integrable Hamiltonian systems. 
These quasi-periodic motions are characterized by frequencies
satisfying $s$ rational relations, with $r=N-s$ ranging between $1$
(periodic motions) and the number $N$ of degrees of freedom
(KAM tori). The number $s$ equals the number of normal frequencies
appearing in the perturbed motions, whereas $r$ is the number of
independent components of the rotation vector.

The analysis of such motions is simpler under
some (generic) non-degeneracy assumptions on the perturbations.
On the contrary the situation becomes immediately very complicated
if no restriction at all is made on the perturbation.
Situations of this kind arise also in similar contexts:
we can mention the conservation of KAM tori under perturbations for
systems of $N$ harmonic oscillators, proved for $N=2$ but conjectured
to hold in general \cita{H}, \cita{R2}, and the stability of
Hill's equation under quasi-periodic perturbations \cita{GBC}.
In the case of lower-dimensional tori, the non-degeneracy
assumption is that the normal frequencies become different from zero
when the perturbation is switched on. If such an assumption is
removed, then only partial results hold, and only in the case
$s=1$, that is only in the case of
one normal frequency \cita{Ch1}, \cita{Ch2}.

Starting from the work of Eliasson \cita{E2}, a new approach to
KAM theory of quasi--integrable Hamiltonian systems arose,
based on the analysis of cancellations in the ``Lindstedt series'' for
the functions mapping unperturbed motions (uniform
rotations, in suitable coordinates) into corresponding perturbed ones.

A convenient way to exploit cancellations to resolve apparent
divergences in the Lindstedt series is through methods inspired by
quantum field theory, consisting in graphical expansions, summation
of classes of diverging subdiagrams, iterative study of the flow of
the effective constants and the possible introduction of
counterterms. With these techniques and ideas, a number of known
results have been reproduced and new ones have been obtained; see
\cita{G1}, \cita{Ge2}, \cita{Ge3} and \cita{GBG} for some reviews.

In this paper we follow the latter approach to investigate the conservation
of $(N-1)$-dimensional tori for systems with $N$ degrees of freedom.
More precisely we consider $N$ degrees of freedom systems described by
analytic Hamiltonians of the form
$$ H(I,\f) = H_{0}(I) + \e f(I,\f) ,
\Eq(1.1) $$
with $(I,\f)\in \DD \times \TTT^{N}$, where $\TTT=\RRR\setminus
2\p\ZZZ$ is the standard torus, $\DD$ is an open subset of
$\RRR^{N}$, $\e$ is a real parameter, and the free Hamiltonian
$H_{0}(I)$ is assumed to be uniformly convex:
$\dpr_{I}^{2} H_{0}(I) \ge C > 0$ for all $I\in\DD$.

\*

\0{\bf Definition 1} (Simple resonance).
{\it A {\rm simple Diophantine resonance}
for the unperturbed system is a
motion taking place on the torus $\{I_0\}\times\TTT^N$ with
$\o_0\defi\dpr_I H_0(I_0)$ satisfying a rational relation
$\o_0\cdot\n_0=0$ for some $\n_0\in\ZZZ^N$ and $|\o_0\cdot\n|> C
|\n|^{-\t}$ for suitable $C,\t>0$ and for {\it all} $0\ne
\n\in\ZZZ^N$ not parallel to $\n_0$.}

\*

Thus, if $\e=0$, the motions with rotation velocity $\o_0$ will
foliate the torus $\{I_0\}\times\TTT^N$ into a one parameter family of
invariant tori of dimension $N-1$.

For $\e\ne0$ invariant tori with dimension $N-1$ run by quasi-periodic
motions with spectrum $\o_0$ will, in general, only continue to exist
``close'' to some of the unperturbed tori $\{I_0\}\times\TTT^N$. The
problem is simpler under non-degeneracy assumptions on the average
$\langle f \rangle$ of the perturbing function $f$ on the
torus $\{I_0\}\times\TTT^N$; it has been studied in \cita{GG1} and
\cita{GG2} with techniques employed, under different assumptions on
the perturbation, in this paper.  Define the average of $f$ on
$\{I_0\}\times\TTT^N$ as
$$ \langle f(\f) \rangle
\defi\lim_{T\to\io} \fra1T\int_0^T f(\f+\o_0\,t)dt .
\Eq(1.2)$$
It depends nontrivially on $\f$ in the sense that in general it is
a non-constant periodic function of $\f$.  This is more
easily visualized in coordinates {\it adapted} to the resonance:
they are defined by a linear canonical transformation
$(I,\f)\defi \big(I_0+S^{-1}(\AA,B), S^T(\aa,\b)\big)$, with $S$ a
non-singular integer components $N\times N$ matrix with determinant $\det
S=1$ such that $\o_0\=S^T(\oo,0)$ with $\oo\in \RRR^{N-1}$,
$|\oo\cdot\nn|>C_0 |\nn|^{-\t_0}$ for {\it all} $\V0\ne \nn\in\ZZZ^{N-1}$
and $\AA\in \RRR^{N-1},B\in\RRR,\aa\in\TTT^{N-1},\b\in\TTT$.

In the coordinates $(\AA,B,\aa,\b)$ the Hamiltonian becomes an
analytic function of $(\AA,B)$, in the domain obtained from $\DD$
under the transformation $S$, and $(\aa,\b)\in\TTT^{N-1}\times\TTT$.
It will be of the form $H(\AA,B,\aa,\b)=H'_0(\AA,B) + \e f(\aa,\b)$,
where $H'_0(\AA,B)=\oo\cdot\AA+H_0(\AA,B)$, with $H_0$
vanishing to second order at $\AA=\V 0, B=0$ and uniformly convex in
the domain where it is defined. Of course the functions
$H,H_{0},f$ have a different meaning with respect to those
appearing in \equ(1.1), but we prefer to use the same notation
for simplicity. For the same reason we still shall use
the notation $(I,\f)$ to denote the new action-angle variables,
by setting $I=(\AA,B)$ and $\f=(\aa,\b)$.

So in the new coordinates the Hamiltonian $H$, the rotation
vector $\Bo\in \RRR^{N-1}$ and the average
$f_{\V0}(\V A,B,\b)$ of the perturbing energy can be supposed to be
such that
$$\eqalign{&
H=\oo\cdot\AA+H_0(\AA,B)+\e f(\AA,B,\aa,\b),\qquad
|\oo\cdot\nn|\ge\fra{C_0}{|\nn|^{\t_0}} \qquad
\forall\ \V0\ne\nn\in\ZZZ^{N-1} , \cr
&f_{\V0}(\AA,B,\b)\defi
\int\fra{d\aa}{(2\p)^{N-1}}\,f(\AA,B,\aa,\b) ,\quad
{\rm if}\quad f(\V A,B,\Ba,\b)
=\sum_{\Bn \in \zzz^{N-1}} e^{i\Bn\cdot\Ba}\, f_{\Bn}(\V A,B,\b) , \cr}
\Eq(1.3) $$
near the unperturbed resonance and {\it without any loss of
generality}.  The corresponding equations of motion  for
$X(t)\=(\AA(t),B(t),\aa(t),\b(t))$ are
$$ \eqalign{
\dot \AA =& \,- \e \dpr_{\aa} f ,  \cr
\dot  B =& \,- \e \dpr_{\b} f ,  \cr
\dot \aa =& \,\oo + \dpr_{\AA}H_{0} + \e \dpr_{\AA} f ,  \cr
\dot \b =& \,\dpr_{ B}H_{0} +
\e \dpr_{ B} f ,  \cr}\qquad{\rm or}\qquad \dot X=
E\dpr_X H(X),\qquad {\rm with} \qquad E=\pmatrix{0&-1\cr1&0\cr} ,
\Eq(1.4) $$
where $E$ is the standard $2N\times2N$ symplectic matrix. We study
the existence of motions which can be described by a constant $\b_0$
and by functions $\AA(\Bps),B(\Bps),{\bf a}(\Bps), b(\Bps)$ of
$\Bps\in\TTT^{N-1}$, which tend to $0$ as $\e\to0$, such that, by posing
$$\eqalign{
& \AA(t)=\AA(\Bps+\oo\,t),\kern22mm B(t)=B(\Bps+\oo\,t),\cr
& \aa(t)=\Bps+\oo\,t+\V a(\Bps+\oo\,t),\qquad
\b(t)=\b_0+\V b(\Bps+\oo\,t) , \cr}
\Eq(1.5)$$
one obtains, for all $\Bps\in\TTT^{N-1}$, solutions to the equations
of motion \equ(1.4). For brevity we shall sometimes write $X(t)$,
instead of $X(\Bps+\Bo t)$, to indicate solutions of \equ(1.4) of the
form \equ(1.5).

Note that, for $\e=0$, the motions \equ(1.5) reduce to
$$ \AA^{(0)}(t) = \V0 , \qquad B^{(0)}(t) = 0  ,
\qquad \aa^{(0)}(t) = \Bps + \Bo t , \qquad \b^{(0)}(t) = \b_{0} ,
\Eq(1.6) $$
where $\Bps\in\TTT^{N-1}$ are arbitrary. The motions
$X^{(0)}(t)=(\AA^{(0)}(t),B^{(0)}(t),\aa^{(0)}(t),\b^{(0)}(t))$
represent the unperturbed resonant motions filling a one parameter
family of $N-1$ dimensional invariant tori (parameterized by $\b_0$).

Given any function $G(\Bps,\e)$ we shall denote by $G_{\nn}$ the $\nn$-th
Fourier component of its Fourier expansion in $\Bps$ and we shall call
$[G]^{k}$ the $k$-th order term obtained by expanding in $\e$ the
function $G$.  Furthermore we shall denote by $[G]^{k}_{\nn}$ the
$\nn$-th Fourier component of the Fourier expansion of $[G]^{k}$
and by $G_{\ne\V0}$ the function $G-G_{\V0}$.

A formal solution of \equ(1.4), \equ(1.5), as a power series in $\e$,
$X(\Bps)=X^{(0)}(\Bps)+\e X^{(1)}(\Bps)+\ldots$, is well known to
exist if $\b_0\in\TTT$ is a stationarity point for the average
$f_{\V0}(\V0,0,\b)$ (\ie a point with such that $\dpr_\b
f_{\V0}(\V0,0,\b_0)=0$) which is not degenerate (\ie
$\dpr_{\b}^{2}f_{\V0}(\V0,0,\b_{0}) \neq 0$); \cfr for instance
\cita{P}, \cita{JLZ} and \cita{GG1}. {\it Here we consider
explicitly the case in which the non-degeneracy condition on
$\dpr_{\b}^{2}f_{\V0}(\V0,0,\b_{0})$ can fail to hold.  This is a case
in which in general no formal solution in powers of $\e$ can be
constructed.}

If $\b_{0}$ is such that $\dpr_{\b}f_{\V0}(\b_{0})=0$, then there
exists a function $X'$ such that $X^{(0)}+\e X'$ solves
\equ(1.4) up to terms of order $\e^{2}$ (excluded). Namely
$X'(\Bps)=(\AA'(\Bps),B'(\Bps),\V a'(\Bps),b'(\Bps))$
is obtained by applying the operator
$(\Bo\cdot\Dpr_\Bps)^{-1}$ to the vector
$$ \Big(\,-\dpr_\f f(\V0,0,\Bps,\b_0)\ ,\ \dpr_{I} f(\V 0,0,\Bps,\b_0)+
\dpr_{I}^2 H_0(\V0,0)\big(\V A'(\Bps), B'(\Bps)\big)\Big) ,
\Eq(1.7) $$
where $\dpr_\f,\dpr_{I}$ denote respectively the derivatives with
respect to the angle and action variables.
This means that one first determines $\V A'(\Bps)$ and $B'(\Bps)$
by solving $(\Bo\cdot\dpr_\Bps)\big(\V A'(\Bps), B'(\Bps)\big)=
-\dpr_\f f(\V0,0,\Bps,\b_0)$; the (otherwise arbitrary) averages
$\V A'_{\V0},B'_{\V 0}$ are fixed by requiring that the operator
$(\Bo\cdot\dpr)^{-1}$ can be applied to the vector formed by the
last $N$ components of \equ(1.7), \ie $\dpr_{I} f_{\V0}(\V0,0,\b_0) +
\dpr^2_{I} H_0(\V0,0) \big(\V A'_{\V0}, B'_{\V 0}\big)=\V0$. 
In this way also $\V a'(\Bps)$ and $b'(\Bps)$ can be obtained.  The
averages of the angle variables will be chosen $\V a'_\V0=\V0$ while
we leave $b'_\V0$ as a free parameter (to be suitably fixed at higher
orders to make the equations \equ(1.4) formally solvable).

There are, however other solutions which are correct up to order $\e^2$
(excluded).  If $\dpr_{\b}^{j}f_{\V0}(\V0,0,\b_{0})=0$ for all $j \le
k_{0}$ and $\dpr_{\b}^{k_{0}+1} f_{\V0}(\V0,0, \b_{0})\neq0$, one can
imagine to add to the free parameter $b_\V0'$ any polynomial in powers
of $\h=|\e|^{\fra1{k_0}}$ of degree $< k_0$. It can be checked (and it
will be explicitly shown in next sections) that, for any choice of
this polynomial, a solution of the equation of motions correct up to
terms of order $\e^2$ (excluded) exists.  This is, ultimately, the
reason why a consistent expansions in powers of $\e$ cannot in general
be continued beyond first order: the consistency condition of the
equations of motion necessary to improve (to second order included)
the solution fixes one coefficient of such (a priori arbitrary)
polynomials in $\h$ and, in general, this really forces the expansion
to be an expansion in powers of $\h$ different from $\h^{k_0}=\e$.

The analysis below shows that an expansion in $\h$ is
actually possible at all orders if the following assumptions are satisfied.

\*
\0{\bf Assumptions.} \\
{\it
{\rm (a)} The constant $a\defi [\dpr_\b f(X^{(0)}+
\e\,X'_0)]^1_{\V0}$ is $a\ne0$.\\
{\rm (b)} The matrix $\dpr^2_{I} H_0(\V0,0)$ is positive definite.\\
{\rm (c)} There is $k_{0}>0$ such that $\dpr_{\b}^{j}f_{\V0}(\V0,0,\b_{0})=0$
for all $j \le k_{0}$ and $c\defi\fra1{k_0!}\dpr_{\b}^{k_{0}+1}
f_{\V0}(\V0,0, \b_{0})\neq0$.
}
\*

The constant $a$ appearing in assumption (a), written explicitly, is
$$ \eqalign{
a\defi &\,
\Big[\dpr_{\b,\f}f(X^{(0)}(\Bps))\cdot(\Bo\cdot\dpr_\Bps)^{-1}\Big(
\dpr_{I}
f(X^{(0)}(\Bps))
+\dpr^2_{I} H_0(\V0,0)\big(\V A'(\Bps), B'(\Bps)\big)\Big)_{\ne\V0}-
\cr
&-\dpr_{\b,I} f(X^{(0)}(\Bps))
\cdot(\Bo\cdot\dpr_\Bps)^{-1}\dpr_\f f(X^{(0)}(\Bps))_{\ne\V0}
\Big]_{\V0} . \cr}
\Eq(1.8)$$
For instance if $f$ depends only on the angle variables $\f$ and
$\dpr^2_I H(\V0,0)={\openone}$ the constant $a$ is
$$ a =\fra12\dpr_\b \sum_\Bn\fra{|\Bn|^2|f_\Bn(\b)|^2+
|\dpr_\b f_\Bn(\b)|^2} {(\oo\cdot\nn)^2}\ \Bigg|_{\b=\b_0} .
\Eq(1.9)$$
The number $k_0$ appearing in assumption (c) 
is a measure of the ``degeneration'' of the resonance,
while the {\it order} of the resonance is the number of rational
relations between the unperturbed frequencies, which is $1$ in our case
as the unperturbed motions have $N-1$ independent frequencies.  The
case $k_0=1$ was considered in \cita{GG2} and we will not consider it
again here. We shall focus on the case $k_0\ge 2$, in which case a
formal power solution in $\e$ to \equ(1.4) does not exist. In fact,
one checks that, as a consequence of assumptions (a) and (c), the
average over $\Bps$ of the r.h.s.  of the second equation in \equ(1.4)
is different from $0$ at order $\e^2$, for any possible choice of
the free parameter $b_\V0'$ introduced after \equ(1.7).

However, under the assumptions above,
it is possible to find a formal solution to \equ(1.4)
such that the average of $b$ is $b_\V0=O(|\e|^{1/k_0})$:
the average of $b$ will be fixed in terms of the constants $a$ and $c$
in such a way that the average over $\Bps$ of the r.h.s. of the
second equation in \equ(1.4) is $0$ at all orders in $|\e|^{1/k_0}$.

{\it The necessity of a fractional powers expansion can be seen by a
heuristic argument sketched in Appendix \secc(A1): the argument also
suggests, as a conjecture, the forthcoming Theorem 1 and
motivates the assumptions {\rm (a)} to {\rm (c)}.}

In Appendix \secc(A1) it is in fact shown that, in the simple case 
$H_0(\V A,B)=\fra12(\V A^2+B^2)$ and $f$ depending only on the angles, 
a canonical transformation (explicitly constructed in 
Appendix \secc(A1)), defined in a neighborhood of 
$\{\V0\}\times\{0\}\times\TTT^{N-1}\times\{\b_0\}$,
maps $\oo\cdot\AA+H_0(\AA,B)+\e f(\aa,\b)$ into 
$$ \eqalign{&
\oo\cdot\AA + H_0(\AA,B) + \fra{\e\, c}{k_0+1}(\b-\b_0)^{k_0+1} +
\e^2\, a \, (\b-\b_0) \, + \cr
& \qquad\qquad + O(\e(\b-\b_0)^{k_0+2}) +
O(\e^2(\b-\b_0)^2)+O(\e {I}^2)+O(\e^3) . \cr}
\Eq(1.10) $$
As discussed in Appendix \secc(A1), the Hamiltonian equations  
corresponding to \equ(1.10) can be consistently solved to order $\e^2$. 
In particular, for some choices of 
the signs of $\e,a,c$, the angle $\b$ admits an approximate 
{\it quadratic} equilibrium point $O(|\e|^{1/k_0})$, whose stability depends
again on the relative signs of $\e,a,c$.
This second order computation suggests the conjecture that the unperturbed
motion $X^{(0)}(t)$ can be continued at $\e\neq 0$, provided the average
of $\b$ is chosen $O(|\e|^{1/k_0})$. The perturbed motion, if existing
beyond second order, will take place
on a torus (a small perturbation of the free one) which we shall call
elliptic or hyperbolic,
depending on the stability of the 
behavior of the linearization of the motion of $\b$
in the vicinity of its equilibrium point:
if the corresponding pair of nonzero Lyapunov exponents
is imaginary then the torus will be said {\it elliptic}, if it is real 
it will be said {\it hyperbolic}.

In the following a {\it sparse Cantor set dense at $0$}
will mean a set $\EE$ contained in an
interval $I=[-\e_0,0]$ or $I=[0,\e_0]$ with an open dense complement
in $I$ and with $0$ as a density point in the sense of Lebesgue
integration. In particular $\EE$ will have positive measure.

\*

\0{\bf Theorem 1.} {\it Consider the system described by the
Hamiltonian \equ(1.3), under the assumptions {\rm (a)},
{\rm (b)} and {\rm (c)}.
There exists $\e_0>0$ such that for $|\e|<\e_0$ the following holds.

(i) If $k_0$ is odd and $c\,\e < 0$ there is at least one hyperbolic
invariant torus of dimension $N-1$
with rotation vector $\Bo$. If $c\,\e > 0$ there is a sparse Cantor set
$\EE\subset [-\e_0,0]$ with the property that for $\e\in\EE$ there is
at least one elliptic torus of dimension $(N-1)$ with rotation vector $\Bo$.

(ii) If $k_0$ is even and $\e \,c$ has the same sign of $-a$,
there is at least one hyperbolic invariant torus of dimension $N-1$
with rotation vector $\Bo$. Moreover, there is a sparse Cantor set $\EE$
dense at $0$ such that, if $\e\in\EE$ and $\e\,c$ has the same sign of $-a$,
there is at least one elliptic invariant
torus of dimension $N-1$ with rotation vector $\Bo$.}

\*

\0{\it Remarks.}
(1) If assumption (c) is violated, since $f$ is analytic in $\b$,
then $f_\V0(\V0,0,\b)\=0$ as a function of $\b$. In this case
we can perform a canonical transformation removing the perturbation at order
$\e$ and casting the Hamiltonian into the form $H'=H'_0+\e^2 f'$, for some
new analytic functions $H'_0$ and $f'$.
If $f'$ satisfies the assumptions above we can apply Theorem 1 to $H'$.
If $H'$  satisfies assumption (c) and violates assumption (a)
we cannot say much. If $H'$ 
violates assumption (c) we can again remove the perturbation
at lowest order through a new canonical transformation and
cast the Hamiltonian into the new form $H''=H''_0+\e^4 f''$
and hope to be able to apply Theorem 1 to $H''$. And so on.
\\
(2) Assumption (a) is essential and, if it does not
hold, our expansion may fail to be convergent. In fact, in some cases
($k_0$ even), it is easy to show that if assumption (a) fails there
cannot be perturbed motions of the form \equ(1.5), see Appendix \secc(A2)
for an example. If $k_0$ is odd, \cita{Ch1} proved existence of
hyperbolic tori even if assumption (a) fails; in these cases we
expect that a new perturbation parameter must be identified.
The heuristic analysis in Appendix \secc(A1)
concretely suggests plausible results to be expected if $a=0$
(under alternative assumptions).
\\
(3) We expect that assumption (b) is not essential and that it could be
weakened into the request that $\dpr^2_I H_0(\V0,0)$ is non-degenerate
and that $\dpr^2_{BB}H_0(\V0,0)\neq 0$.  Certainly the convexity
assumption simplifies some of the estimates (see Appendix \secc(A6))
and we did not attempt to eliminate it.
\\
(4) The only known result on the problems considered here is in \cita{Ch1},
where conservation of $(N-1)$-dimensional ``hyperbolic tori'' is
proved under weaker assumptions, although
we study conservation of both hyperbolic and elliptic
$(N-1)$--dimensional tori under assumptions (a) to (c) above. 
\\
(5) In principle one could proceed in a different way rather
than following our approach. One could first perform the canonical
transformation described in Appendix \secc(A1) and leading to the
Hamiltonian \equ(1.10), then study the system so obtained with other
techniques, such as those in \cita{E1}, \cita{Po}, \cita{R1} or \cita{JLZ}.
To this aim one should use that in the new coordinates the unperturbed
Hamiltonian contains terms of order $\h^{2k_{0}-1}$, while the
perturbation is of order $\eta^{2k_{0}}$, hence has a further $\eta$.
\\
(6) The analysis via Lindstedt series and summations of classes of
diagrams is the main aspect of this work. Also the analyticity
properties in $\eta$ at $\Bo$ fixed are, to our knowledge, new.
\\
(7) Technically the present work is strongly inspired
by \cita{GG1}, \cita{Ge1} and \cita{GG2}.
The proofs that can be taken literally from \cita{GG2} will not be
repeated here: hence familiarity with the latter reference is essential.
\\
(8) The resummed series has manifest holomorphy properties which show
that the set $\EE$ is in the boundary of a complex domain where the
functions $X(\Bps)$ are analytic in $\e$: we do not discuss the
details (see \cita{GG1},\cita{GG2}).

\*

The paper is organized as follows. In Sections \secc(2) and \secc(3)
we prove formal solvability of the equation of motion in power series
in $|\e|^{1/k_0}$ and we describe a graphical representation
of the terms appearing in the formal power series.
Such a representation involves labeled rooted trees, and
is very similar to diagrammatic representations through
Feynman diagrams arising in quantum field theory.
In Sections \secc(4) to \secc(6) we describe an iterative resummation
scheme which eliminates some classes of divergent subdiagrams
and iteratively changes the power expansion, and
we prove convergence of the resummed series.
Some details of the proofs are deferred to the Appendices.

\*\*
\section(2,Lindstedt series)

\0We define $\e = \s \h^{k_{0}}$,
with $\s\in\{\pm1\}$, $\h>0$, and look for a family of formal
solutions of the equations of motion in powers of $\h$ parameterized
by $\Bps\in\TTT^{N-1}$, in the special form $X(t)=(\V A(t),B(t),\Ba(t),\b(t))$:
$$ \eqalign{
\AA(t) & = \sum_{k=k_0}^{\io} \h^{k}
\sum_{\nn\in\zzz^{N-1}} e^{i\nn\cdot(\Bps+ \oo t)} \AA^{(k)}_{\nn} , \cr
 B(t) & = \sum_{k=k_0}^{\io} \h^{k}
\sum_{\nn\in\zzz^{N-1}} e^{i\nn\cdot(\Bps+ \oo t)}  B^{(k)}_{\nn} , \cr
\aa(t) & = \Bps + \oo t + \V a(t), \qquad
\V a(t) = \sum_{k=k_0}^{\io} \h^{k}
\sum_{\nn\not =\V0} e^{i\nn\cdot(\Bps + \oo t)} \V a^{(k)}_\nn , \cr
\b(t) & = \b_{0} + b(t) , \qquad
b(t) =  \sum_{k=k_0}^{\io} \h^{k}
\sum_{\nn\not =\V0} e^{i\nn\cdot(\Bps + \oo t)} b^{(k)}_{\nn} +
\sum_{k=1}^{\io} \h^{k} b^{(k)}_{\V0}, \cr}
\Eq(2.1) $$
where all the involved functions have also a dependence on $\b_{0}$
which has not been made explicit. The formal series \equ(2.1) will be
called {\it Lindstedt series} as it extends the corresponding notions
already used in the theory of quasi-periodic motions on maximal tori.

{\it In the following the average $b^{(k)}_{\V0}$ will be
abbreviated as $\b_k$.}

\*

\0{\it Remarks.} (1) The functions $\AA,B,{\V a}$ and $b-b_\V0$ have been
chosen as power series in $\h$ starting with the $k_0$--th order as
the first non trivial order; ${\V a}$ has been chosen with zero
average (this just corresponds to a redefinition of the origin of
$\TTT^{N-1}$), while the average of $b$ has been chosen as a series in
$\h$ starting with the first order; the coefficients $\b_k\defi
b^{(k)}_\V0$ will be chosen in such a way that the Lindstedt series
admits a formal solution.

\0(2) With the choices in \equ(2.1), equations \equ(1.4) are identically
solved for any order $k<k_0$ in $\h$.  The parameters $\b_k$, $k<k_0$,
are left as free parameters, to be explicitly chosen below.

\*

To write the generic $k$--th order of \equ(1.4) we introduce the
following definitions: given any function $\Bps\to F(X(\Bps))$, let
$[F]^{(k)}$ be the $k$--th order in the Taylor expansion of $F$ in
$\h$ and let $[F]^{(k)}_{\nn}$ be the $\nn$-th Fourier component of the
Fourier expansion of $[F]^{(k)}$. Note that this notation concerns
expansions in $\h$ and is different from the one introduced after
\equ(1.6) which dealt with expansions in $\e$: with the new notation
the quantity denoted $a=\,[\dpr_\b f(X^{(0)})+
\e\,X')]^1_{\V0}$ before \equ(1.8) has to be written  as
$ a\,=\,\s\,[\dpr_\b f(X^{(0)}+\s
\h^{k_0}\,X')]^{(k_0)}_{\V0}$
as the superscript ${}^k$ denotes $k$-th order in $\e$ while ${}^{(k)}$
denotes $k$-th order in $\h$.  Then, if $\nn\neq\V0$ and $k\ge k_0$,
the equations \equ(1.4) become
$$ \eqalign{
\left( i \oo\cdot\nn \right) \AA^{(k)}_{\nn}
& = - \s [\dpr_{\aa} f]^{(k-k_{0})}_{\nn} , \cr
\left( i \oo\cdot\nn \right)  B^{(k)}_{\nn}
& = - \s [\dpr_{\b} f]^{(k-k_{0})}_{\nn} , \cr
\left( i \oo\cdot\nn \right) {\V a}^{(k)}_{\nn}
& = [\dpr_{\AA}H_{0}]^{(k)}_{\nn} +
\s [\dpr_{\AA} f]^{(k-k_{0})}_{\nn} , \cr
\left( i \oo\cdot\nn \right) b^{(k)}_{\nn}
& = [\dpr_{ B}H_{0}]^{(k)}_{\nn} +
\s [\dpr_{ B} f]^{(k-k_{0})}_{\nn} , \cr}
\Eq(2.2) $$
and, since $\Bo$ satisfies the Diophantine property, they can be solved
{\it provided}
$$ \eqalign{
\V0 = & [\dpr_{\aa} f]^{(k-k_{0})}_{\V0}
, \cr
0 =  &[\dpr_{\b} f]^{(k-k_{0})}_{\V0} , \cr
\V0 = & [\dpr_{\AA}H_{0}]^{(k)}_{\V0} +
\s [\dpr_{\AA} f]^{(k-k_{0})}_{\V0}  , \cr
0 =  &[\dpr_{ B}H_{0}]^{(k)}_{\V0} +
\s [\dpr_{ B} f]^{(k-k_{0})}_{\V0}  . \cr}
\Eq(2.3) $$
If assumptions (a), (b) and (c) in Section \secc(1)
are satisfied, such a formal solution
can be shown to exist provided the formal series for the average
$b_\V0$ is suitably chosen.

\*

\0{\bf Lemma 1.} {\it Under the assumptions {\rm (a)},
{\rm (b)} and {\rm (c)}
if $k_0$ is odd, a formal solution of \equ(1.4)
in the form \equ(2.1) always exists; if $k_0$ is even, a formal
solution of \equ(1.4) in the form \equ(2.1) exists if $\s a c<0$.
When a formal solution
exists, $\AA^{(k)},B^{(k)},\V a^{(k)},b^{(k)}$ are uniquely
fixed in the case $k_0$ odd. If $k_0$ is even, there are two
possible such sequences corresponding to the choices $\b_1\=b^{(1)}_\V0=
\pm (-\s a/c)^{1/k_{0}}$.}

\*

\0{\it Proof.} Let $X^{(0)}(\Bps)=(\V0,0,
\Bps,\b_0)$, $\e=\s\,\h^{k_0}$ with $\s=\sign(\e)$ and $\h>0$, and
look for a formal solution $t\to(\V A(t),B(t),\Ba(t),\b(t))$ obtained
by setting $\Bps=\Bps+\Bo t$ in
$$ X(\Bps)= X^{(0)}(\Bps)+\sum_{k=k_0}^\io \h^k
X^{(k)}(\Bps)+\sum_{k=1}^\io\h^k \x^{(k)} ,
\Eq(2.4)$$
where $\x^{(k)}=(\V0,0,\V0,\b_k)$. Set $X^{(h)}\=0$ for $0<h<k_0$ and
$\x^{(0)}=0$.

Suppose inductively that $X^{(h)}$ has been determined for
$h=0,1,\ldots, k-1$, for $k\ge1$. Consider a generic polynomial
$Y(\Bps)\defi \sum_{h=0}^{k-1}\h^h Y^{(h)}(\Bps)$ and suppose,
inductively, that if $Y^{(h)}=X^{(h)}+\x^{(h)}$ the function
$Y(\oo t)$ solves the equation of motion $\dot X=E\dpr_X H(X)$ up to order
$k-1$. This is true for $k=1$.
Let $\D\defi(\Bo\cdot\dpr_{\Bps})$ and
remark that the identities $\ig \sum_{j=1}^{2N}
\dpr_{\ps_i} Y_j\cdot\dpr_{ Y_j} H(Y)\,d\Bps=0$ and $\ig \sum_{j=1}^{2N}
\dpr_{\ps_i}\, Y_j\cdot (E \D Y)_j\,d\Bps=0$
are identities {\it for all periodic functions} $Y(\Bps)$.

Then $0\=\ig \sum_j\dpr_{\ps_i} Y_j\cdot ((E\,\D Y)_j+\dpr_{Y_j} H(Y))\,d\Bps$
and $\dot Y=E\dpr_Y H(Y)+O(\h^k)$ imply (note that $Y$ has
degree $<k$ in $\h$) $\ig \dpr_{\ps_i} Y^{(0)}_j\cdot\,
[\dpr_{ Y_j} H(Y)]^{(k)}\,$ $d\Bps=\V0$, \ie $\ig [\dpr_{\Ba}
H(Y)]^{(k)}\,d\Bps=\V0$ or
$$\big[\dpr_\Ba f\big(\sum_{h=0}^{k-1}\h^h(X^{(h)}+\x^{(h)})\big)
\big]_{\V0}^{(k-k_0)}=\V0 ,
\Eq(2.5)$$
which is the first of the compatibility conditions \equ(2.3).
This leaves $\V a_{\V0}^{(k)}$ as an arbitrary parameter:
we can set $\V a^{(k)}_{\V0}=\V0$.
We also remark that, if $I\=(\V A,B)$,
the last two of the four equations in \equ(2.3) have the form
$$(E\dpr)_\f\dpr_I H_0(\V 0,0)\, I^{(k)}_{\V0}+
F^{(k)}(\{X^{(h)},\x^{(h)}\}_{0\le h\le k-1})=0 .
\Eq(2.6)$$
By assumption (b) in Section \secc(1),
$\dpr^2_I H_0(\V 0,0)\=(E\dpr)_\f\dpr_I H_0(\V 0,0)$ is invertible; hence
to impose the last two compatibility conditions in \equ(2.3),
it suffices to fix $I^{(k)}_{\V0}=-[\dpr^2_I H_0(\V 0,0)]^{-1}
F^{(k)}(\{X^{(h)},$ $\x^{(h)}\}_{0\le h\le k-1})$, which is a known
function of $\{X^{(h)},\x^{(h)}\}_{0\le h\le k-1}$, for any possible
choice of the functions $\{\x^{(h)}\}_{0\le h\le k-1}$.

So, to fulfill the compatibility conditions \equ(2.3) and
to continue the inductive construction of $X^{(k)},\x^{(k)}$,
we are left with imposing the second of \equ(2.3), which can only hold if
$\b_{k-2k_0+1}$ satisfies certain compatibility properties. In fact
if $k<2k_0$ there is no requirement because the necessary condition
that $[\dpr_\b f]_\V0^{(k-k_0)}=0$ is automatically
satisfied by assumption (c).

For $k=2k_0$ the condition $[\dpr_\b f]_\V0^{(k_0)}=0$
can be expressed as follows. Let $c\defi\fra1{k_0!}\dpr^{k_0+1}_\b
f_\V0(\V0,0,\b_0)$ and remark that with the notations leading to
\equ(1.8) it is $X^{(k_0)}\=\s X'$: then the second of \equ(2.3)
becomes
$$\Big[\dpr_{\b X}f(X^{(0)})\cdot X^{(k_0)}+
\fra1{k_0!}\dpr^{k_0+1}_\b f(X^{(0)})\,(\b_1)^{k_0}\Big]_{\V0}
\= \,\s\, a +c\, \b_1^{k_0}\,=\,0 ,
\Eq(2.7)$$
which means
$$\b_1\,=\,\cases { (-a \s/c)^{1/k_0}& if $k_0$ is odd, \cr
\pm (-a \s/c)^{1/k_0} & if $k_0$ is even and
$a \s c< 0$, \cr}\Eq(2.8)$$
which is the compatibility condition to which $\b_1$ must be
subject.

For $k>2 k_0$ the second of \equ(2.3) only involves a sum of quantities
depending on $X^{(h)}$ with $h\le k-k_0$ and on $\x^{(h)}$ with
$h\le k-2 k_0$ {\it with the exception of a single term}, proportional
to $ck_0 \b_1^{k_0-1}\b_{k-2k_0+1}$, involving $\x^{(k-2k_0+1)}$.
Therefore the second compatibility condition in \equ(2.3) can be
fulfilled by properly fixing $\b_{k-2k_0+1}$ in terms of $X^{(h)}$
with $h\le k-k_0$ and of $\x^{(h)}$ with $h\le k-2 k_0$,
{\it provided $\b_1$ exists and $\b_1\ne0$}, \ie {\it provided} $a\ne0$ as assumed here.

This means that if $k_0$ is odd all $\b_k$ are uniquely determined
while if $k_0$ is even there are two possible sequences $\b_k$
depending on the two choices for $\b_1$ in \equ(2.8). \qed

\*

\0{\it Remarks.} (1) For $k_0$ even the choice of $\b_k$
will be possible only if $\e$ has sign $\s$ such that $-\s\,a/c>0$.
\\
(2) With the notation \equ(2.1) one has $X^{(k)}_{\V0}=
(\AA^{(k)}_{\V0},B^{(k)}_{\V0},0,0)$ as $\V a^{(k)}_{\V0}=\V0$
and $b^{(k)}_{\V0}=\xi^{(k)}$.

\*\*
\section(3,Tree formalism and formal series)

\figini{in1}
\8<
\8</punto { 
\8<gsave 3 0 360 newpath arc fill stroke grestore} def>
\8<>
\8</linea { 
\8<gsave 4 2 roll moveto lineto stroke grestore} def>
\8<>
\8<
\8< /origine1assexper2pilacon|P_2-P_1| { >
\8< 4 2 roll 2 copy translate exch 4 1 roll sub >
\8< 3 1 roll exch sub 2 copy atan rotate 2 copy >
\8< exch 4 1 roll mul 3 1 roll mul add sqrt } def >
\8< >
\8< /punta0 {0 0 moveto dup dup 0 exch 2 div lineto 0 >
\8< lineto 0 exch 2 div neg lineto 0 0 lineto fill >
\8< stroke } def >
\8< >
\8< /dirpunta{
\8< gsave origine1assexper2pilacon|P_2-P_1| >
\8< 0 translate 7 punta0 grestore} def >
\8<>
\8< /puntatore {
\8< 4 copy 4 copy 4 copy 4 copy >
\8< dirpunta 4 2 roll moveto lineto stroke} def >
\8< >
\8<
\8<>
\8</L {50} def>
\8</H {3} def>
\8<>
\8</P0 {0 H} def /P1 {L H} def /Pi {L 2 div H} def>
\8<>
\8<P0 Pi linea P1 Pi puntatore Pi P1 linea>
\8<>
\8<P1 punto>
\figfin

\figini{in2}
\8<
\8</puntovuoto { 
\8<gsave 3 0 360 newpath arc stroke grestore} def>
\8<>
\8</linea { 
\8<gsave 4 2 roll moveto lineto stroke grestore} def>
\8<>
\8<
\8< /origine1assexper2pilacon|P_2-P_1| { >
\8< 4 2 roll 2 copy translate exch 4 1 roll sub >
\8< 3 1 roll exch sub 2 copy atan rotate 2 copy >
\8< exch 4 1 roll mul 3 1 roll mul add sqrt } def >
\8< >
\8< /punta0 {0 0 moveto dup dup 0 exch 2 div lineto 0 >
\8< lineto 0 exch 2 div neg lineto 0 0 lineto fill >
\8< stroke } def >
\8< >
\8< /dirpunta{
\8< gsave origine1assexper2pilacon|P_2-P_1| >
\8< 0 translate 7 punta0 grestore} def >
\8<>
\8< /puntatore {
\8< 4 copy 4 copy 4 copy 4 copy >
\8< dirpunta 4 2 roll moveto lineto stroke} def >
\8< >
\8<
\8<>
\8</L {50} def>
\8</H {3} def>
\8<>
\8</P0 {0 H} def /P1 {L H} def /Pi {L 2 div H} def>
\8<>
\8<P0 Pi linea P1 exch 4 sub exch Pi puntatore Pi P1 exch 3 sub exch linea>
\8<>
\8<P1 puntovuoto>
\figfin

\figini{in3}
\8<
\8</punto { 
\8<gsave 3 0 360 newpath arc fill stroke grestore} def>
\8<>
\8</linea { 
\8<gsave 4 2 roll moveto lineto stroke grestore} def>
\8<>
\8<
\8< /origine1assexper2pilacon|P_2-P_1| { >
\8< 4 2 roll 2 copy translate exch 4 1 roll sub >
\8< 3 1 roll exch sub 2 copy atan rotate 2 copy >
\8< exch 4 1 roll mul 3 1 roll mul add sqrt } def >
\8< >
\8< /punta0 {0 0 moveto dup dup 0 exch 2 div lineto 0 >
\8< lineto 0 exch 2 div neg lineto 0 0 lineto fill >
\8< stroke } def >
\8< >
\8< /dirpunta{
\8< gsave origine1assexper2pilacon|P_2-P_1| >
\8< 0 translate 7 punta0 grestore} def >
\8<>
\8< /puntatore {
\8< 4 copy 4 copy 4 copy 4 copy >
\8< dirpunta 4 2 roll moveto lineto stroke} def >
\8< >
\8<
\8<>
\8</L {50} def>
\8</H {4} def>
\8<>
\8</P0 {0 H} def /P1 {L H} def /Pi {L 2 div H} def>
\8<>
\8<P0 Pi linea P1 Pi puntatore Pi P1 linea>
\figfin

\0Given that the formal Lindstedt series is well defined, by proceeding
as in \cita{G2} a graphical representation of the contributions
to the perturbative series will be introduced.  The idea to get the
rules explained below is to start by representing the \rhs of
\equ(2.2) as a power series in $\h$ by simply expanding the functions
$\dpr_\f f,\dpr_{I}H$ in their arguments around $X^{(0)}(\Bps)$
and then each argument again in powers until
one obtains a power series in $\h$.

Since the arguments of $f$ depend on the $X^{(h)}$ with $h<k$ we
obtain recursively a natural representation in terms of
trees. Consider $(X^{(k)}_\Bn)_\g$ where $\g$ ranges in the symbols list
$$\AAA\defin\{A_1,\ldots,A_{N-1},B,\a_1,\ldots,\a_{N-1},\b\}\;.\Eq(3.1)$$
This notation is more convenient than the alternative one which would
simply label the components with a label $\g=1,2,\ldots,N,N+1,\ldots,
2N$ because the first $N-1$ components play a very different role than
the $N$-th or the remaining ones. It can be disturbing as the labels
in \equ(3.1) have the meaning of canonical variables when they do not
appear as labels: hence the reader should keep this in mind in what
follows.  Occasionally, only in cases of possible confusion, we shall
use the labels $1,\ldots,2N$ instead of \equ(3.1).

It will be convenient to call the first $N$ labels {\it action
components} and the last $N$ {\it angle components}; the first $N-1$
of the two groups will be called, respectively, {\it fast actions} and
{\it fast angles} component labels while the last will be called a
{\it slow action} or, respectively, {\it slow angle} component label. When,
occasionally, it will turn out to be useful, we shall denote the $N$
action labels simply by $I$ and the angle labels by $\f$.  If
$\g\in\AAA$ and $\g$ is one of the first $N$ labels (\ie it is an
action component), then $\g+N$ will indicate the corresponding label
of the angle components and vice versa if $\g\in\AAA$ is an angle
component then $\g-N$ will denote the corresponding label in the group
of action components.
With a slight abuse of notation, given $\g\in\AAA$, we shall write
$\g \in I$ if $\g$ is an action component and $\g\in \f$ if it is
an angle component. If $\g,\g'\in \AAA$ then $\d_{\g\g'}$ denotes
the Kronecker delta; we shall also use the notation
$$ \d_{\g\f} = \cases{
1, & if $\g\in \f$ , \cr 0, & if $\g\in I$ , \cr} \qquad
\d_{\g I} = \cases{
1, & if $\g\in I$ , \cr 0, & if $\g\in \f$ . \cr}
\Eq(3.2) $$
Looking at \equ(2.2) we see that, for $\Bn\ne\V0$ and $k\ge k_0$,
we can write it as
$$\eqalignno{
(X^{(k)}_\Bn)_\g=\fra{\d_{\g\g'}}{i\Bo\cdot\Bn}
&\Big[\s\sum_{p\ge 1}
\sum_{\Bn_0+\Bn_1+\ldots+\Bn_p=\Bn\atop k_1+\ldots k_p=k-k_0}
\fra1{p!} (E\dpr)_{\g'} \dpr_{\g_1\ldots \g_p}f_{\Bn_0}(\V0,0,\b_{0})
\prod_{n=1}^p (X^{(k_n)}_{\Bn_n}+\x^{(k_n)}\d_{\Bn_n\V0})_{\g_n}+
\cr
&+\sum_{p\ge 1}\sum_{\Bn_1+\ldots+\Bn_p=\Bn\atop k_1+\ldots k_p=k}
\fra1{p!}
(E\dpr)_{\g'}\dpr_{\g_1\ldots \g_p} H_0(\V0,0)\prod_{n=1}^p
(X^{(k_n)}_{\Bn_n})_{\g_n}\Big] . &\eq(3.3)\cr} $$
Here the symbols $\dpr_\g$ have to be interpreted
as derivatives of $f_\Bn$ with respect to action arguments of $f$ if
$\g$ is an action component, as derivatives with respect to $\b$
if $\g=\b$ and as multiplications by $i\Bn_i$ if $\g=\a_i$,
$i=1,\ldots,N-1$. {\it A summation convention is conveniently adopted for
pairs of repeated component labels.} In the second line no
$\x^{(k_n)}$ appears because $\x^{(h)}$ has only angle components but
$H_0$ does not depend on the angles.

For $\Bn=\V0$ and $\g\in I$ (i.e. $\g$ an action component;
\cfr Remark (2) at the end of Section \secc(2)),
we use \equ(2.6) and write
$$\eqalignno{
&(X^{(k)}_{\V0})_\g=- \left( \dpr^2_I H_0(\V0,0)^{-1} \right)_{\g,\g'-N}
\d_{\g'\f}\Big[\sum_{p=2}^{\io} \sum_{k_1+\ldots k_p=k\atop
\Bn_1+\ldots+\Bn_p=\V0} \fra{1}{p!}(E\dpr)_{\g'}
\dpr_{\g_1\ldots \g_p} H_0(\V0,0)\prod_{n=1}^p
 (X^{(k_n)}_{\Bn_n})_{\g_n}+
\cr
&+\s\sum_{p=1}^{\io} \sum_{k_1+\ldots k_p=k-k_0\atop
\Bn_0+\Bn_1+\ldots+\Bn_p=\V0} \fra{1}{p!}
(E\dpr)_{\g'}\dpr_{\g_1\ldots \g_p}
f_{\Bn_0}(\V0,0,\b_{0}) \prod_{n=1}^p
 (X^{(k_n)}_{\Bn_n}+\x^{(k_n)}\d_{\Bn_n\V0})_{\g_n}
\Big]\, .
& \eq(3.4) \cr} $$

For $\Bn=\V0$ and $\g$ an angle label the \equ(2.2) are identically
satisfied if the $\x^{(h)}$ are determined as prescribed in
Section \secc(2). The equation that fixes $\x^{(h)}$ is again a relation
of the type of \equ(3.3) and \equ(3.4) as we shall discuss in detail later.

Proceeding as in \cita{G2} and \cita{GG1} we represent $(X^{(h)}_\Bn)_\g$
as \
${}_\g$\includegraphics{in1.ps}\kern27pt\raise7pt\hbox{$\Bn$}\kern23pt${}^{(k)}$
and $\x^{(h)}\=(\V0,0,\V0,$ $\b^{(k)})$ as
${}_\g$\includegraphics{in2.ps}\kern25pt\raise7pt\hbox{$\V0$}\kern25pt${}^{(k)}$
and we realize that \equ(3.3) and \equ(3.4) can be very conveniently
(as it turns out) represented by graphs of the type represented in Figure 1.

\*
\figini{fig1}
\8<
\8</punto { 
\8<gsave 3 0 360 newpath arc fill stroke grestore} def>
\8</puntino { 
\8<gsave 2 0 360 newpath arc fill stroke grestore} def>
\8<>
\8</puntovuoto { 
\8<gsave 3 0 360 newpath arc stroke grestore} def>
\8<>
\8</puntox {
\8</y2 exch def /x2 exch def /y1 exch def /x1 exch def /x exch def>
\8<x1 x2 x1 sub x mul add>
\8<y1 y2 y1 sub x mul add} def>
\8<>
\8</linea { 
\8<gsave 4 2 roll moveto lineto stroke grestore} def>
\8<>
\8</tlinea {gsave 4 2 roll moveto lineto [4 4] 2 >
\8<setdash stroke [] 0 setdash grestore} def >
\8<>
\8<
\8< /origine1assexper2pilacon|P_2-P_1| { >
\8< 4 2 roll 2 copy translate exch 4 1 roll sub >
\8< 3 1 roll exch sub 2 copy atan rotate 2 copy >
\8< exch 4 1 roll mul 3 1 roll mul add sqrt } def >
\8< >
\8< /punta0 {0 0 moveto dup dup 0 exch 2 div lineto 0 >
\8< lineto 0 exch 2 div neg lineto 0 0 lineto fill >
\8< stroke } def >
\8< >
\8< /dirpunta{
\8< gsave origine1assexper2pilacon|P_2-P_1| >
\8< 0 translate 7 punta0 grestore} def >
\8<>
\8< /puntatore {
\8< 4 copy 4 copy 4 copy 4 copy >
\8< dirpunta 4 2 roll moveto lineto stroke} def >
\8< >
\8<
\8<>
\8</L {50} def>
\8</LL {70} def>
\8</H {90} def>
\8<>
\8</P0 {0 H 2 div} def>
\8</P1 {L H 2 div} def>
\8</P2 {L LL add H} def>
\8</P3 {L LL add 0} def>
\8</P4 {1 5 div P2 P3 puntox} def>
\8</P5 {2 5 div P2 P3 puntox} def>
\8</P6 {3 5 div P2 P3 puntox} def>
\8</P7 {4 5 div P2 P3 puntox} def>
\8<>
\8</Q0 {.5 P0 P1 puntox} def>
\8</Q2 {.45 P2 P1 puntox} def>
\8</Q3 {.45 P3 P1 puntox} def>
\8</Q4 {.45 P4 P1 puntox} def>
\8</Q5 {.45 P5 P1 puntox} def>
\8</Q6 {.4 P6 P1 puntox} def>
\8</Q7 {.45 P7 P1 puntox} def>
\8<>
\8<P0 P1 linea >
\8<P2 P1 linea>
\8<
\8<P4 P1 tlinea>
\8<
\8<P5 P1 linea>
\8<.05 P6 P1 puntox P1 linea>
\8<.05 P7 P1 puntox P1 tlinea>
\8<.04 P3 P1 puntox P1 linea>
\8<>
\8<P1 puntino>
\8<P2 punto>
\8<P4 punto>
\8<P5 punto>
\8<P6 puntovuoto>
\8<P7 puntovuoto>
\8<P3 puntovuoto>
\8<>
\8<P1 Q0 dirpunta>
\8<P2 Q2 dirpunta>
\8<P6 Q6 dirpunta>
\8<P3 Q3 dirpunta>
\8<P5 Q5 dirpunta>
\figfin

\*
\eqfig{140pt}{90pt}
{\ins{0pt}{42pt}{$\st \g$}
\ins{45pt}{63pt}{$\st\d_0$}
\ins{45pt}{55pt}{$\st\Bn_0$}
\ins{23pt}{57pt}{$\Bn$}
\ins{37pt}{43.5pt}{$\st \g'$}
\ins{50pt}{34pt}{$\st \g_{p+q}$}
\ins{57pt}{62pt}{$\st \g_1$}
\ins{127pt}{100pt}{$\st(k_1)$}
\ins{129pt}{91pt}{}
\ins{83pt}{80pt}{$\st\Bn_1$}
\ins{129pt}{72pt}{}
\ins{127pt}{63pt}{$\st(k_p)$}
\ins{129pt}{54pt}{}
\ins{127pt}{44pt}{$\st(k_{p+1})$}
\ins{129pt}{35pt}{}
\ins{80.5pt}{47.5pt}{$\st\Bn_p$}
\ins{82pt}{39pt}{$\st\V0$}
\ins{127pt}{6pt}{$\st(k_{p+q})$}
\ins{129pt}{-3pt}{}
\ins{85pt}{13pt}{$\st\V 0$}
}{fig1}{}

\*\*
\line{\vtop{\line{\hskip2.0truecm\vbox{\advance\hsize by -4.1 truecm
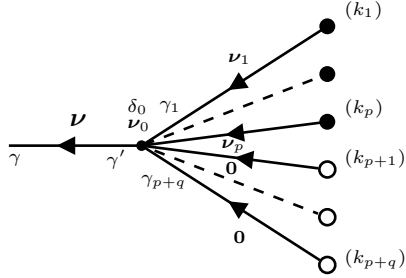
\0{{\css Figure 1.} {\nota Graphical representation of equations \equ(3.3)
and \equ(3.4). For $\st\Bn\neq\V0$ the graph represents \equ(3.3),
while for $\st\Bn=\V0$ it represents \equ(3.4).\vfil}} } \hfill} }}
\*\*

The label $\d_0=0,1$, that we call the {\it degree label} of the node
to which it is associated, identifies the terms in \equ(3.3) or
\equ(3.4) which contain inside the square brackets $H_0$ (then
$\d_0=0$) or $f$ (then $\d_0=1$). The other label attached to the
first node indicates the harmonic $\Bn_0$ (mode label)
selected in the terms with
$f$ (\ie with $\d_0=1$) and we take $\Bn_0=\V0$ if $\d_0=0$ (because
$H_0$ does not depend on the angles).  A further component label $\g'$
is attached to the right extreme of the line exiting the central node
in Figure 1 and it indicates the derivative $(E\dpr)_{\g'}$
in \equ(3.3) and \equ(3.4). We call a component label a ``right'' or ``left''
component label if it is attached at the beginning or at the end of
the (oriented) line.

The component labels attached to the first node determine
the components of the tensors defined by the derivatives of
$f_{\Bn_0}$ or of $H_0$. The labels attached to the left extreme of
each endline (to the right of the bifurcation point in the figure)
determine which component of $(X^{(k_n)}_{\Bn_n})_{\g_n}$ is taken in
the products of $X$'s in \equ(3.3) or \equ(3.4). Finally the root line
symbolizes the factor that is outside the square brackets in the
expression \equ(3.3) or in the expression
\equ(3.4): it will be called {\it propagator} of the
line \ ${}_\g$\kern-6pt\hbox{\includegraphics{in3.ps}\kern27pt%
\raise7pt\hbox{$\Bn$}\kern11pt${}_{\g'}$\hglue10pt}. Hence the
propagator of the line will be the matrix
$$\widetilde
g_{\g\g'}(\Bn)\ = \cases{
(i\Bo\cdot\Bn)^{-1}\,\d_{\g\g'} ,
& if $\Bn\ne\V0$, \cr
- \left( \dpr^2_I H_0(\V0,0)^{-1} \right)_{\g,\g'-N}
\d_{\g I}\d_{\g'\f} ,
& if $\Bn=\V0$, \cr}
\Eq(3.5) $$
and $\nn$ will be called the {\it momentum} of the line.

Finally the endpoints are divided into endpoints representing
$X^{(h)}$ for some $h$, marked by bullets, and others representing
$\x^{(h)}$ marked by white disks that we shall call {\it leaves}.

It follows from the analysis of Section \secc(2) that,
setting $\b_1$ equal to the constant in \equ(2.8) and
replacing $\fra{1}{(k_0-1)!}\dpr_\b^{k_0+1}f_{\V0}(\V0,0.\b_{0})$
with $k_0\, c$ by the definition of $c$ before \equ(2.7),
the coefficients $\b_h$ with $h\ge2$ can be derived
in terms of $\b_1$ from the relation
$$ \b_{k-2k_0+1}=\fra1{k_0\, c\,\b_1^{k_0-1}}
\sum_{s\ge1} \mathop{{\sum}^*}_{k_1+\ldots+k_s=k-k_0
\atop \Bn_0+\Bn_1+\ldots+\Bn_s=\V0}\fra1{s!} (E\dpr)_B\dpr_{\g_1\ldots \g_s}
f_{\Bn_{\V0}}\prod_{m=1}^s
(X^{(k_m)}_{\Bn_m}+\x^{(k_m)}\d_{\Bn_m\V0} )_{\g_m},
\Eq(3.6)$$
where the derivatives of $f_{\Bn_0}$ are, as above, evaluated at
$(\V0,0,\b_{0})$ and the $*$ on the sum recalls that we are excluding
from the sum the contribution equal to $-k_0\,
c\,\b_1^{k_0-1}\b_{k-2k_0+1}$.

At this point we can repeat the construction and replace each endpoint
of Figure 1 ending in a node or in a leaf with a node into which merge
several lines each of which comes out of a node or a leaf with some
labels $(h),\Bn,\g$ representing $(X^{(h)}_\Bn)_\g$ for some
$h,\Bn,\g$ or, in the case of leaves, $(\x^{(h)})_\g$.

Using also the relations \equ(3.4) and \equ(3.6), the construction can
be iterated until we are left with a tree $\th$ whose endpoints either carry
a degree label $1$ or are leaves representing $\b_1$: we denote by
$L(\th)$ the set of endpoints representing such leaves.  The
constraint \equ(3.6) can be automatically implied by imagining that
the propagator of a line $\ell$ with momentum $\Bn$ is
$$ g_{\g\g'}(\Bn) = \cases{
(i\Bo\cdot\Bn)^{-1} \,\d_{\g\g'} , & if $\Bn\ne\V0$, \cr
\d_{\g\g'} , & if $\Bn =\V0, \, \vvv\in L(\th)$, \cr
- {\displaystyle
\left( \dpr^2_I H_0(\V 0,0)^{-1} \right)_{\g,\g'-N}
\d_{\g I} \d_{\g'\f} +
\fra{\d_{\g\b}\d_{\g' B}}{\s\h^{2k_0-1}{k_0\,c\,\b_1^{k_0-1}}} } , &
if $\Bn=\V0, \, \vvv\not\in L(\th)$, \cr}
\Eq(3.7) $$
where $\vvv$ is the node preceding $\ell$ on $\t$.  
We can write $(X_\Bn)_\g$ as a formal power series in $\h$, whose terms
can be computed in terms of tree values.
Let a tree $\th$ be a tree diagram with nodes which look
like the node drawn in Figure 1 and let
$V(\th)$, $L(\th)$ and $\L(\th)$ denote the sets of nodes, leaves
and lines of $\th$, respectively. The {\it tree value} $\Val(\th)$
will be a monomial in $\h$ obtained by multiplying\\
(1) a factor
$$\eqalign{
[F(\nn_{\vvvvv})]_{\g'_{\ell_{\vvvvv}},\Bg(\vvvv)}
\, = \, & \s\h^{k_0}(E\dpr)_{
\g'_{\ell_{\vvvvv}}}
\dpr_{\g_{\vvvvv 1}\ldots \g_{\vvvvv p_{\vvvvv}}}
f_{\Bn_{\vvvvv}}(\V0,0,\b_{0}),\qquad {\rm if\ } \d_{\vvvv}=1 , \cr
[F(\nn_{\vvvvv})]_{\g'_{\ell_{\vvvvv}},\Bg(\vvvv)}
\, = \, & (E\dpr)_{\g'_{\ell_{\vvvvv}}}
\dpr_{\g_{\vvvvv 1}\ldots \g_{\vvvvv p_{\vvvvv}}}
H_0(\V0,0),\qquad \kern13mm{\rm if\ } \d_{\vvvv}=0 , \cr}
\Eq(3.8) $$
per each node $\vvv\in V(\th)$ into which merge $p_{\vvvv}\ge1$ lines
carrying component labels $\Bg(\vvv)=(\g_{\vvvv 1} \ldots
\g_{\vvvv p_{\vvvv}})$ and emerges a line $\ell_{\vvvv}$
carrying a label $\g'_{\ell_{\vvvvv}}$;
\\
(2) a factor $\s\h^{k_0}(E\dpr)_{\g'_{\ell_{\vvvvv}}} f_{\Bn_{\vvvvv}}
(\V0,0,\b_{0})$ per each endnode $\vvv\not\in L(\th)$
(note that necessarily $\d_{\vvvv}=1$);
\\
(3) a factor $\h \b_1$ per each leaf $\vvv\in L(\th)$ (note that
necessarily $\nn_{\vvvv}=\V0$ and $\g_{\ell_{\vvvvv}}=B$);
\\
(4) a factor $g_{\g_{\ell}\g'_{\ell}}(\Bn)$ per line
$\ell\in \L(\th)$, given by \equ(3.7).

Note that the construction described above and in Section \secc(2)
forbids  the presence in the tree diagrams of some configurations of
nodes. More precisely, calling {\it trivial} the nodes $\vvv$
with $p_{\vvvv}=1$  and $\nn_{\vvvv}=\V0$ and $b$--{\it trivial}
the nodes $\vvv$ with $p_{\vvvv}=k_0$, $\nn_{\vvvv}=\V0$,
$\g'_{\ell_{\vvvvv}}=B$ and immediately preceded by at least
$k_0-1$ leaves, the following configurations of nodes:
\\
{(i)} {\it trivial nodes with $\d_{\vvvv}=0$ and the entering line
with $\V0$ momentum} and
\\
{(ii)} {\it $b$--trivial nodes with the exiting line with $\V0$
momentum and all the entering lines
with left component labels equal to $\b$}
\\
are not allowed, in the sense that $\Val(\th)=0$ if $\th$ contains such
configurations of nodes.
The reason why the trees containing such configurations of
nodes are forbidden is a consequence of the use of the relations
\equ(3.4) and \equ(3.6). Trees with no forbidden node will be
called {\it allowed trees}.
\*

The result is that the perturbed motion runs on a $(n-1)$-dimensional
torus whose equations are formally written
as a sum of values of (allowed) tree diagrams, computable by using
rules very similar to those listed in \cita{GG1}, with the value of an
allowed tree $\th$ defined as
$$\Val(\th)={1\over |\L(\th)|!}\Big(\prod_{\vvvv\in V(\th)}
\e^{\d_{\vvvvv}}\Big) (\h\b_1)^{|L(\th)|}
\Big(\prod_{\vvvv\in V(\th)}F_{\vvvv}\Big)
\Big(\prod_{\ell\in \L(\th)}G_\ell\Big) ,
\Eq(3.9)$$
where $G_\ell\defi g_{\g_\ell\,\g'_\ell}(\Bn_\ell)$ with
$\nn_\ell=\nn_{\ell_{\vvvvv}}=\sum_{\wwww\le \vvvv}\nn_{\vvvv}$, while
$F_{\vvvv}$ is the node factor (a tensor) defined in items (1) and (2)
above. In the product all the labels are summed over, except for the
root label $\g_{\ell_0}$, where $\ell_0$ is the line entering the
root. We call $\Th^{o}_{k,\nn,\g}$ the set of trees with {\it degree}
$|L(\th)|+k_0 \sum_{\vvvv\in V(\th)}\d_{\vvvv}-
(2k_0-1)\sum_{\ell\in\L(\th)} \d_{\Bn_\ell\V0} \d_{\g_\ell\b}
\d_{\g'_{\ell }B } =k$, $\nn_{\ell_0}=\nn$ and $\g_{\ell_0}=\g$ (these
are trees whose value is proportional to $\h^k$). As in \cita{GG2}, we
denote by $\Th_{k,\nn,\g}$ the set of trees with $k$ nodes, and with
labels $\nn_{\ell_0}=\nn$ and $\g_{\ell_0}=\g$ associated with the
root line.

The definitions above are given so that the formal
series for $X_\g$ is given by the sum
$$ X^{(0)}_{\g} + \sum_{\Bn\in\zzz^{N-1}}
{\rm e}^{i\Bn\cdot (\pps + \Bo t)} \sum_{k=1}^{\io}
\sum_{\th\in\Th_{k,\nn,\g}} \Val(\th) .
\Eq(3.10)$$

The above rules give, order by order in powers of $\h$, the solution
of the perturbed equations of motion. In particular one can check that
the trees contributing a monomial in $\h$ of degree $k\ge1$
to the conjugating function have a number of lines that is bounded
above and below  proportionally to $k$. This is a property
extensively used in the convergence analysis, for instance
to show that the number of non-numbered trees of degree $k$ is
bounded by a constant to the power $k$; \cfr \cita{GG2} for details.
An explicit bound is $\ge k/k_0$ and $\le 3k_0 k$, see Appendix \secc(A3).

\*\*
\section(4, Elimination of the trivial nodes)

\0The problem of proving convergence of the series just defined is very
similar to that treated in \cita{GG2}. As in \cita{GG2} the difficulty is
that even exploiting the cancellations analogous to those of the
maximal tori case, we are left with tree graphs containing chains of
subdiagrams with one entering and one exiting lines, carrying the same
momentum, that we call ``self--energy (sub)diagrams''. Naively such
subdiagrams are the source of bad bounds on the $k$--th the
contribution proportional to $\h^k$ to the
series.  Proceeding as in in \cita{GG2} we iteratively resum such
chains into ``renormalized propagators'' and change step by step the
structure of the perturbation series. At each step we define different
rules to compute the tree values: we assign at each line $\ell$ a
scale label $[n_\ell]$, with $n_{\ell}\ge -1$, depending on the size
of its propagators, and, at the $n$--th step ($n=0,1,\ldots$),
we will not allow trees containing chains of self-energy diagrams
on scale $\ge n-1$; at the same time we will assign to each line
a propagator different from that in \equ(3.7), depending on the
value of its scale label, as explained below.

\*
\figini{fig2}
\8<
\8</punto { 
\8<gsave 3 0 360 newpath arc fill stroke grestore} def>
\8<>
\8</puntino { 
\8<gsave 1.5 0 360 newpath arc fill stroke grestore} def>
\8<>
\8</puntovuoto { 
\8<gsave 3 0 360 newpath arc stroke grestore} def>
\8<>
\8</puntox {
\8</y2 exch def /x2 exch def /y1 exch def /x1 exch def /x exch def>
\8<x1 x2 x1 sub x mul add>
\8<y1 y2 y1 sub x mul add} def>
\8<>
\8</linea { 
\8<gsave 4 2 roll moveto lineto stroke grestore} def>
\8<>
\8</tlinea {gsave 4 2 roll moveto lineto [4 4] 2 >
\8<setdash stroke [] 0 setdash grestore} def >
\8<>
\8<
\8< /origine1assexper2pilacon|P_2-P_1| { >
\8< 4 2 roll 2 copy translate exch 4 1 roll sub >
\8< 3 1 roll exch sub 2 copy atan rotate 2 copy >
\8< exch 4 1 roll mul 3 1 roll mul add sqrt } def >
\8< >
\8< /punta0 {0 0 moveto dup dup 0 exch 2 div lineto 0 >
\8< lineto 0 exch 2 div neg lineto 0 0 lineto fill >
\8< stroke } def >
\8< >
\8< /dirpunta{
\8< gsave origine1assexper2pilacon|P_2-P_1| >
\8< 0 translate 7 punta0 grestore} def >
\8<>
\8< /puntatore {
\8< 4 copy 4 copy 4 copy 4 copy >
\8< dirpunta 4 2 roll moveto lineto stroke} def >
\8< >
\8<
\8<>
\8</L {50} def>
\8</LL {70} def>
\8</H {90} def>
\8<>
\8</B0 {0 H 2 div} def>
\8</B1 {L H 2 div} def>
\8</B2 {L 2 mul H 2 div} def>
\8</C0 {.5 B0 B1 puntox} def>
\8</C1 {.5 B1 B2 puntox} def>
\8<>
\8<B0 B1 linea >
\8<B2 B1 linea>
\8<B1 C0 dirpunta>
\8<B2 C1 dirpunta>
\8<
\8<B1 puntino>
\8<>
\8<130 0 translate>
\8<>
\8</P0 {0 H 2 div} def>
\8</P1 {L H 2 div} def>
\8</P2 {L LL add H} def>
\8</P3 {L LL add 0} def>
\8</P4 {1 5 div P2 P3 puntox} def>
\8</P5 {2 5 div P2 P3 puntox} def>
\8</P6 {3 5 div P2 P3 puntox} def>
\8<
\8<>
\8</Q0 {.5 P0 P1 puntox} def>
\8</Q2 {.45 P2 P1 puntox} def>
\8</Q3 {.45 P3 P1 puntox} def>
\8</Q4 {.45 P4 P1 puntox} def>
\8</Q5 {.45 P5 P1 puntox} def>
\8</Q6 {.4 P6 P1 puntox} def>
\8<>
\8<P0 P1 linea >
\8<.05 P6 P1 puntox P1 linea>
\8<.05 P5 P1 puntox P1 tlinea>
\8<.05 P4 P1 puntox P1 tlinea>
\8<.04 P3 P1 puntox P1 linea>
\8<.04 P2 P1 puntox P1 linea>
\8<P2 puntovuoto  P4 puntovuoto P5 puntovuoto P6 puntovuoto >
\8<>
\8<P1 puntino>
\8<>
\8<P1 Q0 dirpunta>
\8<P2 Q2 dirpunta>
\8<P6 Q6 dirpunta>
\8<P3 Q3 dirpunta>
\figfin

\eqfig{250pt}{95pt}
{
\ins{-15pt}{50pt}{(a)}
\ins{22pt}{58pt}{$\st\Bn$}
\ins{70pt}{58pt}{$\st\Bn$}
\ins{152pt}{56pt}{$\st\Bn$}
\ins{112pt}{50pt}{(b)}
\ins{178pt}{58pt}{$\st1$}
\ins{47pt}{58pt}{$\st \d$}
\ins{46pt}{40pt}{$\V0$}
\ins{176pt}{40pt}{$\V0$}
\ins{33pt}{38pt}{$\st\g$}
\ins{58pt}{41pt}{$\st\g'$}
\ins{168pt}{40pt}{$\st\g$}
\ins{184pt}{35pt}{$\st\g'$}
\ins{186pt}{61pt}{$\st\g_1$}
\ins{211pt}{77pt}{$\st\V0$}
\ins{255pt}{92pt}{$\st1$}
\ins{255pt}{79pt}{$\st2$}
\ins{255pt}{40pt}{$\st k_0-1$}
\ins{255pt}{6pt}{$\st k_0$}
\ins{216pt}{37pt}{$\st\V0$}
\ins{211pt}{15pt}{$\st\Bn$}
}
{fig2}{}

\*\*
\line{\vtop{\line{\hskip2.0truecm\vbox{\advance\hsize by -4.1 truecm
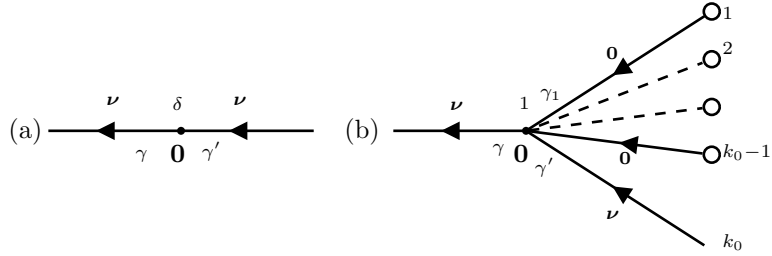
\0{{\css Figure 2.} {\nota Examples of lowest order self-energy clusters
on scale $[-1]$. In (a) one has $\g=B$ and $\g' \in I$
or vice versa, while in (b) one has $\g=B$ and $\g'=\b$.
The order label $\d$ in the first graph can be $\d=0,1$.
The leaves in the second graph represent factors $\b_1$ so that they
contribute to the order of the self-energy clustser a power
$\h^{k_0-1}$.\vfil}} } \hfill} }}
\*\*


The first step is the removal of the trivial nodes and of a class of
few other subgraphs that can be present
in {\it allowed trees}, see Section \secc(3), whose value is not $0$.
Assign, in a way similar to that described in \cita{GG2},
a scale label $[-1]$ to the $\V0$--momentum propagator and
let $\SS^\RR_{k,-1}$ be the set of self-energy clusters $T$ on
scale $[-1]$ and degree $k$
(\ie such that $k=|L(T)|+\sum_{\vvv\in V(T)}\d_{\vvvv} \,k_0
-(2k_0-1)\sum_{\ell\in\L(T)} \d_{\Bn_\ell\V0} \d_{\g_\ell\b}
\d_{\g'_{\ell }B }$), consistently with the notion of degree 
defined after \equ(3.9).

A {\it cluster on scale $[-1]$} is either a single node or a maximal
connected set of nodes and lines, such that all the lines are on scale
$[-1]$; the lines of non-zero momentum that enter or exit the nodes of
the cluster are called {\it external lines}.  A {\it self-energy
cluster on scale $[-1]$} is defined as a cluster on scale $[-1]$ such
that there are only two external lines (one exiting and one entering),
and they are connected to the same node of the cluster. This means
that a self-energy cluster is formed by a node $\vvv$ with $\V0$-mode
together with all lines (if any) of momentum $\V0$, hence on scale
$[-1]$, which are linked to $\vvv$ by a path consisting of lines of
momentum $\V0$; see Figure 2. {\it A cluster on scale $[-1]$ with only
one exiting and one entering line, which are not connected to the same
node, will not be considered a self-energy cluster on scale $[-1]$.}

The self-energy value, i.e. the value $\VV_{T}(\h)$ of the self-energy
cluster $T$ on scale $[-1]$, is defined, similarly to \equ(3.9), as
$$\VV_{T}(\h)={1\over |\L(T)|!}
\Big(\prod_{\vvvv\in V(T)}\e^{\d_{\vvvvv}}\Big) (\h\b_1)^{|L(T)|}
\Big(\prod_{\vvvv\in V(T)}F_{\vvvv}\Big)
\Big(\prod_{\vvvv\in \L(T)}G_\ell\Big) ,
\Eq(4.1)$$
where $V(T)$, $L(T)$ and $\L(T)$ are the set of nodes,
leaves and lines, respectively, contained inside $T$.
Then we can define
$$\MM^{[0]}(\h) = \sum_{k=k_0}^{\io}\sum_{T \in\SS^\RR_{k,-1} }
\VV_{T}(\h) ,
\Eq(4.2)$$
where the sum runs over all the self-energy clusters on scale $[-1]$.

\*

\0{\it Remarks.}
(1) The clusters $T$ in \equ(4.1) and \equ(4.2) with $k=k_0$ correspond
to the trivial nodes appearing in the tree expansion of previous
section. In the next section the general notion of cluster will be
introduced.
\\
(2) The clusters on scale $[-1]$ must contain only lines on scale
$[-1]$ and there are infinitely many of them and \equ(4.2) can be
illustrated by the two clusters in Figure 2. The self-energy values
$\VV_T(\h)$ corresponding to the graphs in Figure 2 
give the lowest order(in powers of $\h$) contributions to different
entries of the matrix $\MM^{[0]}$,
as discussed in the caption of Figure 2.
\\
(3) Unlike the case in \cita{GG2}, the equation \equ(4.1), defining
the matrix $\MM^{[0]}(\h)$, is really an infinite series; but it is
still convergent, if $\h$ is sufficiently small. Convergence
of the series defining $\MM^{[0]}$ is a straightforward 
consequence of the fact that a self--energy cluster of degree $k$ has 
a number of lines bounded by constant times $k$, see Appendix \secc(A3),
and of the fact that the propagators of the lines with $\V0$-momentum
can be bounded by an $O(1)$ constant if the line is preceded by a leaf
or if $\g=I_i,\g'=\f_j$ and can be bounded by an $O(1)$ constant times
$\h^{-2k_0+1}$ if the line is not preceded by a leaf and $\g=\b,\g'=B$.

\*

By construction, $\MM^{[0]}(\h)$ is real and has the following 
special structure:
$$ \MM^{[0]}(\h) = \sum_{k = k_0}^{\io}
\sum_{T\in\SS^\RR_{k,-1}}\VV_T= \pmatrix {Q&R\cr P&-Q^\dagger\cr}
\Eq(4.3)$$
where the superscript ${}^\dagger$ denotes Hermitian conjugation and
 $P,Q,R$ are $N\times N$ matrices which to lowest order in $\h$
have the form (with a natural meaning of the symbols, in agreement
with our convention on the component labels \equ(3.1))
$$ P= \dpr_{I}^2 H_0+\e \dpr^2_{I} f_{\V0} , \qquad
Q=\pmatrix{0\cr - \e \dpr_{\b I} f_{\V0}\cr} , \qquad
R=\pmatrix{0&0\cr 0 & -\fra{\e\h^{k_0-1}}{(k_0-1)!}
\dpr^{k_0+1}_\b f_{\V0}\,\b_1^{k_0-1} \cr} ,
\Eq(4.4) $$
where $f_{\V0}$ has $(\V0,0,\b_0)$ as arguments.  The complete
expression, including the higher orders in $\h$, 
simply replaces the {\it non-zero}
terms in \equ(4.4) by convergent series that we can write
$$ P = \dpr_{I}^2 H_0+\e \lis M_{II} , \qquad
Q =\pmatrix{0\cr - \e \lis M_{\b I} \cr} , \qquad
R=\pmatrix{0&0\cr 0 & - \e\h^{k_0-1} \lis M_{\b \b} \cr} ,
\Eq(4.5) $$
where the entries $\lis M_{\g\g'}(\h)$ are bounded uniformly in $\h$,
for $\h$ small enough. They can be computed order by order
using the rules of the last section.
{\it The vanishing entries of the matrix $\MM^{[0]}(\h)$ remain
zero to all orders}: a property which simply follows from the definition
of value of a self--energy cluster and from the remark that a derivative
$\dpr_\g$ with $\g=\aa$ acting on $f_\V0$ must be interpreted as a
multiplication by $\V0$. Note that, since $P,Q,R$ are real and $P,R$ are
symmetric, $\MM^{[0]}(\h)$ satisfies the following symmetry properties:
$$ E\MM^{[0]}(\h)E=\big[\MM^{[0]}(\h)\big]^T,
\qquad \big[\MM^{[0]}(\h)\big]^*=\MM^{[0]}(\h) , \Eq(4.8)$$
where $*$ denotes complex conjugation and $T$ transposition. A  
consequence of \equ(4.8) is that $\MM^{[0]}E$ is Hermitian.

Now, we formally resum the chains of self--energy clusters on scale
$[-1]$ into the new propagator
$$ g^{[\ge 0]}(x;\h)={1\over ix- \MM^{[0]}(\h)} ,
\Eq(4.6)$$
where $x\neq 0$. This means that we modify the tree expansion
described in previous section: the new expansion will involve only
trees not containing self--energy graphs on scale $[-1]$ (and in
particular containing neither trivial nor $b$--trivial nodes) and
with the propagator of the lines with non-zero momentum replaced by
\equ(4.6). It is not clear at all that the new resummed series is
well defined: on the contrary it will become clear that it is
affected by divergence problems similar to those of the original
series. In some sense, we just eliminated a few of the possible source
of problems (that are actually an infinite class of divergent
sub--diagrams).  However, the idea is to begin by eliminating this
first few sources of problems and then, step by step, iteratively
eliminate one after the other all possibile sources of problems (first
the less ``dangerous'' and then the more and more dangerous ones).

Certainly we must at least suppose that $ix-
\MM^{[0]}(\h)$ can be inverted: otherwise the values of the trees
representing the new series might even be meaningless!  To give a
meaning to $(ix-\MM^{[0]}(\h))^{-1}$ it is sufficient to impose $\det(ix
-\MM^{[0]}(\h))\neq 0$ for $x\neq 0$, by eliminating a denumerable dense
set of values of $\h$. One can compute the determinant of $(ix
-\MM^{[0]}(\h))$, finding $\det(ix -\MM^{[0]}(\h)))=-(ix)^{2(N-1)}
\cdot \Big[x^2+\e\h^{k_0-1}\lis M_{\b\b}\big(\dpr^2_{BB}H_0+\e\lis
M_{BB}\big) +\e^2 |\lis M_{\b B}|^2\Big]$, so that the condition
of invertibility of $(ix -\MM^{[0]}(\h))$ for $x\neq 0$ becomes
$$x^2+\e\h^{k_0-1}\lis M_{\b\b}\big(\dpr^2_{BB}H_0+\e\lis M_{BB}\big)
+\e^2 |\lis M_{\b B}|^2\neq 0\Eq(4.7)$$
for all $x=\oo\cdot\nn$, $\nn\in\ZZZ^{N-1}$. In computing $\det(ix
-\MM^{[0]}(\h))$ the property $\lis M_{\b B}=-\lis M_{B\b}\in \RRR$
has been used.

If $\h$ is chosen according to \equ(4.7), we have that the norm of 
$g^{[\ge 0]}(x;\h)$ is equal to the eigenvalue of
$\big[ix-\MM^{[0]}(\h) \big]E$ with smallest absolute value: this is 
because $\|ix-\MM^{[0]}(\h)\|=\|(ix-\MM^{[0]}(\h))E\|$ and, as remarked after 
\equ(4.8), $(ix-\MM^{[0]}(\h))E$ is Hermitian. Here and in the following 
we use the uniform norm: given a $2N\times 2N$ complex matrix $g$, we 
define $\| g \| = \sup_{z\in \ccc^{2N} , |z|=1} | g \, z|$,
where $|z|=\sum_{i=1}^{2N}|z_i|$.

An approximate computation of the eigenvalues of $(ix-\MM^{[0]}(\h))E$, 
see Appendix \secc(A4), implies the following result.

\*

\0{\bf Lemma 2.} {\it There exists an $O(1)$ constant $\r>0$ such
that if $x^2>\r\h^{2k_0-1}$ then the resummed propagator
$g^{[\ge 0]}(x;\h)$ in \equ(4.6) can be bounded as:
$$ \left| g^{[\ge 0]}(x;\h) \right|\le \max\Big\{
{2\m_N^{(0)}\over x^2} , {2 \over \m_{1}^{(0)}} \Big\} ,
\Eq(4.10)$$
where $\m_{1}^{(0)}$ and $\m_{N}^{(0)}$ are, respectively, the
minimum and maximum eigenvalues of $P$, see \equ(4.3) and \equ(4.5).}

\*

{}From now on we shall proceed following \cita{GG2}.
First of all we assume that $|\e|$ is in an interval
$(\lis\e/4,\lis\e]$ such that, by setting
$\lis\e=\lis\h^{k_{0}}$, we can define the integer $n_{0}$ through
$$ C_{0}^{2}2^{-2(n_{0}+1)} < \r\lis\e\,\lis\h^{k_{0}-1}
\le C_{0}^{2}2^{-2n_{0}} ,
\Eq(4.11)$$
with $\r$ as in Lemma 2.

In the first range of scales (in which $x^2 \ge 2C_{0}^{2}2^{-2n_{0}}$)
the small denominators can be bounded by the ``classical''
small divisor $x^2$ and we proceed as described in
the section ``Non--resonant resummations'' of \cita{GG2}.

For smaller scales we shall see below that the small divisor will
bounded below by an $O(1)$ constant times $\min\{x^2,
|x^2+\e\h^{k_0-1}\l^{[n]}(x;\e)| \}$, with $\l^{[n]}(x;\e)$ a suitable
$O(1)$ function. However here the distinction between the {\it
hyperbolic} case ($\e\h^{k_0-1}\l^{[n]}(x;\e)> 0$) and the {\it
elliptic} one ($\e\h^{k_0-1}\l^{[n]}(x;\e)< 0$) has to be made: in
the hyperbolic case the small divisors will be always bounded by an
$O(1)$ constant times $x^2$, even for $x^2\le O(\e\h^{k_0-1})$, while
in the elliptic case we will have to proceed differently, essentially
as described in the section ``Infrared resummations'' of \cita{GG2}.

\*\*
\section(5, Multiscale analysis and non resonant resummation)

\0The resummations  will be defined via trees with no
self--energy clusters on scale $[-1]$ and with lines bearing further
labels. Moreover the definition of propagator will be changed, hence
the values of the trees will be different from the ones in Section
\secc(3): they are constructed recursively as follows.

We introduce a {\it multiscale decomposition} (see \cita{GG2}):
we call $\psi(D)$ a $C^{\io}$ non-decreasing compact support function
defined for $D\ge0$,
$$ \psi(D) =
1 , \quad {\rm for} \quad D \ge C_{0}^2,\qquad \psi(D) =
0 , \quad {\rm for} \quad D \le C_{0}^2/4, \Eq(5.1) $$
where $C_{0}$ is the Diophantine constant of $\oo$, and
let $\chi(D)=1-\ps(D)$.
Define also $\psi_{n}(D)=\psi(2^{2n}D)$ and
$\chi_n(D)=\chi(2^{2n}D)$ for all $n\ge 0$.
Hence $\psi_{0}=\psi,\,\chi_{0}=\chi$ and
$$ 1\=\ps_n(D(x))+\chi_n(D(x)),\qquad\hbox{for all}\ n\ge0 ,
\Eq(5.2)$$
for all choices of the function $D(x)\ge0$: in particular for
$D(x)=x^2$, that we shall now use.

A simple way to represent the value of a tree as sum of many terms
is to make use of the identity in \equ(5.2).
The resummed propagator $g^{[\ge 0]}(x;\h) \defi
(ix-\MM^{[0]}(\h))^{-1}$ of each line with non-zero momentum (hence
with $x\neq0$) is written as
$$ g^{[\ge 0]}(x;\h) = \psi_{0}(x^2) \, g^{[\ge 0]}(x;\h) +
\chi_{0}(x^2)\,g^{[\ge 0]}(x;\h) \defi
g^{[0]}(x;\h) + g^{\big\{\ge 1\big\}}(x;\h) ,
\Eq(5.3)$$
and we note that $g^{[0]}(x;\h)$ vanishes if $x^2$
is smaller than $(C_{0}/2)^2$.

If we replace each $g^{[\ge0]}(x;\h)$ with the sum in \equ(5.3) then
the value of each tree with $k$ nodes is split as a sum of up
to $2^k$ terms which can be identified by affixing
on each line with momentum $\nn\ne\V0$ a label $[0]$ or $\big\{\ge 1\big\}$.
Further splittings of the tree values can be achieved as follows.

\*

\0{\bf Definition 2} (Propagators).
{\it Let $n_0$ be an integer, and for $1 \le p < n_0$,
give $2N\times 2N$ matrices $\MM^{[p]}(x;\h)$
satisfying the symmetry properties
$$E\MM^{[p]}(x;\h)E=\big[\MM^{[p]}(-x;\h)\big]^T,
\qquad \big[\MM^{[p]}(x;\h)\big]^*=
\MM^{[p]}(-x;\h) . \Eq(5.4)$$
Let $\MM^{[0]}(x;\h)\=\MM^{[0]}(\h)$ and
$\MM^{[\le n]}(x;\h)=\sum_{p=0}^{n} \MM^{[p]}(x;\h)$.
Define for $n_0 \ge n\ge 1$ the {\rm propagators}
$$\eqalign{
& g^{[n-1]}(x;\h)\defi
\fra{\psi_{n}(x^2)\prod_{m=0}^{n-1}
\chi_m(x^2)}{
ix-\MM^{[\le n-1]}(x;\h)}, \cr
& g^{\big\{\ge n\big\}}(x;\h)\defi
\fra{\prod_{m=0}^{n-1}\chi_m(x^2)}
{ix-\MM^{[\le n-1]}(x;\h)} ,
\cr
& g^{[\ge n]}(x;\h)\defi \fra{\prod_{m=0}^{n-1}
\chi_m(x^2)}{
ix-\MM^{[\le n]}(x;\h)} , \cr}
\Eq(5.5) $$
and $g^{[0]}(x;\h)=\psi_{0}(x^2)\,(ix-\MM^{[0]}(\h))^{-1}$.
We call the labels $[n],\{\ge n\},[\ge n]$ scale labels.}\\ 

\0{\it Remarks.}  (1) The matrices $\MM^{[p]}(x;\h)$ will be defined
recursively under the requirement that the functions $(\AA,B,\V a,b)$
defining the parametric equations \equ(2.1) of the invariant torus will be
expressed in terms of trees whose lines carry scale labels indicating
that their values are computed with the propagators in \equ(5.5).
\\
(2) To have the propagators in \equ(5.5) well defined, we
have to eliminate for each value of $p$ a denumerable set of $\e$'s,
by imposing that $(ix-\MM^{[\le p]}(x;\h))$ be invertible,
in analogy with \equ(4.7).
\\
(3) So far $n_0$ can be any integer number. It will be fixed as
prescribed after Lemma 2, in such a way that a bound like \equ(4.10)
will hold for all propagators $g^{[p]}$ with $p < n_0$.

\*

To define recursively the matrices we introduce the notions of
clusters and of self-energy clusters of a tree whose lines and
nodes carry the same labels  introduced so far and {\it in addition}
each line carries a scale label which can be either $[-1]$,
if the momentum of the line is zero, or $[p]$,
with $0\le p< n_0$, or $[\ge n_0]$, with $n_0$ the same integer
appearing in the statement of Definition 2 (still to be suitably fixed).
Given a tree $\th$ decorated in this way we give
the following definition, for $n< n_0$.

\*

\0{\bf Definition 3} (Clusters).
{\it (i) A {\rm cluster} $T$ on scale $[n]$, with $0 \le n$,
is a maximal set of nodes and lines connecting them
with propagators on scales $[p]$, $p \le n$, one of which, at least,
on scale exactly $[n]$. We denote with $V(T)$, $L(T)$ and $\L(T)$ the
set of nodes, the set of leaves and the set of lines, respectively,
contained in $T$. The number of nodes in $T$ 
will be denoted by $k_{T}$.\\
(ii) The $m_{T}\ge 0$ lines entering the cluster $T$ and the possible
line coming out of it (unique if existing at all) are called the
{\rm external lines} of the cluster $T$.\\
(iii) Given a cluster $T$ on scale $[n]$, we shall
call $n_{T}=n$ its scale.}

\*

\0{\it Remarks.}  (1) The clusters on scale $[-1]$ were defined
before \equ(4.1): they can contain either
only lines with scale $[-1]$ or no line at all (\ie they can contain
just a single node).\\ 
(2) Here $n < n_{0}$. However the definition above is given in
such a way that it will extend unchanged when also scales equal to
or larger than $n_{0}$ will be introduced.\\
(3) The clusters of a tree can be regarded as sets of lines
hierarchically ordered by inclusion and have hierarchically ordered
scales.\\
(4) A cluster $T$ is not a tree (in our sense); however we can
uniquely associate a tree with it by adding the entering and the
exiting lines and by imagining that the lower extreme of the exiting
line is the root and that the highest extremes of the entering lines
are nodes carrying a mode label equal to the momentum flowing into
them (\cfr \cita{GG2}, Figure 3).

\*

\0{\bf Definition 4} (Self-energy clusters).
{\it A {\rm self-energy cluster} on scale $[n]$, with $n \ge 0$,
of a tree $\th$ will be any cluster $T$ on scale $[n]$
with the following properties:\\
(i) $T$ has only one entering line
$\ell_{T}^{2}$ and one exiting line $\ell_{T}^{1}$;\\
(ii) $\sum_{\vvvv\in V(T)} \nn_{\vvvv} = \V0$;\\
(iii) there is no line on scale $[-1]$ along the path connecting
$\ell_{T}^{2}$ to $\ell_{T}^{1}$.\\
We call  $k_{T}$ 
the number of nodes in $V(T)$.}
\*

\0{\it Remarks.}
(1) The essential property of a self-energy cluster is that it has
necessarily just one entering line and one exiting line, and both have
{\it equal momentum} (because $\sum_{\vvvv\in V(T)}\nn_{\vvvv}=\V0$).
Note that both scales of the external lines of a self-energy cluster $T$
are strictly larger than the scale of $T$ regarded as a cluster,
but they can be different from each other by just one unit.
\\ 
(2) The self-energy clusters on scale $[-1]$ were defined before
\equ(4.1).
\\ 
(3) For $n\ge 0$, the number of nodes of any self-energy cluster on
scale $[n]$ is $\ge2$, and the corresponding degree is $\ge 2k_{0}$.
This can be seen as follows. Call $T_{0}$ the connected subset of $T$
containing no line on scale $[-1]$ and containing the two nodes
to which the external lines are attached. Then $T_{0}$ must
have at least two nodes with $\d=1$, and $T \setminus T_{0}$
is the union of subtrees with positive degree.
\\ 
(4) The clusters which satisfy properties (i) and (ii), but
not (iii) are not considered self-energy clusters.  The same happened
for the self-energy clusters on scale $[-1]$.  The reason for such a
definition is that the cancellation mechanisms that will imply the
bounds needed on the derivatives of the self-energy values (see next
definition and Lemma 3 below) can be derived only under such an extra
condition.  On the other hand the clusters which verify only the first
two properties, but contain lines on scale $[-1]$ along the path
connecting the external lines, require no resummation, and can be
dealt with in the same way as the other clusters which are not
self-energy clusters. We note that in \cita{GG2} property (iii)
was explicitly required only for self-energy clusters on scale $[-1]$
but it was not mentioned any more for the others:
the proofs in \cita{GG2}, however, implicitly used property
(iii) also for the self--energy clusters on scale $>-1$.

\*

\0{\bf Definition 5} (Renormalized trees).
{\it Let $\Th^{\RR}_{k,\nn,\g}$ be the set of trees with degree $k$
(see comments after \equ(3.9)),
root line momentum $\nn$ and root label $\g$ which contain no self-energy
clusters. Such trees will be called {\rm renormalized trees}.}

\*

\0{\bf Definition 6} (Self--energy matrices).
{\it (i) We denote with $\SS^{\RR}_{k,n}$ the set of self-energy clusters
with degree $k$ and scale $[n]$ which do not contain other
self-energy clusters; we call them {\rm
renormalized self-energy clusters} on scale $n$.
\\
(ii) Given a self-energy cluster $T\in\SS^{\RR}_{k,n}$
we shall define the {\rm self-energy value} of $T$ as the
matrix\footnote{${}^1$}{\nota This is a matrix because the self-energy
cluster inherits the labels $\g,\g'$ attached to the endnode of the
entering line and to the initial node of the exiting line.}
$$ \VV_{T}(\oo\cdot\nn;\h) =
{1\over |\L(T)|!}\Big(\prod_{\vvvv\in V(T)}\e^{\d_{\vvvvv}}\Big)
(\h\b_1)^{|L(T)|}
\Big(\prod_{\vvvv\in V(T)}F_{\vvvv}\Big)
\Big(\prod_{\ell\in \L(T)}g_\ell^{[n_\ell]}\Big)
\Eq(5.6) $$
where $g^{[n_{\ell}]}_{\ell}=g^{[n_{\ell}]}(\oo\cdot\nn_{\ell};\h)$.
Note that, necessarily, $n_\ell\le n$. The $k_T-1$ lines of the
self-energy cluster $T$ will be imagined as distinct and to carry
a number label ranging in $\{1,\ldots, k_T-1\}$.\\
(iii) The {\rm self-energy matrices}
$\MM^{[n]}(x;\h)$, $n\ge 1$, will be defined recursively as
$$ \MM^{[n]}(x;\h) = \Big(\prod_{p=0}^{n-1}
\chi_{p}(x^2)\Big) \sum_{k=2 k_0}^{\io}
\sum_{T \in \SS^{\RR}_{k,n-1} } \VV_{T}(x;\h)
\defi \Big(\prod_{p=0}^{n-1} \chi_{p}(x^2)\Big) M^{[n]}(x;\h) ,
\Eq(5.7)$$
where the self-energy values are evaluated by means of the
propagators on scales $[p]$, with $p=-1,0,\ldots,n$.}

\*

The definition \equ(5.7) makes sense because we have already
defined the propagators on scale $[0]$ and the matrices
$\MM^{[0]}(x;\h)\=\MM^{[0]}(\h)$ (\cfr Definition 2).
Of course we have still to check that the series converges.

\*

With the above new definitions let
$h_{\nn,\g}$ with $\g=\AA,B,$ $\Ba,\b$ be the values of
$\AA_\nn,B_\nn,{\bf a}_\nn,b_\nn$. We have the formal identities
$$ h_{\nn,\g} = \sum_{k=1}^{\io} \sum_{\th \in
\Th^\RR_{k,\nn,\g}} \Val(\th),
\Eq(5.8) $$
where we have redefined the {\it value} of a tree $\th \in
\Th^\RR_{k,\nn,\g}$ as
$$ \Val(\th) = {1\over |\L(\th)|!}\Big(\prod_{\vvvv\in V(\th)}
\e^{\d_{\vvvvv}}\Big) (\h\b_1)^{|L(T)|} \Big( \prod_{\ell\in \L(\th) }
g^{[n_{\ell}]} (\oo\cdot\nn_{\ell};\h) \Big)
\Big( \prod_{\vvvv\in V(\th)} F_{\vvvv} \Big) ,
\Eq(5.9) $$
with $[n_\ell]=[-1],[0],\ldots,[n_{0}-1],[\ge n_{0}]$.
Note that  \equ(5.8) is not a power series in $\h$.

The statement in \equ(5.8) is checked to be
an identity between formal series (as in the corresponding check in
\cita{GG2}).

As a first step to bypass the formal level, the series \equ(5.7) defining
$M^{[n]}(x;\h)$ has to be shown to be really convergent.
This will be true because in the evaluation of $M^{[n]}(x;\h)$ {\it
the only involved propagators have scales $[p]$ with $p\le n-1$} so
that, see the factors $\psi_{n}(x^2),\chi_{n}(x^2)$ in \equ(5.5),
their denominators not only do not vanish but have controlled sizes
that can be bounded below proportionally to $x^{2}$ by \equ(4.10),
\ie simply by a constant times $C_{0}^2|\nn|^{-2\t_{0}}$.
Using this fact one can actually show that the matrices
$M^{[n]}(x;\h)$ are well defined and satisfy
symmetry properties similar to \equ(5.4).

Furthermore {\it cancellations} similar to the maximal KAM tori cancellations
hold even in this case so that $M^{[n]}(x;\h)$ has a special structure,
as described in the following Lemma (see Appendices \secc(A5) and
\secc(A6) for a proof).

\*

\0{\bf Lemma 3.}  {\it Let $\lis\e<\e_0$ with $\e_0$ small enough,
and $\e\in I(\lis\e)=(\lis\e/4,\lis\e]$. Define also $\lis\h$
through $\lis\e=\lis\h^{k_{0}}$.
If $1 \le n < n_0$, with $n_{0}$ defined in \equ(4.11)),
the following properties hold.\\
(i) The series defining the matrices $\MM^{[\le n]}(x;\h)$,
$x=\oo\cdot\nn$, 
converge and the matrices satisfy the
same symmetry properties noted for $\MM^{(0)}$
$$E\MM^{[\le n]}(x;\h)E=\big[\MM^{[\le n]}(-x;\h)\big]^T,
\qquad \big[\MM^{[\le n]}(x;\h)\big]^*=
\MM^{[\le n]}(-x;\h),
\Eq(5.11)$$
where $E$ is the $2N\times 2N$ symplectic matrix, \equ(1.4).
Hence, by \equ(5.11), $(ix -\MM^{[\le n]})E$ is Hermitian. \\
(ii) The matrix $\MM^{[\le n]}(x;\h)$ can be written in the form
$$\pmatrix{Q^{[\le n]}(x;\h) & R^{[\le n]}(x;\h) \cr
P^{[\le n]}(x;\h) & - Q^{[\le n]\dagger}(x;\h) \cr} ,
\Eq(5.12) $$
where, if $\chi_{n-1}(x^2) \neq 0$,
$$ \eqalign{
P^{[\le n]}(x;\h) & =
\pmatrix{\dpr^2_{\AA\AA} H_0 + \e \lis M_{\AA\AA}^{[\le n]}(x;\h) & 
\dpr^2_{\AA B} H_0+\e 
\lis M_{\AA B}^{[\le n]}(x;\h) \cr
\dpr^2_{B\AA} H_0+\e \lis M_{B\AA }^{[\le n]}(x;\h)& \dpr^2_{BB} H_0 +
\e \lis M_{BB}^{[\le n]}(x;\h) \cr} = [P^{[\le n]}(x;\h)]^\dagger , \cr
& \cr
Q^{[\le n]}(x;\e) & =
\pmatrix{
ix\e^2\lis M_{\aa\AA}^{[\le n]}(x;\h) & ix\e^2 \lis M_{\aa B}^{[\le n]}
(x;\h)\cr \e\lis M_{\b\AA}^{[\le n]}(x;\h) &
\e \lis M_{\b B}^{[\le n]}(x;\h)\cr} , \cr
& \cr
R^{[\le n]}(x;\e) & =
\pmatrix{x^2\e^2\lis M_{\aa\aa}^{[\le n]}(x;\h) & ix\e^2
\lis M_{\aa\b}^{[\le n]}(x;\h) \cr
-ix\e^2 \lis M_{\b\aa}^{[\le n]}(x;\h) & \e\h^{k_0-1}
\lis M_{\b\b}^{[\le n]} (x;\h) \cr} = [R^{[\le n]}(x;\h)]^\dagger , \cr}
\Eq(6.0)$$
(iii) The entries $\lis M_{\g,\g'}^{[\le n]}$ are such that
the corrections $\lis M_{\g,\g'}^{[n]}(x;\h)=\lis M_{\g,\g'}^{[\le n]}(x;\h) - 
\lis M_{\g,\g'}^{[\le n-1]}(x;\h)$ are bounded, 
uniformly in $x$ and $\e$ for $\e$
small enough, by $B \, e^{-\k_1 2^{n/\t}}$
for $n< n_0$ and for suitable $n_0$-independent constants
$B,\k_1,\t>0$; one can take $\t=\t_{0}$\\
(iv) One has
$$ \|\dpr_x \MM^{[\le n]}(x,\h)\|\le B \e^2 , \qquad
\|\dpr_\e \big[ \MM^{[\le n]}(x,\h)
- \MM^{[0]}(x,\h) \big] \| \le B \, \e ,
\Eq(5.15) $$
where the derivatives must be interpreted in the sense of Whitney and
the constants $B,\k_1,\t>0$ can be taken the same as in item (iii).}

\*

\0{\it Remarks.}
(1) The symmetry property \equ(5.11) is proved in Appendix \secc(A5).
Note that at the first step $\MM^{[0]}(x;\h)$ satisfies it, see \equ(4.8).\\
(2) The key property in \equ(6.0) is that some entries
of $Q^{[\le n]}(x;\h)$ and $R^{[\le n]}(x;\h)$
are proportional to $x$ or $x^2$: this is proved by exploiting
cancellations among families of self--energy clusters, as described
in detail in Appendix \secc(A6); note that the single self--energy clusters
contributing to $\MM^{[\le n]}(x;\h)$ do not have in general the structure
in \equ(6.0) and only their sum has.

\*

A crucial technical point in the proof of Lemma 3
is the fact that, if the scale $n_\ell$ of a line $\ell$ is smaller than 
$n_0$, as defined in \equ(4.11), then the corresponding propagator
admits a bound that is qualitatively the same as \equ(4.10).
More precisely the following result holds.

\*

\0{\bf Lemma 4.} {\it Let $\e\in (\lis\e/4,\lis\e]$.
The propagator $g^{[0]}(x;\h)$ admits the same bound as
$g^{[\le 0]}(x;\h)$ in \equ(4.10).
For $1\le n< n_0$, with $n_0$ given by \equ(4.11), 
the propagator on scale $[n]$ can be bounded by
$$ \left| g^{[n]}(x;\h) \right| \le {C \over x^2} ,
\Eq(5.16)$$
for some positive constant $C$.}

\*

The proof of Lemma 4 proceeds as that of Lemma 2, in Appendix
\secc(A4), and we do not repeat it here.  Once the bound \equ(5.16) is
established, the proof of convergence is the same as the one discussed
in Appendix A3 of \cita{GG2} and we do not repeat it here. Item (ii)
of Lemma 3 simply follows from convergence and from the remark that
$\MM^{[\le n]}(x;\h)-\MM^{[0]}(x;\h)$ is of order $\e^2$
(\cfr Remark (3) after Definition 4).
The bounds in items (iii) and (iv) of Lemma 3 also follow
from the proof of convergence, see Appendix A3 of \cita{GG2}.

We have therefore constructed a new representation of the formal
series for the parametric equations for the invariant torus: in it
only trees with lines carrying a scale label $[-1],[0],\ldots,[n_0-1]$
or $[\ge n_0]$ and {\it no self--energy clusters} are present (note
that, so far, self--energy clusters may have only scales $[n]$
with $n<n_0$). The above lemma will be the starting block
of the construction that follows.

\*\*
\section(6, Renormalization: the infrared resummation)

\0From the proof of Lemma 2 it is clear that for $x^2\le \r\h^{2k_0-1}$ 
and $n\ge n_0$ it will not be possible to bound $g^{[n]}(x;\h)$ 
by a constant times ${x^{-2}}$.

So, the first problem to face when reaching scales $n\ge n_0$ is the
computation of the eigenvalues of $(ix-\MM^{[\le n]}(x;\h))E$, in terms of
which an estimate of the size of the propagator $g^{[n]}$ can be deduced.

If $\MM^{[\le n]}(x;\h)$ does have the structure in \equ(6.0) then an 
approximate computation of the eigenvalues of $(ix-\MM^{[\le n]})E$ 
leads to the following Lemma, proved in Appendix \secc(A7).

\*

\0{\bf Lemma 5.}
{\it Let $n\ge n_0$ and let us assign a matrix
$\MM^{[\le n]}(x;\h)$ satisfying the properties in \equ(6.0)
admitting right and left derivatives with respect to $x$ and $\e$,
bounded as the derivatives in \equ(5.15), and having the structure
described by \equ(6.0), with the entries
$\lis M_{\g,\g'}^{[\le n]}(x;\h)$ such that the corrections
$\lis M_{\g,\g'}^{[n]}(x;\h)=\lis M_{\g,\g'}^{[\le n]}(x;\h) - 
\lis M_{\g,\g'}^{[\le n-1]}(x;\h)$ can be bounded (as in item (iii) of 
Lemma 3) by $|\lis M_{\g,\g'}^{[n]}(x;\h)| \le B e^{-\k_{1}
2^{n/\t_{1}}}$, for suitable constants $B$, $\k_{1}$ and $\t_{1}$. Then
the uniform norm of $(ix-\MM^{[\le n]}(x;\h))$ can be bounded below by
$$ \| ix-\MM^{[\le n]}(x;\h)\| \ge {1\over 4\m}
\min \{x^2,|x^2-\l^{[n]}(x;\h)|\} ,
\Eq(6.1)$$
where $\m=\m_N^{(0)}$ is the largest eigenvalue of $P^{[0]}$ and
$\l^{[n]}(x;\h)=\ell^{[n]}(x;\h)\e\h^{k_0-1}$
is a real function with $\ell^{[n]}(x;\h)=c k_0\dpr^2_{B}
H_0\b_1^{k_0-1}\big(1+O(\h)\big)$.
Furthermore $\l^{[n]}(x;\h)$ is right and left
differentiable in $\e$ and $x$ and the
derivatives satisfy the following dimensional bounds:
$$ C^{-1} \h^{k_0-1} \le |\dpr_\e^\pm\l^{[n]}(x;\h)|
\le C \h^{k_0-1} , \qquad
|\dpr_x^\pm\l^{[n]}(x;\h)|\le C\h^{k_0} ,
\Eq(6.2)$$
for some positive constant $C$.}

\*

\0{\it Remarks.} (1) The bound \equ(6.1) suggests to replace the classical
small divisor $x^2$ used in previous sections by the ($n$--dependent) 
quantity $\min \{x^2,|x^2-\l^{[n]}(x;\h)|\}$, that is essentially
what we shall do below. In particular we shall replace the argument $x^2$ 
of the support functions $\psi,\c$ by a quantity $\D^{[n]}(x;\h)$,
behaving like $\min \{x^2,|x^2-\l^{[n]}(x;\h)|\}$,
which will be the measure of the strength of the resonance for scales 
larger than $n_0$.
With this choice we shall in general introduce a singularity in the definition
of the propagators and self--energy matrices on scales $n\ge n_0$: this is 
due to the presence of a minimum and of an absolute value in the
definition of $\D^{[n]}(x;\h)$. This is ultimately 
why in \equ(6.2) the right and left derivatives appear, rather than the 
plain derivatives, as in \equ(5.15) above.
This could be avoided by using a smoothed version of the quantity
$\D^{[n]}(x;\h)$ introduced below, but we shall not discuss it here.\\
(2) The bound on the derivatives of $\l^{[n]}(x;\h)$
with respect to $\e$ follows from the expression of $\MM^{[0]}(\h)$,
in \equ(4.3) to \equ(4.5), and from the bounds \equ(5.15),
which allow to control the corrections. On the contrary
the bound on on the derivatives of $\l^{[n]}(x;\h)$
with respect to $x$ does not follow directly from \equ(5.15),
and it is explained in Appendix \secc(A7).

\*

Depending on the sign of
$\e$, from now on, the analysis changes qualitatively. 
If $\e$ is such that $c\e\b_1^{k_0-1}<0$ 
(so that $\l^{[n]}(x;\h)$ in \equ(6.1) is negative), 
the minimum in \equ(6.1) is always realized by the classical small divisor
$x^2$. This implies that in this range of scales it is possible to
proceed in the same way discussed above in Section \secc(5). We do not
repeat the details, and we concentrate on the opposite case, namely
$\e$ with the sign such that $\l^{[n]}(x;\h)> 0$,
which presents new difficulties.

In this case convergence problems can still arise from the propagators
$g^{[\ge n_0]}(x;\h)$, which become uncontrollably large for
$x=\oo\cdot\nn$ close to the eigenvalue $\l^{[n]}(x;\h)$ in \equ(6.1).

We introduce a sequence of {\it self--energies} $\ul{\l}^{[n]}(\h)$,
\cita{GG2}, representing the locations of the singularities of
$ix-\MM^{[\le n]}(x;\h)$ and, correspondingly, we introduce a sequence of
{\it propagator divisors} $\Delta^{[n]}(x;\h)$, which will give a bound for
the size of the propagator on scale $n\ge n_0$.  Then, we modify the
scale decomposition, measuring the strength of the singularity in
terms of $\Delta^{[n]}(x;\h)$ rather than, as done in Section \secc(5),
in terms of the classical small divisor $x^2$.

The choice of the scale decomposition will be done in such a way that
the dimensional bounds for the propagators will be the same as for
those with scales $n<n_0$: this will reduce the analysis of
the infrared resummations to the same convergence proof discussed 
in Appendix A3 of \cita{GG2}.

We introduce the following definition.
\*

\0{\bf Definition 7} (Self-energies and propagator divisors).
{\it Let the function $\l^{[n]}(x;\h)$ be as in Lemma 5.
\\
(i) The sequence of {\rm self-energies}
$\ul{\l}^{[n]}(\h)$ is defined for $n\ge n_0$ by
$$ \ul{\l}^{[n]}(\h)\defi
\l^{[n]} \Big(\Sqrt{\ul{\l}^{[n-1]}(\h)},\h\Big), \qquad
\ul{\l}^{[n_0]}(\h)\defi \l^{[n_0]}(0;\h) ,
\Eq(6.3)$$
provided $\ul{\l}^{[n]}(\h)\ge0$, $n\ge n_0$.
\\
(ii) The {\rm propagator divisors} are defined for $n\ge n_0$ by
$$ \D^{[n]}(x;\h) \defi \min\{x^2,\big| x^2 - \ul{\l}^{[n]}(\h) \big|\} .
\Eq(6.4) $$
}

By repeating the analysis of Section \secc(5) we can represent the 
function $X(\pps)$ via sums of values of
trees whose lines can carry scale labels
$[-1],[0],\ldots,[n_0-1],[n_0], [n_0+1],\ldots$ and which contain no
self-energy clusters (\ie they are renormalized trees;
see Definition 5 in Section \secc(5)).
The new propagators will be defined by the same procedure
used to eliminate the self-energy clusters on scales $[n]$ with $n\le
n_0-1$. However for $n\ge n_0$ the scale of a line will be determined
in terms of the recursively defined $\D^{[n]}(x;\h)$, \equ(6.4),
rather than in terms of $x^2$, see \equ(5.2).

Set $X_{n_0-1}(x)\defi
\prod_{m=0}^{n_0-1}\chi_m(x^{2})$, $Y_{n}(x;\h)\defi
\prod_{m=n_0}^{n}\chi_{m}(\D^{[m]}(x;\h))$ for $n\ge n_0$ and
$Y_{n_0-1}\=1$. The definition of the new propagators will be
$$\eqalign{
g^{[n_0]} \defi & \kern3mm
X_{n_0-1}(x)\, \psi_{n_0}(\D^{[n_0]}(x;\h))\,
(x^2-\MM^{[\le n_0]}(x;\h))^{-1} , \cr
g^{[n_0+1]} \defi & \kern3mm
X_{n_0-1}(x)\,\chi_{n_0} (\D^{[n_0]}(x;\h))\,
\psi_{n_0+1}(\D^{[n_0+1]}(x;\h)) \,
(x^2-\MM^{[\le n_0+1]}(x;\h))^{-1} , \cr
&\ldots\cr
g^{[n]} \defi & \kern3mm
X_{n_0-1}(x)\,Y_{n-1}(x;\h)\,\psi_{n}(\D^{[n]}(x;\h))
\,(x^2-\MM^{[\le n]}(x;\h))^{-1} , \cr}
\Eq(6.5) $$
and so on, using indefinitely the identity $1\=\psi_n(\D^{[n]}(x;\h))+
\chi_n(\D^{[n]}(x;\h))$ to generate the new propagators.

\*

In this way we obtain a formal representation of $X(\pps)$ as a sum of
tree values in which only renormalized trees appear
and in which each line $\ell$ carries a {\it scale label}
$[n_\ell]$. This means that we can formally write $X(\pps)$ as in
\equ(5.8), with $\Val(\th)$ defined according to \equ(5.9), but now
the scale label $[n_{\ell}]$ is such that $n_{\ell}$ can assume all
integer values $\ge -1$, and no line carries a scale label like
$[\ge n]$: {\it only scale labels $[n]$ are possible}.

We can summarize the discussion above in the following definition.

\*

\0{\bf Definition 8} (Propagators and self--energy matrices).
{\it Given a sequence of $2N\times 2N$ matrices $\MM^{[\le m]}(x;\h)$,
$m\ge1$, let $\MM^{[n]}(x;\h)=\MM^{[\le n]}(x;\h)-\MM^{[\le
n-1]}(x;\h)$ with  $\MM^{[\le0]}(x;\h)\=\MM^{[0]}(\h)$ (see \equ(4.2)),
so that $\MM^{[\le n]}(x;\h)=\sum_{m=0}^{n}\MM^{[m]}$ $(x;\h)$. Setting
$\D^{[n]}(x;\h)\=x^2$ if $n< n_0$, define for all $n\ge 0$
$$ g^{[n]}(x;\h)=
\fra{\psi_{n}(\D^{[n]}(x;\h))\prod_{m\ge 0}^{n-1}
\chi_m(\D^{[m]}(x;\h))}{x^2-\MM^{[\le n]}(x;\h)} .
\Eq(6.6)$$
(for $n=0$ this means $\psi_0(x^2)\,(ix^2-\MM^{[0]}(\h))^{-1}$).
We say that $g^{[n]}_{\ell}= g^{[n]}(\oo\cdot\nn_{\ell};\e)$ is a
propagator on scale $[n]$.
The matrices $\MM^{[m]}(x;\h)$ will be defined as in Section \secc(5) for
$n< n_0$ and will be defined recursively also for $n\ge n_0$ 
in terms of the self-energy clusters $\SS^\RR_{k,n-1}$
introduced in Definition 4, Section \secc(5), setting for $n>
n_0$ (\cfr \equ(5.7))
$$ \MM^{[n]}(x;\h)= \Big(\prod_{m=0}^{
n-1}\chi_m(\D^{[m]}(x;\h))\Big)\sum_{k=2}^{\io}
\sum_{T \in \SS^{\RR}_{k,n-1} } \VV_{T}(x;\h) ,
\Eq(6.7)$$
where the self-energy values $\VV_{T}(x;\h)$ are evaluated by means
of propagators on scales less than $[n]$. Note that
we have already defined (consistently with \equ(6.7)))
the matrices $\MM^{[\le n]}$ with $n< n_0$ and
the propagators on scale $[-1],[0],\ldots,$ $[n_0-1]$
(so that \equ(6.6) defines also $g^{[n_0]}(x;\h)$).}

\*
Of course the above definition makes sense only if the series
\equ(6.7) can be shown to be convergent for all $n$. For this purpose
an inductive assumption on the propagators on the scales
$[m]$, $0 \le m < n$ is necessary.

\*

\0{\bf Inductive assumption.}
{\it Let $n_0$ be fixed as in Lemma 3.
\\
(i) For $0\le m\le n-1$ the matrices $\MM^{[m]}(x;\h)$
are defined by convergent series for all $|\e|\in I(\lis\e)=
[\lis\e/4,\lis\e]$ and, for all $x$, they satisfy the properties (i) and (iv)
of Lemma 3. Moreover they can be represented as in \equ(5.12), with 
$P^{[\le n]},Q^{[\le n]},R^{[\le n]}$ as in \equ(6.0) and the entries
$\lis M_{\g'\g}^{[\le n]}$ bounded as described after \equ(6.0).
\\
(ii) There exist $K>0$ and open sets $\EE^o_m$, $m=0,\ldots, n$, with
$\EE^o_{m}\subset I(\lis\e)$, such that, defining recursively
$\ul{\l}^{[m]}(\h)$ in terms of $\ul{\l}^{[m-1]}(\h)$ for $m=
n_0,\ldots,n-1$ by (i) in Definition 7 above, while setting
$\ul{\l}^{[m]}(\h)\=0$ for $m=0,\ldots,n_0-1$,
and defining $\t_{1} \defi \t_{0}(1+\d_1) + N$, with $\d_1>0$,
one has for $\e\not\in \EE^o_m$
$$ \eqalign{&
\G^{[m]}(x;\h) \defi
\Big| |x|-\Sqrt{\ul{\l}^{[m]}(\h)} \Big|\ge
2^{-\fra12 m} \fra{C_{0}}{|\nn|^{\t_{1}}} , \cr
&|\EE_m^o|\le K 2^{-\fra12 m} \lis\e \, \lis\h^{\d_1(k_0-1/2)} , \cr}
\Eq(6.8) $$
with $\lis\e=\lis\h^{k_{0}}$, for all $m\le n-1$ and all $x$.}

\*

\0{\it Remark.} As in \cita{GG2}, a key point is checking
that $\D^{[n]}(x;\h)$ can be used to
bound below the denominators of the non-vanishing propagators
on scale $[n]$. If $x$ has scale $[n]$, with $n\ge n_0$, one has
$$\eqalign{
& \Big|x^{2}- \l^{[n]}(x;\h) \Big| \ge
\Big| x^{2} - \ul{\l}^{[n]}(\h) \Big| -
\Big| \ul{\l}^{[n]}(\h) - \l^{[n]}(x;\h) \Big| \cr
& \qquad \ge \fra12\Big| x^{2}- \ul{\l}^{[n]}(\h) \Big| +
2^{-(n+2)}C_{0}-\Big| \l^{[n]}
\big( \Sqrt{\ul{\l}^{[n-1]}(\h)},\h \big) -
\l^{[n]}(x;\h) \Big| \cr
& \qquad \ge \fra12 \Big| x^{2} - \ul{\l}^{[n]}(\h) \Big| \qquad
\tto \qquad \|x^2-\MM^{[n]}(x,\h)\|\ge \fra1{8\m}
\,\D^{[n]}(x;\h) , \cr}
\Eq(6.9) $$
having used the lower cut-off $\psi_{n}(\D^{[n]}(x;\h))$ in the
propagator (see \equ(6.5)) to obtain the first two terms in the second
line while the upper cut-off $\chi_{n-1}(\D^{[n-1]}(x;\h))$ has been
used to obtain positivity of the difference between the second and
third terms in the second line, after applying the second inequality
in \equ(6.2), so that the last term in the second line of \equ(6.9)
can be bounded above proportionally to $\e 2^{-n} C_{0}$.

\*

The inequality \equ(6.9) allows us a complete word by word reduction
of the proof of the inductive assumption above 
to the corresponding inductive assumption of \cita{GG2}. 
The symbols here and in \cita{GG2} have been
chosen to coincide so that the analysis in Section 6 and Appendix A3
of \cita{GG2} can be taken over and reinterpreted, with no change, as
proofs of the above inductive hypothesis, 
apart for the check of the measure $\EE^0_m$ in \equ(6.8), which 
in the present case is slightly different from the corresponding
computation in \cita{GG2}. The estimate \equ(6.8)
of the measure of $\EE^0_m$ is explicitly given in Appendix \secc(A8).

We can summarize the previous discussion into the following Lemma.

\*

\0{\bf Lemma 6.} {\it There is $\e_{0}$ small enough
such that for $\lis\e < \e_{0}$ and $\e\in I(\lis\e)$, if the inductive
hypothesis is assumed for $\,0\le m\le n-1$ then it holds for $m=n$.}

\*

The series for $X(\pps)$ is now fully ``renormalized'' and its terms are
well defined for $\e$ in a set $\EE$ whose measure is large (near $0$).
The series is expressed as a sum of renormalized tree values of tree
graphs without any self--energy graph. Therefore the series is
convergent (by Siegel's lemma); see the corresponding discussion 
in Appendix A3 of \cita{GG2}.

In the derivation geometric series with ratio $z>1$ have been
considered and the rule $\sum_{k=0}^\io z^k=(1-z)^{-1}$ has been
repeatedly used: this is not mathematically rigorous and
therefore an a posteriori check must be made that the function
$X(\pps)$, via \equ(1.5), actually does satisfy the
equations \equ(1.4). The check, however, is a repetition of the
corresponding (essentially algebraic) check in Appendix A5 of \cita{GG2}.

Therefore the proof of Theorem 1 is complete

\*\*
\section(7, Conclusions.)

\0We conclude by mentioning some problems that seem to us of interest.

\*

\0(1) First we note that the proof of Theorem 1 also provided some
informations about the analyticity properties in $\h$, hence in $\e$,
of the surviving lower-dimensional tori; \cfr \cita{GG1} and
\cita{GG2} for further details. As in the quoted papers,
it can be interesting to further investigate such properties,
and related ones, such as Borel summability.
\\
(2) The uniqueness of the resonant tori appears to be a hard problem.
Even a proof of Borel summability would not resolve the problem as
there could be solutions which are not Borel summable. Just as in the
maximal tori case analyticity is not sufficient to guarantee
uniqueness. The very recent \cita{BT} does not settle the question,
as it proves uniqueness of the $C^{\io}$ diffeomorphism
mapping the unperturbed invariant tori which are conserved into
those which are explictly constructed, but it dose not
eliminate the possibility that other nearby quasi-periodic motions
with the same rotation vectors exist.
\\
(3) Another open problem concerns the possibility of removing the
assumptions we made in this paper. We have been able to do this in
several particular cases but we did not find a satisfactory general
formulation: for instance can something be said in general in the case
in which the average of the perturbation vanishes identically? 
\\
(4) Finally, it would be interesting to understand what happens
in the case of $n$-dimensional tori, with $n$ strictly less than $N-1$
(that is in the case of several normal frequencies),
by relaxing the assumptions on the perturbing potential
with respect the results existing in literature, as done here and
in Cheng's paper \cita{Ch1} and \cita{Ch2} for $n=N-1$.
This is a substantially more difficult endeavour with respect to
that considered here and in Cheng's quoted papers.

\*\*
\appendix(A1, Heuristic analysis leading to fractional series)

\0Consider the simple case $H(\V A,B,\aa,\b)=\Bo\cdot\V A+
\fra12(\V A^2+B^2)+\e f(\Ba,\b)$, and let assumptions (a) and (c)
in Section \secc(1) be satisfied by $f$.
Let us consider the generating function
$$\G(\V A',B',\aa,\b)=\Ba\cdot\V A'+ \b B'+\e \Psi_1(\V A',B',\Ba,\b)+
\e^2 \Psi_2(\V A',B',\Ba,\b)+\e^2\aa\cdot\zzzzz+\e^2\b\r ,
\Eqa(A1.1)$$
where, if $\D\defi \Bo\cdot\dpr_{\Ba}$,
$$\Psi_1(\V A',B',\aa,\b)\defi -\D^{-1}\big[1-(\V
A'\cdot\dpr_\Ba\,+\,B'\dpr_\b)\,\D^{-1}\big] f_{\neq \V0}(\Ba,\b),
\Eq(A1.2)$$
and, if $\Phi(\V A',B',\Ba,\b)=\fra12
\big[(\dpr_\Ba\Ps_1)^2+ (\dpr_\b\Ps_1)^2\big]$,
$$ \eqalign{&\Psi_2(\V A',B',\aa,\b)\defi -\D^{-1}\big[1-(\V
A'\cdot\dpr_\Ba\,+\,B'\dpr_\b)\,\D^{-1}\big]
\Phi_{\neq \V0}(\V A',B',\Ba,\b) ,\cr
&\zzzzz=- \dpr_{\V A'}\Phi_\V0(\V A',0,\b_0)\big|_{\V A'=\V0} ,
\qquad\r=-\dpr_{B'}\Phi_\V0(\V0,B',\b_0)\big|_{B'=0} .\cr}
\Eqa(A1.3)$$
The canonical map $(\V A,B,\Ba,\b)\otto(\V A',$ $B',\Ba',\b')$
generated by $\G(\V A',B',\aa,\b)$ transforms the Hamiltonian $H(\V
A,B,\aa,\b)$ into
$$ \eqalignno{
H'(\V A',B',\aa',\b') & = \Bo\cdot\V A' + \fra12({\V A'}^2+{B'}^2)+
\e f_\V0(\b')-&\eqa(A1.4) \cr
&-{\e}^2\dpr_{\b'}f_\V0(\b')\cdot\D^{-2}\dpr_{\b'}f_{\neq \V0}
(\aa',\b')+{\e}^2\Phi_\V0(\V0,0,\b')+O(\e {I'}^2)+O(\e^3)\;.\cr}
$$
Writing the Hamiltonian equations generated by \equ(A1.4), we realize that
in a small neighborhood of $\b_0$ the evolution equation for
$\b'$ takes the form:
$$\ddot\b'=- c \e(\b'-\b_0)^{k_0} -\e^2 a+\e^2 a' (\b-\b_0)
+ O\big(\e(\b'-\b_0)^{k_0+1}\big)
+ O\big(\e^2(\b'-\b_0)^2\big)
+O(\e^3)\;,\Eqa(A1.5)$$
where $a=\dpr_{\b'}\Phi_\V0(\V0,0,\b_0)$ is
the same constant defined in \equ(1.9) and $a'$ is a suitable constant.

Therefore, under the assumption that both $a$ and $c$ are non-zero,
the perturbed equilibrium of the angle $\b'$ is equal,
{\it up to high order corrections}, to $\b_0'=\b_0+\d\b_0$, with
$$\d\b_0=\cases { (-a \e/c)^{1/k_0}& if $k_0$ is odd, \cr
\pm (-a \e/c)^{1/k_0} & if $k_0$ is even and
$a \e/ c< 0$. \cr}
\Eqa(A1.6)$$
If $k_0$ is odd and $c\e>0$ the approximate perturbed equilibrium
point $\b_0'$ is quadratically stable, while if $c\e<0$ $\b_0'$ is
quadratically unstable: hence we shall say that the resonant
invariant torus is elliptic if $c\e>0$ and hyperbolic if $c\e<0$.
The center of oscillations is displaced by $O(\e^{1/k_0})$ so that a
fractional series in powers of $\h=\e^{1/k_0}$ has to be expected (at best).

If $k_0$ is even the unperturbed equilibrium point $\b_0$ can be continued
into a perturbed equilibrium $\b_0'=\b_0+\d\b_0$ (with $\d\b_0$ vanishing as
$\e\to 0$) {\it only if} $ a \e c <0$. In this case, if $c\e\d\b_0>0$
the approximate perturbed equilibrium point $\b_0'$
is quadratically stable, while if $c\e\d\b_0<0$
$\b_0'$ it is quadratically unstable: hence we shall say that
the resonant invariant torus is elliptic if $c\e\d\b_0>0$ and hyperbolic if
$c\e\d\b_0<0$. Given the results known for elliptic and hyperbolic
resonances Theorem 1 becomes a natural conjecture.

If $c\neq0$ and $a=0$ the theory depends on the corrections in \equ(A1.5):
for instance if $a'\neq0$ a natural conjecture is that the new equilibrium
point for $\b'$ is approximately $\b'_0+\d\b_0$,
with $\d\b_0$ satisfying the equation $(\d\b_0)^{k_0-1}=\e a'/c$,
whenever this equation is solvable. The stability of this equilibrium
point can be again analyzed in terms of the relative signs
of $\e,a,c$. If $a' =0$ one expects that the higher order terms
have to be studied in details.

\*\*
\appendix(A2, On assumption {(a)})

\0Here we want to show that assumption (a) is not a
purely technical assumption; that is we want to shows that generically
a Hamiltonian of the form \equ(1.3) violating assumption (a) cannot
admit quasi-periodic motions of codimension 1 continuously connected
to an unperturbed motion of the form \equ(1.6) in the limit $\e\to 0$.
We show this by producing an explicit example, given by
an Hamiltonian of the form \equ(1.3), satisfying assumptions (c)
and (b) and violating assumption (a), for which we can prove absence
of quasi-periodic motions of codimension 1 tending to
an unperturbed motion of the form \equ(1.6) as $\e\to 0$.

The counterexample is given by the following Hamiltonian
$$ H = A + \fra{1}{2} A^{2} + \fra{1}{2} B^{2} +
\e \left( \fra{\sin^3\b}{3} + \sin\b\cos\a \right) ,
\Eqa(A2.1) $$
where $A,B\in\RRR^1$ and $\a,\b\in\TTT^1$. If $\e=0$ the
Hamiltonian \equ(A2.1) is strictly convex (then it satisfies
assumption (b)) and it admits the unperturbed periodic motions
$$ A(t)=0 ,\qquad \a(t)=t , \qquad B(t) =0 , \qquad \b(t) =\b_0 ,
\Eqa(A2.2) $$
parametrized by the choice of $\b_0\in\TTT^1$. Choosing $\b_0=0$,
we see that $f_\V0(0)=\dpr_\b f_\V0(0)=\dpr_\b^2 f_\V0(0)=0$
and $\dpr_\b^3 f_\V0(0)=2\neq 0$ (so that assumption (c) is satisfied with
$k_0=2$). Using the fact that $f$ is independent of the action variables
and that $f(\a,\b=0)=\dpr^2_\b f(\a,\b=0)\=0$, we see that
the quantity $a$ defined in \equ(1.8) is identically $0$: this means that
assumption (a) is violated in the case under analysis.

We now investigate the possible existence of periodic motions of the form
$$\a(t)=t+a(t;\e) , \qquad A(t)=\dot a(t;\e) ,\qquad
\b(t)=\h(\e)+b(t;\e) , \qquad  B(t) =\dot b(t;\e) ,
\Eqa(A2.3)$$
where $\h(\e)$ is a continuous function of $\e$ such that
$\lim_{\e\to 0}\h(\e)=0$ and $a(t;\e),b(t;\e)$ are $0$ average
periodic functions of $t$ of period $2\p$.
We want to show that it is impossible that a motion of the form
\equ(A2.3) satisfies the Hamiltonian equations of motion
$$ \eqalign{
& \ddot a(t;\e)=\e \sin(t+a(t;\e))\sin(\h+b(t;\e))\;,\cr
& \ddot b(t;\e)=-\e \big[\sin^2(\h+b(t;\e))\cos(\h+b(t;\e))
+\cos(t+a(t;\e))\cos(\h+b(t;\e))\big] . \cr}
\Eqa(A2.4)$$
In fact a $0$ average periodic solution to \equ(A2.4)
is necessarily of the form
$$\eqalign{
& a(t;\e) = -\e\big(\h\sin t+\fra{\e}{4}\sin t\cos t\big) +
\e^{2} \, O(\e^2+\h^2) , \cr
& b(t;\e) = \e\cos t + \e \, O(\e^2+\h^2) . \cr}
\Eqa(A2.5)$$
Averaging over $t$ the second of \equ(A2.4) we find
$$ \int_0^{2\p}\fra{{\rm d}t}{2\p}
\left( \cos(t+a(t;\e))\cos(\h+b(t;\e)) +
\sin^2(\h+b(t;\e))\cos(\h+b(t;\e) ) \right) = 0 .
\Eqa(A2.6)$$
Now, using the expressions \equ(A2.5), we see that \equ(A2.6) is equal
to $\h^2+\fra{\e^2}{2}$, plus terms of order at least
$\h^{4} + \e^{4}$, and this leads to a contradiction.

\*\*
\appendix(A3, Counting the number of trees lines)

\0We want to show that the number of lines
of a tree contributing to $\AA^{(k)}$, $B^{(k)}$, $\aaa^{(k)}$
or $b^{(k)}$ can be bounded, above and below, by an $O(1)$ constant times $k$.
A lower bound $k/k_0$ is an immediate consequence of the definitions.

We proceed by induction, using that:
\\
(1) the trees contributing to $\AA^{(k_0)}$ and $B^{(k_0)}$
have one line (see \equ(2.2)); those contributing to 
$\AA^{(k_0+1)}$ and $B^{(k_0+1)}$ have two lines;
\\
(2) the trees contributing to $\V a^{(k_0)}_\nn$ and $b_\nn^{(k_0)}$,
with $\nn\neq\V0$, have at most two lines and those contributing to
$\V a^{(k_0+1)}_\nn$ and $b_\nn^{(k_0+1)}$, with $\nn\neq\V0$,
have at most three lines;
\\
(3) $b^{(1)}_\V0\=\b_1$ is represented by a trivial tree (\ie a tree
with one line) and the trees contributing to $b^{(2)}_\V0\=\b_2$ have at most
$k_0+2$ lines.

We introduce the following definitions:\\
$B_k(I)$ is the maximum number of lines
of the trees contributing to $\AA^{(k)},
B^{(k)}$;\\
$B_k(\f)$ is the maximum number of lines
of the trees contributing to $\V a^{(k)}_\nn,
b^{(k)}_\nn$, $\nn\neq\V0$;\\
$B_k(b_\V0)$ is the maximum number of lines of the trees contributing to
$b^{(k)}_\V0\=\b_k$.

We shall make the following inductive assumption:
$$ \eqalign{
&B_k(I)\le 3k_0(k-k_0+1)-4k_0-2 \hskip2.truecm k\ge k_0+1 , \cr
&B_k(\f)\le 3k_0(k-k_0+1)-4k_0-1 \hskip2.truecm k\ge k_0+1 , \cr
 &B_{k-k_0+1}(b_\V0)\le 3k_0(k-k_0+1)-4k_0
\hskip1.7truecm k\ge k_0+1 . \cr}
\Eqa(A3.1)$$
By the remarks above and the fact that $k_0\ge 2$ it follows that
at the first step (that is $k=k_0+1$)
the inequalities above are verified.

Assume inductively the inequalities in \equ(A3.1) for
$k_0+1 \le k\le h-1$ and let us prove them for $k=h\ge k_0 +2$.
Let us start with the third inequality. We call $\vvv_0$ the first node
preceding the root. By the rules explained in Section \secc(3) we
have that the sum of the orders of the $s=s_{\vvvv_{0}}$ subtrees
entering $\vvv_0$ must be $h$. So, using the
inductive assumptions, and calling (see Figure 3):

\*

\0$s_I$ the number of subtrees of type $I^{(k)}, k\ge k_0+1$,
entering $\vvv_0$;\\
$s_I'$ the number of subtrees of type $I^{(k_0)}$
entering $\vvv_0$;\\
$s_\f$ the number of subtrees of type $\f^{(k)}_\nn, k\ge k_0+1,
\nn\neq\V0$, entering $\vvv_0$;\\
$s_\f'$ the number of subtrees of type $\f^{(k_0)}_\nn,\n\neq\V0$,
entering $\vvv_0$;\\
$s_0$ the number of subtrees of type $b^{(k)}_\V0, k\ge 2$,
entering $\vvv_0$;\\
$s_0'$ the number of subtrees of type $b^{(1)}_\V0$
entering  $\vvv_0$,

\*
\figini{fig3}
\8<
\8</punto { 
\8<gsave 3 0 360 newpath arc fill stroke grestore} def>
\8</puntino { 
\8<gsave 2 0 360 newpath arc fill stroke grestore} def>
\8<>
\8</puntovuoto { 
\8<gsave 3 0 360 newpath arc stroke grestore} def>
\8<>
\8</puntox {
\8</y2 exch def /x2 exch def /y1 exch def /x1 exch def /x exch def>
\8<x1 x2 x1 sub x mul add>
\8<y1 y2 y1 sub x mul add} def>
\8<>
\8</linea { 
\8<gsave 4 2 roll moveto lineto stroke grestore} def>
\8<>
\8</tlinea {gsave 4 2 roll moveto lineto [4 4] 2 >
\8<setdash stroke [] 0 setdash grestore} def >
\8<>
\8<
\8< /origine1assexper2pilacon|P_2-P_1| { >
\8< 4 2 roll 2 copy translate exch 4 1 roll sub >
\8< 3 1 roll exch sub 2 copy atan rotate 2 copy >
\8< exch 4 1 roll mul 3 1 roll mul add sqrt } def >
\8< >
\8< /punta0 {0 0 moveto dup dup 0 exch 2 div lineto 0 >
\8< lineto 0 exch 2 div neg lineto 0 0 lineto fill >
\8< stroke } def >
\8< >
\8< /dirpunta{
\8< gsave origine1assexper2pilacon|P_2-P_1| >
\8< 0 translate 7 punta0 grestore} def >
\8<>
\8< /puntatore {
\8< 4 copy 4 copy 4 copy 4 copy >
\8< dirpunta 4 2 roll moveto lineto stroke} def >
\8< >
\8<
\8<.8 .8 scale>
\8</L {50} def>
\8</LL {70} def>
\8</H {90} def>
\8<>
\8</P0 {0 H 2 div} def>
\8</P1 {L H 2 div} def>
\8</P2 {L LL add H} def>
\8</P3 {L LL add 0} def>
\8</P4 {1 5 div P2 P3 puntox} def>
\8</P5 {2 5 div P2 P3 puntox} def>
\8</P6 {3 5 div P2 P3 puntox} def>
\8</P7 {4 5 div P2 P3 puntox} def>
\8<>
\8</Q0 {.5 P0 P1 puntox} def>
\8</Q2 {.45 P2 P1 puntox} def>
\8</Q3 {.45 P3 P1 puntox} def>
\8</Q4 {.45 P4 P1 puntox} def>
\8</Q5 {.45 P5 P1 puntox} def>
\8</Q6 {.4 P6 P1 puntox} def>
\8</Q7 {.45 P7 P1 puntox} def>
\8<>
\8<P0 P1 linea >
\8<P2 P1 linea>
\8<P4 P1 linea>
\8<P5 P1 linea>
\8<.05 P6 P1 puntox P1 linea>
\8<.05 P7 P1 puntox P1 linea>
\8<.04 P3 P1 puntox P1 linea>
\8<P1 Q0 dirpunta>
\figfin

\eqfig{104pt}{75pt}
{
\ins{103pt}{76pt}{$s_I$}
\ins{103pt}{64pt}{$s'_I$}
\ins{103pt}{47pt}{$s_\f$}
\ins{103pt}{34pt}{$s'_\f$}
\ins{103pt}{17pt}{$s_0$}
\ins{103pt}{6pt}{$s'_0$}
\ins{36pt}{31pt}{$\vvv_0$}
}
{fig3}{}

\*\*
\line{\vtop{\line{\hskip2.0truecm\vbox{\advance\hsize by -4.1 truecm
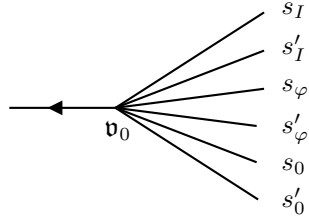
\0{{\css Figure 3.} {\nota The lines entering $v_0$
represent {\it symbolically} the bundle of subtrees entering
$\vvvv_0$ and with root lines of types $I^{(k)},I^{(k_0)},
\f_\Bn^{(k)},\f_\Bn^{(k_0)}$, with $k\ge k_0+1$, and $\b_k,\b_1$,
with $k\ge2$. The number of subtrees in each bundle is indicated by
the right labels.
\vfil}} } \hfill} }}
\*\*

\0$B_{h-k_0+1}(b_\V0)$ can be bounded by the maximum
over the choices of $s_I,s_I',s_\f,s_\f',s_0,s_0'$ of:
$$ 1+ 3k_0h -\Big\{(3k_0^2+k_0+2)s_I+(3k_0^2-1)s_I'
+(3k_0^2+k_0+1)s_\f+(3k_0^2-2)s_\f'+4k_0s_0+(3k_0-1)s_0'\Big\} . 
\Eqa(A3.2)$$
Let us first consider the case $s_I+s_I'+s_\f+s_\f'\ge 1$.\\
In this case, if $s=1$, since $h\ge k_0+2$, we must have that
either $s_\f$ or $s_I$ is $=1$, and, in both cases, we can bound
\equ(A3.2) by $1+ 3k_0h-(3k_0^2+k_0+1)$, that is the desired bound;
if, on the contrary, $s\ge 2$, then \equ(A3.2) can be bounded by $1+
3k_0h -(3k_0^2-2)-(3k_0-1)\le 3k_0h -3k_0^2-k_0$.\\ Let us now
consider the case $s_I+s_I'+s_\f+s_\f'=0$, in which case $s=s_0+s_0'$
and necessarily $s\ge k_0$ (note that in this case the $s+1$
derivatives in \equ(3.6) must be derivatives with respect to $\b$ so
that $s\ge k_0$). If $s\ge k_0+1$, then
\equ(A3.2) can be bounded by $1+ 3k_0h -(k_0+1)(3k_0-1)\le 3k_0h -3k_0^2-k_0$.
If $s=k_0$, by the rules in Section \secc(3)
it must be $s'_0\le k_0-2$, so that \equ(A3.2) can be
bounded by $1+ 3k_0h -4k_0(k_0-s_0')-(3k_0-1)s_0'
\le 1+ 3k_0h -4k_0^2+(k_0-2)(k_0+1)$
that implies the desired inductive bound.

Let us now consider the first of the three inequalities in \equ(A3.1).
By the rules explained in Section \secc(3) we have that the
sum of the degrees of the $s$ subtrees entering $\vvv_0$ is 
$h-k_0$ or $h$ if, respectively, $\d_{\vvvv_0}=1,0$. 
If $\d_{\vvvv_0}=1$, $B_{h}(I)$ can be bounded by the maximum
over the choices of $s_I,s_I',s_\f,s_\f',s_0,s_0'$ of
$1+3k_0(h-k_0)-\{\cdots\}$, where the brackets include the same expression
as in \equ(A3.2), and we can bound this number by
$1+3k_0h-3k_0^2-(3k_0-1)$ that is even better than the desired bound.
If $\d_{\vvvv_0}=0$, one must have $s=s_I+s_I'\ge 2$ (as $s=1$
corresponds to a not allowed tree graph) and
$B_{h}(I)$ can be bounded by $1+3k_0h-2(3k_0^2-1)$, that is even
better than the desired bound.

We finally consider the second of the three inequalities in \equ(A3.1).
By the rules explained in Section \secc(3) we have
that the sum of the orders of the $s$ subtrees entering
$\vvv_0$ must be $h-k_0$ or $h$, if, respectively, $\d_{\vvvv_0}=1,0$..
If $\d_{\vvvv_0}=1$ the number of lines of the corresponding trees
can be bounded in the same way as we proceeded above for the
first inequality in \equ(A3.1).
If $\d_{\vvvv_0}=0$, only lines of type $I$ can enter $\vvv_0$,
and then the corresponding trees have a number of lines bounded
by $1+3k_0h-\{(3k_0^2+k_0+2)s_I+(3k_0^2-1)s_I'\}$. So, if $s_I\ge 1$
the desired bound follows. If $s_I=0$, since $h\ge k_0+2$,
we must have $s_I'\ge 2$, and again the bound follows.

\*\*
\appendix(A4, Proof of Lemma 2)

\0To estimate the eigenvalues of $(ix-\MM^{[0]}(\h))E$, we
start by rewriting $(ix-\MM^{[0]}(\h))E$ in the form
$$(ix-\MM^{[0]}(\h))E=\pmatrix{-R&-ix+Q\cr ix+Q^\dagger& P\cr} .
\Eqa(A4.1))$$
Note that, by the strict convexity of the Hessian of $H_0$
if $\e$ is small enough, $P$ admits $N$ positive eigenvalues
$0<\m_1^{(0)}\le\cdots\le\m_N^{(0)}\=\m$.
Let $v_\a(x;\h)=\pmatrix{u\cr v\cr}$ be an eigenvector of
$(ix-\MM^{[0]})E$ with eigenvalue $\a$ (here $u$ and $v$ are two
column vectors of dimension $N$):
$$ \pmatrix{-R&-ix+Q\cr ix+Q^\dagger& P\cr}
\pmatrix{u\cr v\cr}=
\pmatrix{-R u+(-ix+Q)v\cr (ix+Q^{\dagger}) u+P v}=
\a\pmatrix{u\cr v\cr} .
\Eqa(A4.2)$$
Now, either $|\a|\ge \m_{1}^{(0)}/2$ or $\fra{P}2<P-\a<\fra{3P}2$. 
In the latter case $(P-\a)$ is invertible, $2/(3\m_N^{(0)}) \le
(P-\a)^{-1}\le 2/\m_1^{(0)}$, and we can rewrite \equ(A4.2) as
$$v=-(P-\a)^{-1}(ix+Q^{\dagger}) u,\qquad\quad -Ru-(-ix+Q)
(P-\a)^{-1}(ix+Q^{\dagger}) u= \a u .
\Eqa(A4.3)$$
The second equation in \equ(A4.3) is of the form
$$\Big[-x^2(P-\a)^{-1}+O(\e x)+O(\e\h^{k_0-1})\Big]u=\a u,
\Eqa(A4.4)$$
so that $|\a|\ge \fra2{3\m}x^2+ O(\e x)+O(\e\h^{k_0-1})$. If
$x^2\ge \r|\e|\h^{k_0-1}$,  for $\r$ large enough, the latter estimate
implies $|\a|\ge \fra1{2\m}x^2$, and Lemma 2 is proven.

\*\*
\appendix(A5, Symmetry properties of the self-energy matrices)

\0In this section we discuss the symmetry properties 
\equ(5.11) of $\MM^{[\le n]}(x;\h)$.

We begin with proving \equ(5.11) for $\MM^{[n]}(x;\h)$.
We inductively suppose that the same symmetry properties hold
for both $\MM^{[p]}(x;\h)$ and $g^{[p]}(x;\h)$, for $0\le p<n$,
$$ E g^{[p]}(x;\h) E=[g^{[p]}(-x;\h)]^{T} ,
\qquad [g^{[p]}(x;\h)]^*=g^{[p]}(-x;\h) ,
\Eqa(A5.1)$$
where the $*$ denotes complex conjugation and $T$ transposition.
Note that if $p=0$, $\MM^{[0]}(\h)$ and  $g^{[0]}(x;\h)$ satisfy the
desired symmetry properties, as discussed in Section \secc(4).

Consider the representation of $\MM^{[n]}(x;\h)$ given by \equ(5.6) and
\equ(5.7): in the following confusion between $T$ denoting
transposition and $T$ denoting a self--energy cluster will be avoided
by renaming the cluster $T$ with a new symbol $D$.  Note that
$F^{*}(\nn_{\vvvv})=F(-\nn_{\vvvv})$, so that,
using the second of \equ(A5.1), $[\VV_D(x;\h)]^{*}$
can be written as\footnote{${}^2$}{\nota We
stress that we use the convention that the
internal $\g$ indices of $D$ are summed over, while the momentum
labels are not summed over; in this way the value $\VV_D$ explicitly
depends only on: the momenta of the internal nodes, the external
momentum and the two $\g$ labels associated to the incoming and
outcoming lines.}
$$\eqalignno{
[\VV_D(x;\h)]^*&={(\h\b_1)^{|L(D)|}\over |\L(D)|!}
\Big(\prod_{\vvvv\in V(D)}\e^{\d_{\vvvvv}}\Big)
\Big(\prod_{\vvvv\in V(D)}F^{*}(\nn_{\vvvv})\Big)
\Big(\prod_{\ell\in \L(D)} [g^{[n_\ell]}(x_\ell;\h)]^{*} \Big)=
          &\eqa(A5.2)\cr
&={(\h\b_1)^{|L(D)|}\over |\L(D)|!}
\Big(\prod_{\vvvv\in V(D)}\e^{\d_{\vvvvv}}\Big)
\Big(\prod_{\vvvv\in V(D)}F(-\nn_{\vvvv})\Big)
\Big(\prod_{\ell\in \L(D)}g^{[n_\ell]}(-x_\ell;\h)\Big)=\VV_{D^*}(-x;\e) ,
\cr} $$
where $D^*$ is a cluster topologically equivalent to $D$, with
opposite mode labels (clearly the correspondence $D\otto D^*$ is
one--to--one). Summing over the choices of $D$ (see \equ(5.7)) the
second of \equ(5.11) follows.

To prove the first of \equ(5.11),
it is convenient to shorten notations; see Figure 4 for a pictorial
representation of the symbols. Given a self--energy cluster
$D$, let us denote by $\vvv$ and $\vvv'$ the
nodes such that the exiting line of $D$ comes out from $\vvv'$ and
the entering line of $D$ enters $\vvv$.
\\
(1) Let $\vvv'\=\vvv_1,\ldots,\vvv_{n}\=\vvv$ be the nodes on
the path $\LL\=\LL(D)$ joining $\vvv$ to $\vvv'$, and $\Bn_j$
be their mode labels; let also $\ell_j=(\vvv_{j+1}\vvv_j)$
be the line joining the two successive nodes $\vvv_{j+1}$ and
$\vvv_j$ in $\LL$; let $\g'_j,\g_j$ be the component labels
at the beginning of the line in $\LL$ exiting from $\vvv_j$ and at the
end of the line in $\LL$ entering the node $\vvv_j$ (they are,
respectively, the lines $\ell_{j-1}$ and $\ell_j$). Recall that,
by definition of self--energy cluster, any propagator $g_j$
associated to the lines $\ell_j\in\LL$ has a scale $[n_{\ell_j}]$,
with $n_{\ell_j}\ge 0$, \ie
there cannot be $\V0$-momentum propagators along the path $\LL$.
\\
(2) Let $\Bth_j$ be the (possibly empty) family of subtrees of $D$ with
root in $\vvv_j$ and with no line in common with $\LL$
(not represented in Figure 4: $\Bth_j$ can be imagined
to be a set of trees with the root line ending in $\vvv_j$ or,
possibly, $\Bth_j$ may consist just in $\vvv_j$).
\\
(3) Let $N_j$ be the $N\times N$ symmetric matrix $\dpr_{\g'_j\g_j}
f_{\Bn_j}(\b_0,\Bth_j)$, where $f_{\Bn_j}(\b_0,\Bth_j)$ is a function
depending only on the set of trees $\Bth_j$ and on $f_{\Bn_j}(\b)$ 
(it can be read off \equ(5.6) and \equ(5.7)) 
and $\dpr_{\g'},\dpr_\g$ must be interpreted as explained after
\equ(3.2).
\\
(4) The momentum flowing in a generic line $\ell$ of the graph $D$
will be the sum of the momentum $\Bn^0_{\ell}$ that would flow on the
line if the entering line momentum $\Bn$ was $\V0$ plus $\Bn$ if the
line is on the path $\LL$; then the propagator matrix 
$g_j$ associated to the line $\ell_j$ of $\LL$ is equal to
$g^{[n_{\ell_j}]}(x^0_j+x;\h)$, where $x^0_j$ is the scalar product of
$\Bo\cdot \Bn^0_{\ell_j}$ and $x=\Bo\cdot\Bn$.

\*
\figini{fig4}
\8<
\8</origine1assexper2pilacon|P_2-P_1| { >
\8<4 2 roll 2 copy translate exch 4 1 >
\8<roll sub 3 1 roll exch sub 2 copy atan >
\8<rotate 2 copy exch 4 1 roll mul 3 >
\8<1 roll mul add sqrt } def >
\8<>
\8</punta0 { 
\8<0 0 moveto dup dup 0 exch 2 div lineto 0 >
\8<lineto 0 exch 2 div neg lineto >
\8<0 0 lineto fill stroke } def >
\8<>
\8</freccia0 { 
\8<0 0 moveto 0 2 copy lineto stroke exch 2 >
\8< div exch translate 7 punta0 >
\8<stroke } def >
\8<>
\8</freccia{ 
\8<gsave origine1assexper2pilacon|P_2-P_1| >
\8<freccia0 grestore} def >
\8<>
\8</tlinea {gsave 4 2 roll moveto lineto [4 4] 2 >
\8<setdash stroke [] 0 setdash grestore} def >
\8</linea { 
\8<gsave 4 2 roll moveto lineto stroke grestore} def>
\8<>
\8</punto { 
\8<gsave 1.5 0 360 newpath arc fill stroke grestore} def>
\8<>
\8</TR {/FO exch def /Times-Roman findfont FO scalefont setfont} def>
\8</SI {/FO exch def /Symbols findfont FO scalefont setfont} def>
\8</TRB {/FO exch def /Times-Roman-Bold findfont FO scalefont setfont} def>
\8</SIB {/FO exch def /Symbols-Bold findfont FO scalefont setfont} def>
\8<>
\8<
\8<>
\8</L {50} def>
\8</H {10} def>
\8</R {10}def>
\8<>
\8</P0 {0 H} def>
\8</P1 {L H} def>
\8</P2 {2 L mul H} def>
\8</P3 {3 L mul H} def>
\8</P4 {4 L mul H} def>
\8</P5 {5 L mul H} def>
\8</P6 {6 L mul H} def>
\8<>
\8<P1 punto P2 punto P3 punto P4 punto P5 punto >
\8<P1 P0 freccia>
\8<P2 P1 freccia>
\8<P3 P2 freccia>
\8<P4 P3 tlinea>
\8<P5 P4 freccia>
\8<P6 P5 freccia>
\8<>
\8</HH {2 H mul} def>
\8</H1 {P1 HH add} def>
\8</H0 {P1 HH sub} def>
\8</K1 {P5 HH add} def>
\8</K0 {P5 HH sub} def>
\8<>
\8<H0 K0 linea H1 K1 linea>
\8<P1 HH 90 270 arc stroke>
\8<P5 HH 270 450 arc stroke>
\figfin

\*
\eqfig{300pt}{20pt}{
\ins{10pt}{24pt}{$\ell_0$}
\ins{280pt}{24pt}{$\ell_n$}
\ins{280pt}{2pt}{$\Bn$}
\ins{10pt}{2pt}{$\Bn$}
\ins{37pt}{6pt}{$\vvv'=\vvv_1$}
\ins{100pt}{2pt}{$\vvv_2$}
\ins{100pt}{2pt}{$\vvv_2$}
\ins{235pt}{2pt}{$\vvv=\vvv_n$}
\ins{40pt}{22pt}{$\st\g'$} 
\ins{53pt}{20pt}{$\st\g_1$} 
\ins{70pt}{24pt}{$\ell_1$} 
\ins{89pt}{22pt}{$\st\g'_2$} 
\ins{102pt}{20pt}{$\st\g_2$} 
\ins{190pt}{2pt}{$\vvv_{n-1}$} 
\ins{240pt}{22pt}{$\st\g'_n$} 
\ins{253pt}{20pt}{$\st\g$} 
}{fig4}{}

\*\*
\line{\vtop{\line{\hskip2.0truecm\vbox{\advance\hsize by -4.1 truecm
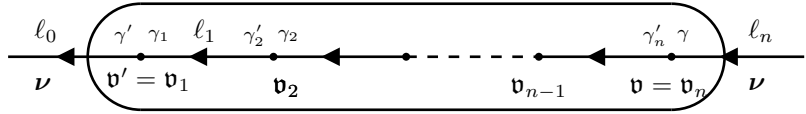
\0{{\css Figure 4.} {\nota A self-energy cluster $D$
and the path $\LL$ connecting its external lines.
The subtrees internal to $D$ with root on $\LL$ are
not drawn.\vfil}} } \hfill} }}
\*\*


Then the matrix $\VV_D(x;\h)$, see \equ(A5.1), can be concisely
written as
$$ \big(\VV_D(x;\h)\big)_{\g'\g}\,=\,
\Big(E N_1 g_1 E N_2 g_2\ldots g_{n-2}E N_{n-1} g_{n-1} E
N_n \Big)_{\g'\g} .
\Eqa(A5.3)$$
Interchanging the external lines (\ie having $\vvv'$ as
entering node and $\vvv$ as exiting node) generates a new self--energy
cluster $D'$ in which the momenta flowing in the lines of the
subtrees $\Bth_j$ with roots on the nodes $\vvv_j\in\LL$ are unchanged
while the momentum on the line $\ell_j\in\LL$ changes from
$\Bn^0_{\ell_j}+\Bn$, with $\Bn^0_{\ell_j}$ equal to the momentum which
would flow on $\ell_j$ if the external momentum $\Bn$ was set equal
to $\V0$, to a new value $-\Bn^0_{\ell_j}+\Bn$.

The matrix $ \VV_{D'}(x;\h)$ can therefore be written
$$ \big(\VV_{D'}(x;\h)\big)_{\g'\g}\,=\, \big(E N_n g_{n-1}' \,E N_{n-1}
g'_{n-2} \ldots g_{2}' \, E N_2\,g'_{1}\, E\, N_1\big)_{\g'\g} ,
\Eqa(A5.4)$$
where $g'_j=g^{[n_{\ell_{j}}]}(-x^0_{\ell_j}+x;\h)$.
Inserting $\openone\=E^TE$ after $N_n,\ldots,N_2$ and using the symmetry 
of $N_1,\ldots,N_n$, the inductive
validity of the first of \equ(A5.1) and the transposition rules for
matrix products, we get
$$ \left( \VV^{T}_{D'}(x;\h) \right)_{\g'\g}\,=\,
\Big(N_1 g''_1 E N_2 g''_2\ldots
g''_{n-2}E N_{n-1} g''_{n-1} E N_n E^T \Big)_{\g'\g} ,
\Eqa(A5.5)$$
where $g''_j\=g^{[n_{\ell_{j}}]}(x^0_{\ell_j}-x;\h)$. Hence
$E\VV^T_{D'}(x;\h)E\=\VV_D(-x;\h)$ completing the
inductive proof of \equ(A5.1),
because $\MM^{[n]}(x;\h)$ is defined as a sum over $D$ of the values
$\VV_D$ (see \equ(5.7)).

\*

\0{\it Remark}. From \equ(5.11) and \equ(A5.1) follows that
$g^{[\ge n]}(x;\h)$ satisfies the same symmetry properties
as $\MM^{[n]}(x;\h)$:
$$E\,g^{[\ge n]}(x;\h)\,E= [g^{[\ge n]}(-x;\h)]^{T} , \qquad
[g^{[\ge n]}(x;\h)]^* = g^{[\ge n]}(-x;\h) .
\Eqa(A5.6)$$
%

\*\*
\appendix(A6, Cancellations)

\0In this section we discuss the cancellations needed to show that 
$\MM^{[\le n]}(x;\h)$ has the structure described in \equ(5.12)
and \equ(6.0). In particular we want to show that the elements
$\big(\MM^{[n]}(x;\h)\big)_{\g'\g}$
with either $\g'=A_{i}$ or $\g=\a_{j}$ are proportional to $x$, while 
the elements with $\g'=A_{i}$ and $\g=\a_{j}$ are proportional to $x^2$, 
for $x^2\le \r\h^{2k_0-1}$. 

As in \cita{GG2}, the contributions to $\MM^{[n]}(x;\h)$
coming from clusters such that
$$\sum_{\vvvv\in V(T)}|\nn_{\vvvv}|\ge
(C_0/2^6|x|)^{1/\t_0}
\Eqa(A6.1)$$
require no cancellations: in fact
the exponential decay as $\Bn_{\vvvv}\to\io$ of the node functions implies
that the contributions coming from clusters satisfying \equ(A6.1) are
smaller than $Cx^2$, for some constant $C>0$.

When \equ(A6.1) does not hold we need to exploit cancellations,
because a priori there is no reason for some of the entries of
$\MM^{[n]}(x;\h)$ to be proportional to $x$ or $x^2$ for small $x$.
Following the same strategy used in
Appendix A2 of \cita{GG2} (and of Appendix A4 of \cita{GG1}),
the necessary cancellations can be checked to
occurr by collecting clusters violating condition \equ(A6.1) into
families. Such cancellations are due to the same mechanism pointed out
first in \cita{G1}, \cita{G2} and \cita{CF}.
The following analysis follows \cita{GM}.

Given a self--energy cluster $D$ on scale $[n-1]$, 
let us call $D_0$ the connected subset 
of $D$ containing no line on scale $[-1]$ and containing 
the nodes $\vvv$ and $\vvv'$ to which the entering and exiting lines
are attached. By definition $\sum_{\vvvv\in D_0}
\nn_{\vvvv}=\V 0$. Note also that, if the path $\LL$ 
connecting $\vvv$ and $\vvv'$ is non-trivial (i.e. $\vvv\neq \vvv'$),
$\LL$ is completely contained into $D_0$; moreover, if $D$ does
not contain lines on scale $[-1]$, then $D_0\=D$.
We define the family $\FF\=\FF_D$ 
as the set of self--energy clusters obtained from $D$ by shifting the 
entering and exiting lines by reattaching them to the nodes of $D_0$
in all possible ways. We then consider the sum 
$$\VV_\FF(x;\h)=\sum_{D\in\FF}\VV_D(x;\h) ,
\Eqa(A6.2)$$
contributing to the r.h.s. of \equ(5.7). When summing $\VV_\FF(x;\h)$
over all distinct families of clusters on scale $[n-1]$,
we recover the quantity $M^{[n]}(x;\h)$ defined in \equ(5.7).
To prove that $\MM^{[\le n]}$ has the structure described
by \equ(5.12) and \equ(6.0), we proceed in the following way.
We first note that, by the same analysis of Section 6
of \cita{GG2}, $\VV_\FF(x;\h)$ is differentiable with respect to $x$ in 
the sense of Whitney; then, we show that $\big(\VV_\FF(0;\h)
\big)_{\g'\g}=0$ if $\g'=A_i$ or $\g=\a_j$ and 
$\big(\dpr_x\VV_\FF(0;\h)\big)_{\g'\g}=0$ if $\g'=A_i$ and $\g=\a_j$.

Let us begin with proving that $\big(\VV_\FF(0;\h)\big)_{\g'\g}=0$
if $\g'=A_i$ or $\g=\a_j$, and to be definite let us assume that $\g=\a_j$
for some $j=1,\ldots,N-1$. Recall that, employing the notation 
of Appendix \secc(A5), we can write each matrix $\VV_D(x;\h)$ 
appearing in the sum \equ(A6.2) in the form \equ(A5.3). By the explicit
expression of \equ(A5.3) and of the matrices $N_j$, we see that, when 
computing $\VV_D(x;\h)$ in $x=0$, we find $(\VV_D(0;\h))_{\g'\a_j}=
O(\FF,\vvv',\h)_{\g'}\,(\Bn_{\vvvv})_j$ for some function
$O(\FF,\vvv',\h)_{\g'}$ {\it independent} of the choice of the node
$\vvv$ to which the entering line with label $\g=\a_j$ is attached to.  

If we change $D$ within $\FF$ by attaching the entering line to all 
possible nodes in $D_0$ (keeping $\vvv'$ fixed), we see that the global
contribution from such self--energy clusters is equal to 
$O(\FF,\vvv',\h)_{\g'}\cdot$ $\cdot\sum_{\vvvv\in D_0}
(\Bn_{\vvvv})_j$, which is $0$ 
by the property $\sum_{\vvvv\in D_0}(\Bn_{\vvvv})_j=\V 0$. 
So, if $\g=\a_j$, the proof of the fact that $\big(\VV_\FF(0;\h)
\big)_{\g'\g}=0$ is complete. In the case that $\g'=A_i$
the proof is completely analogous and we do not repeat it here.

Let us now turn to the proof of the fact that, if $\g'=A_i$ and $\g=\a_j$,
then $\big(\dpr_x\VV_\FF(0;\h)\big)_{\g'\g}=0$. 
First note that, by the explicit
form \equ(A5.3) of the value $\VV_D(x;\h)$ and in particular 
by the fact that $\VV_D(x;\h)$ depends on $x$ only through the 
propagators along the path $\LL$, we find that, when 
differentiating $\VV_D(x;\h)$ with respect to $x$, 
the effect of the derivative is as follows:
$$ \dpr_x\VV_D(x;\h)=\sum_{i=1}^{n-1}
\Big( EN_1g_1\cdots g_{i-1}EN_i \Big) \dpr_xg_i
\Big( EN_{i+1}g_{i+1}
\cdots g_{n-1}EN_n\Big) .
\Eqa(A6.3) $$
Given the line $\ell_i\in D_0$, we call $D_1$ and $D_2$ the two connected 
distinct subsets of $D_0$ obtained from $D_0$ by detaching the line $\ell_i$
and such that $D_1$ contains the node $\vvv'$
to which the line exiting from $D$ is attached, while $D_2$ contains
the node $\vvv$ to which the line entering $D$ is attached.
$D_1$ and $D_2$ are two subgraphs of $D$ with two external lines,
one coinciding with one of the external lines of $D$,
the other coinciding with $\ell_i$, see Figure 5.

\*
\figini{fig5}
\8<
\8</punto { 
\8<gsave 1.5 0 360 newpath arc fill stroke grestore} def>
\8</linea { 
\8<gsave 4 2 roll moveto lineto stroke grestore} def>
\8</tlinea {gsave 4 2 roll moveto lineto [4 4] 2 >
\8<setdash stroke [] 0 setdash grestore} def >
\8<>
\8</origine1assexper2pilacon|P_2-P_1| {
\8<4 2 roll 2 copy translate exch 4 1 roll sub >
\8<3 1 roll exch sub 2 copy atan rotate 2 copy >
\8<exch 4 1 roll mul 3 1 roll mul add sqrt } def >
\8<>
\8</ellisse0 {
\8<exch dup 0 moveto 0 1 360 {>
\8<3 1 roll dup 3 2 roll dup 5 1 roll exch>
\8<5 2 roll  3 1 roll dup 4 1 roll sin mul>
\8<3 1 roll exch cos mul exch lineto} for stroke pop pop} def>
\8<>
\8</ellisse {
\8<gsave origine1assexper2pilacon|P_2-P_1| pop>
\8<ellisse0 grestore} def>
\8<>
\8</puntox {
\8</y2 exch def /x2 exch def /y1 exch def /x1 exch def /x exch def>
\8<x1 x2 x1 sub x mul add>
\8<y1 y2 y1 sub x mul add} def>
\8<>
\8</origine1assexper2pilacon|P_2-P_1| { >
\8<4 2 roll 2 copy translate exch 4 1 >
\8<roll sub 3 1 roll exch sub 2 copy atan >
\8<rotate 2 copy exch 4 1 roll mul 3 >
\8<1 roll mul add sqrt } def >
\8<>
\8</punta0 { 
\8<0 0 moveto dup dup 0 exch 2 div lineto 0 >
\8<lineto 0 exch 2 div neg lineto >
\8<0 0 lineto fill stroke } def >
\8<>
\8</freccia0 { 
\8<0 0 moveto 0 2 copy lineto stroke exch 2 >
\8< div exch translate 7 punta0 >
\8<stroke } def >
\8<>
\8</freccia{ 
\8<gsave origine1assexper2pilacon|P_2-P_1| >
\8<freccia0 grestore} def >
\8<>
\8<>
\8</TR {/FO exch def /Times-Roman findfont FO scalefont setfont} def>
\8</SI {/FO exch def /Symbols findfont FO scalefont setfont} def>
\8</TRB {/FO exch def /Times-Roman-Bold findfont FO scalefont setfont} def>
\8</SIB {/FO exch def /Symbols-Bold findfont FO scalefont setfont} def>
\8<>
\8</espressioner{
\8</y exch def /x exch def>
\8<gsave 10 TRB>
\8<x y moveto exch show x 6 add y 3 sub moveto>
\8<8 TRB>
\8<show grestore} def>
\8<>
\8</L {40} def>
\8</H {20} def>
\8</Ro {30} def>
\8</Rv {H} def>
\8<>
\8</P0 {0 H} def>
\8</Pvp {L H} def Pvp punto>
\8</Pw {L 2 mul H} def Pw punto>
\8</Pwp  {L 3 mul H} def Pwp punto>
\8</Pv {L 4 mul H} def Pv punto>
\8</P1 {L 5 mul H} def>
\8<>
\8</C1 {0.5 Pw Pvp puntox} def >
\8</C2 {0.5 Pwp Pv puntox} def>
\8<>
\8<Ro Rv C1 P1 ellisse>
\8<Ro Rv C2 P1 ellisse>
\8<>
\8<Pvp P0 freccia>
\8<Pwp Pw freccia>
\8<P1 Pv freccia>
\8<Pvp Pw tlinea >
\8<Pwp Pv tlinea>
\8<>
\8<>
\8<(D)(1) 0.5 Pvp Pw puntox 10 add  espressioner>
\8<(D)(2) 0.5 Pv Pwp puntox 10 add  espressioner>
\8<10 TRB 0.5 P0 Pvp puntox 10 add exch 5 sub exch moveto (x) show>
\8<10 TRB 0.5 P1 Pv puntox 10 add moveto (x) show>
\8<10 TR 0.5 Pw Pwp puntox 10 add exch 11 sub exch moveto (x  + x) show>
\8<8 TR 0.5 Pw Pwp puntox 12 add exch 6 sub exch moveto (0) show>
\8<8 TR 0.5 Pw Pwp puntox 6 add exch 5 sub exch moveto (i) show>
\8<newpath>
\8<8 TR Pv  10 sub exch 3 sub exch moveto (v) show>
\8<8 TR Pwp 10 sub exch 3 sub exch moveto (w') show>
\8<8 TR Pw  10 sub exch 3 sub exch moveto (w) show>
\8<8 TR Pvp 10 sub moveto (v') show>
\figfin

\eqfig{200pt}{45pt}{}{fig5}{}

\*\*
\line{\vtop{\line{\hskip2.0truecm\vbox{\advance\hsize by -4.1 truecm
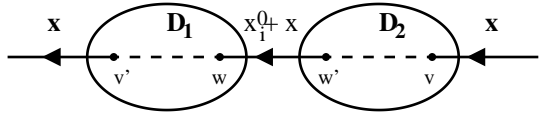
\0{{\css Figure 5.} {\nota Subdiagrams $D_{1}$ and $D_{2}$
in a self-energy cluster $D$. The drawn lines connected
to $\vvvv$ and $\vvvv'$ are the external lines of $D$.\vfil}} } \hfill} }}
\*\*


If we define the values of $D_1,D_2$ by formulae analogous to \equ(A5.3),
we can rewrite each term under the sum in \equ(A6.3) as
$\VV_{D_1}(x;\h)\dpr_xg^{[n_{\ell_i}]} (x^0_i+x;\h)\VV_{D_2}(x;\h)$,
where, with the same notations of Appendix \secc(A5),
we defined $x^0_i=\oo\cdot\nn^0_{\ell_i}$,
with $\Bn^0_{\ell_i}$ equal to the momentum which would flow
on $\ell_i$ if the external momentum $\Bn$ was set equal to $\V0$.

In particular let us consider
the value $(\VV_{D_1}(0;\h)\dpr_xg^{[n_{\ell_i}]}(x^0_i;\h)
\VV_{D_2}(0;\h))_{A_i\a_j}$ at $x=0$, that is one of 
the contributions we are interested in, and again note that 
by the very definition of $\VV_{D_1}$ and $\VV_{D_2}$, it holds
that $\big(\VV_{D_1}(0;\h)\big)_{A_i\g_1}=(\nn_{\vvvv'})_i\O(\www)_{\g_1}$
and $\big(\VV_{D_2}(0;\h)\big)_{\g_2\a_j}=\O(\www')_{\g_2}(\nn_{\vvvv})_j$,
where $\www,\www'$ are the two nodes where $\ell_i$
enters into and exists from, see Figure 5,
and $\O(\www)_{\g_1}$, $\O(\www')_{\g_2}$ are two
functions depending on the structure of $D_1$ and $D_2$,
{\it but not on the choices of $\vvv',\vvv$} within $D_1,D_2$. 

This implies that if we change $D$ within $\FF$ by attaching 
the entering line to all possible nodes in $D_2\cap D_0$ and 
by attaching the exiting line to all possible nodes in $D_1\cap D_0$, 
keeping $\ell_i$ fixed, the sum of the contributions of the form  
$(\VV_{D_1}(0;\h)\dpr_xg^{[n_{\ell_i}]}
(x^0_i;\h)\VV_{D_2}(0;\h))_{A_i\a_j}$ 
from this class of graphs is equal to 
$$ \Big(\sum_{\vvvv'\in D_1\cap D_0}(\nn_{\vvvv'})_i\Big)
\O(\www)_{\g_1}(\dpr_xg^{[n_{\ell_i}]}(x^0_i;\h))_{\g_1\g_2}
\O(\www')_{\g_2}
\Big(\sum_{\vvvv\in D_2\cap D_0}(\nn_{\vvvv})_j\Big) .
\Eqa(A6.4)$$
Note that, given any graph $D$ with the structure in Figure 5, 
the graph $D'$ in which the entering and exiting lines are
interchanged (so that the external lines enter in $\vvv'$ and exit from
$\vvv$, see Figure 6) the momentum through the line $\ell_i$ changes
from $\Bn_{\ell_i^0}+\Bn$ to $-\Bn_{\ell_i^0}+\Bn$. 

\*
\figini{fig6}
\8<
\8</punto { 
\8<gsave 1.5 0 360 newpath arc fill stroke grestore} def>
\8</linea { 
\8<gsave 4 2 roll moveto lineto stroke grestore} def>
\8</tlinea {gsave 4 2 roll moveto lineto [4 4] 2 >
\8<setdash stroke [] 0 setdash grestore} def >
\8<>
\8</origine1assexper2pilacon|P_2-P_1| {
\8<4 2 roll 2 copy translate exch 4 1 roll sub >
\8<3 1 roll exch sub 2 copy atan rotate 2 copy >
\8<exch 4 1 roll mul 3 1 roll mul add sqrt } def >
\8<>
\8</ellisse0 {
\8<exch dup 0 moveto 0 1 360 {>
\8<3 1 roll dup 3 2 roll dup 5 1 roll exch>
\8<5 2 roll 3 1 roll dup 4 1 roll sin mul>
\8<3 1 roll exch cos mul exch lineto} for stroke pop pop} def>
\8<>
\8</ellisse {
\8<gsave origine1assexper2pilacon|P_2-P_1| pop>
\8<ellisse0 grestore} def>
\8<>
\8</puntox {
\8</y2 exch def /x2 exch def /y1 exch def /x1 exch def /x exch def>
\8<x1 x2 x1 sub x mul add>
\8<y1 y2 y1 sub x mul add} def>
\8<>
\8</origine1assexper2pilacon|P_2-P_1| { >
\8<4 2 roll 2 copy translate exch 4 1 >
\8<roll sub 3 1 roll exch sub 2 copy atan >
\8<rotate 2 copy exch 4 1 roll mul 3 >
\8<1 roll mul add sqrt } def >
\8<>
\8</punta0 { 
\8<0 0 moveto dup dup 0 exch 2 div lineto 0 >
\8<lineto 0 exch 2 div neg lineto >
\8<0 0 lineto fill stroke } def >
\8<>
\8</freccia0 { 
\8<0 0 moveto 0 2 copy lineto stroke exch 2 >
\8<div exch translate 7 punta0 >
\8<stroke } def >
\8<>
\8</freccia{ 
\8<gsave origine1assexper2pilacon|P_2-P_1| >
\8<freccia0 grestore} def >
\8<>
\8<>
\8</TR {/FO exch def /Times-Roman findfont FO scalefont setfont} def>
\8</SI {/FO exch def /Symbols findfont FO scalefont setfont} def>
\8</TRB {/FO exch def /Times-Roman-Bold findfont FO scalefont setfont} def>
\8</SIB {/FO exch def /Symbols-Bold findfont FO scalefont setfont} def>
\8<>
\8</espressioner{
\8</y exch def /x exch def>
\8<gsave 10 TRB>
\8<x y moveto exch show x 6 add y 3 sub moveto>
\8<8 TRB>
\8<show grestore} def>
\8<>
\8<
\8</L {40} def>
\8</H {20} def>
\8</Ro {30} def>
\8</Rv {H} def>
\8<>
\8</P0 {Pw pop 0} def>
\8</Pvp {L H} def Pvp punto>
\8</Pw {L 2 mul H} def Pw punto>
\8</Pwp {L 3 mul H} def Pwp punto>
\8</Pv {L 4 mul H} def Pv punto>
\8</P1 {L 5 mul H} def>
\8</P11 { Pwp pop 0} def>
\8</C1 {0.5 Pw Pvp puntox} def >
\8</C2 {0.5 Pwp Pv puntox} def>
\8<>
\8<Ro Rv C1 P1 ellisse>
\8<Ro Rv C2 P1 ellisse>
\8<
\8<P0 Pvp freccia>
\8<Pw Pwp freccia>
\8<Pv P11 freccia>
\8<Pvp Pw tlinea >
\8<Pwp Pv tlinea>
\8<>
\8<(D')(1) 0.5 Pvp Pw puntox 10 add espressioner>
\8<(D')(2) 0.5 Pv Pwp puntox 10 add espressioner>
\8<10 TRB 0.5 P0 Pvp puntox 8 sub exch 8 sub exch moveto (x) show>
\8<10 TRB 0.5 P11 Pv puntox 8 sub exch 3 add exch moveto (x) show>
\8<10 TR 0.5 Pw Pwp puntox 10 add exch 11 sub exch moveto (-x + x) show>
\8<7 TR 0.5 Pw Pwp puntox 12 add exch 3 sub exch moveto (0) show>
\8<8 TR 0.5 Pw Pwp puntox 6 add exch 2 sub exch moveto (i) show>
\8<newpath>
\8<8 TR Pv 10 sub exch 3 sub exch moveto (v) show>
\8<8 TR Pwp 10 sub exch 3 sub exch moveto (w') show>
\8<8 TR Pw 10 sub exch 3 sub exch moveto (w) show>
\8<8 TR Pvp 10 sub moveto (v') show>
\figfin

\eqfig{200pt}{45pt}{}{fig6}{}

\*\*
\line{\vtop{\line{\hskip2.0truecm\vbox{\advance\hsize by -4.1 truecm
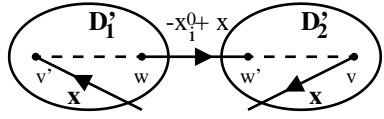
\0{{\css Figure 6.} {\nota The self-energy cluster $D'$
obtained from $D$ by interchanging the external lines of $D$:
the exiting line is attached to a node $\vvvv \in D_{2}'$ and the entering
line is attached to a node $\vvvv \in D_{1}'$.\vfil}} } \hfill} }}
\*\*


At $x=0$, if we change $D$ into $D'$,  
the value $\VV_{D_1}(0;\h)\dpr_xg^{[n_{\ell_i}]}
(x^0_i;\h)\VV_{D_2}(0;\h)$ is changed into
$\VV_{D_2'}(0;\h)$ $\dpr_xg^{[n_{\ell_i}]}
(-x^0_i;\h)\VV_{D_1'}(0;\h)$. In analogy with the 
results of Appendix \secc(A5), employing the symmetry relations \equ(A5.1),
we find $\VV_{D_2'}(0;\h)=E [\VV_{D_2}(0;\h)]^{T}E$ and
$\VV_{D_1'}(0;\h)=E[\VV_{D_1}(0;\h)]^{T}E$. Using also the symmetry
$E\dpr_x g^{[n_{\ell_i}]}(-x^0_i;\h)E=-\dpr_x [g^{[n_{\ell_i}]}]^{T}
(x^0_i;\h)$, we find that the 
element $(A_i,\a_j)$ of the contribution corresponding to
the graph in Figure 6 can be written as
$$\eqalign{&
\Big(\VV_{D_2'}(0;\h)\dpr_xg^{[n_{\ell_i}]}
(-x^0_i;\h)\VV_{D_1'}(0;\h)\Big)_{A_i\a_j}=
-\Big(E[\VV_{D_2}(0;\h)]^{T}\dpr_x [g^{[n_{\ell_i}]}]^{T}
(x^0_i;\h)\VV_{D_1}^T(0;\h)E\Big)_{A_i\a_j}
=\cr
&=-\Big(E\VV_{D_1}(0;\h)\dpr_x g^{[n_{\ell_i}]}
(x^0_i;\h)\VV_{D_2}(0;\h)E\Big)_{\a_j A_i}=
-\Big(\VV_{D_1}(0;\h)\dpr_x g^{[n_{\ell_i}]}
(x^0_i;\h)\VV_{D_2}(0;\h)\Big)_{A_j \a_i} .\cr}
\Eqa(A6.5)$$
Repeating the discussion leading to \equ(A6.4) we see that, if we sum 
\equ(A6.5) over the graphs obtained by attaching the entering line to all
possible nodes $\vvv'\in D_1'\cap D_0$ and attaching the exiting line
to all possible nodes $\vvv\in D_2'\cap D_0$, we get
$$-\Big(\sum_{\vvvv'\in D_1'\cap D_0}(\nn_{\vvvv'})_j\Big)
\O(\www)_{\g_1}(\dpr_xg^{[n_{\ell_i}]}(x^0_i;\h))_{\g_1\g_2}
\O(\www')_{\g_2} \Big(\sum_{\vvvv\in D_2'\cap D_0}(\nn_{\vvvv})_i\Big) .
\Eqa(A6.6) $$
Now, using the fact that $\sum_{\vvvv\in D_0}\nn_{\vvvv}=\V0$ and
that $D_1',D_2'$ and $D_1,D_2$ are topologically equivalent,
we also find that 
$$ \sum_{\vvvv'\in D_1'\cap D_0}(\nn_{\vvvv'})_j =
- \sum_{\vvvv\in D_2\cap D_0}(\nn_{\vvvv})_j , \qquad 
\sum_{\vvvv\in D_2'\cap D_0}(\nn_{\vvvv})_i =
- \sum_{\vvvv'\in D_1\cap D_0}(\nn_{\vvvv'})_i ,
\Eqa(A6.7) $$
so that \equ(A6.4) and \equ(A6.6) sum up to 0.

This completes the proof of the fact that
$$\sum_{D\in \FF} \VV_D(\e;0)_{A_i \a_j}=0,\qquad \dpr_x
\sum_{D\in \FF} \VV_D(\e;x)_{A_i \a_j}\,\big|_{x=0}=0 .
\Eqa(A6.8)$$

\*

\0{\it Remark.} It can be checked that 
the symmetry \equ(A5.1) ``only'' implies that, if we define 
$C_{ij}(x)\defi$ 
$\sum_{D\in \FF} \VV_D(\e;x)_{A_i \a_j}$, it holds 
$C_{ij}(x)=C_{ji}(-x)=C_{ji}^*(x)$: 
therefore the cancellation to second order (in $x$), expressed by
\equ(A6.7), {\it is not a parity cancellation} (unless $i=j$ where by the
self-adjointness of the matrix $C$ it follows that $C_{ii}(x)$ is real
and, hence, even in $x$). If the coefficients
$f_{\Bn}(\b)$ are assumed real then the self-adjointness property
\equ(A5.1) implies that $C_{ij}(x)$ is even in $x$ and the
second order cancellation at $x=0$ is an obvious parity cancellation
and the proof above is greatly simplified, as remarked in the simple
cases considered in \cita{G1} and \cita{G2}.

\*

Having proved that $\big(\MM^{[n]}(0;\h)\big)_{\g'\g}=0$ 
if either $\g'=A_i$ or $\g=\a_j$ and that 
$\big(\dpr_x\MM^{[n]}(0;\h)\big)_{\g'\g}=0$ 
if $\g'=A_i$ and $\g=\a_j$, it has still to be shown that the 
elements $\big(\MM^{[n]}(x;\h)\big)_{\g'\g}$ 
(with the suitable choices of $\g'$ and $\g$) satisfy appropriate
bounds once the factors $x$ determining the order of zero at $x=0$ are
extracted. From convergence one expects that the bounds on the rests 
of the Taylor series in $x$ around $x=0$ should
still be proportional to $\e^2$.

This is in fact true, {\it under the condition that \equ(A6.1) does
not hold}: if \equ(A6.1) does not hold, then changing the nodes
where the external lines are attached to {\it does not change the
scale} of the internal lines. And this implies that the bounds on
the derivatives with respect to $x$ are qualitatively the same as the bounds 
on $\MM^{[n]}(x;\h)$ and boundedness of the entries
of $\lis M^{[\le n]}$ follows, see \equ(6.0).

\*\*
\appendix(A7, Bounds on the propagator)

\0In this section we want to get 
a lower bound for the norm of $(ix-\MM^{[n]}(x;\h))E$, by an
approximate computation of its eigenvalue which is lowest in absolute
value, for $x^2\le \r \h^{2k_0-1}$, with $\r$ 
the same constant appearing in Lemma 3. Moreover we want to prove 
dimensional bounds for the
derivatives in $x$ and $\e$ of the approximate lowest eigenvalue.

Write $(ix-\MM^{[\le n]}(x;\h))E$ as $(ix-\MM^{[\le n]}(x;\h))E=
\pmatrix{-R^{[\le n]}(x;\h)&-ix+Q^{[\le n]}(x;\h)\cr
ix+Q^{[\le n]\dagger}(x;\h)& P^{[\le n]}(x;\h)}$, where $P^{[\le n]},
Q^{[\le n]}, R^{[\le n]}$ are the matrices defined in \equ(6.0).

The $N$ eigenvalues of $P^{[\le n]}(x;\h)$ are of order $1$ and
positive; call them $\m_1^{(n)}\le\ldots\le\m_N^{(n)}$. Since $P^{[\le
n]}(x;\h)$ differs from $P^{[0]}(x;\h)$ by $O(\e^2)$, the eigenvalues
$\m_i^{(n)}$ can be written as $\m_i^{(n)}=\m_i^{(0)}(1+O(\e^2))$.

To estimate the smallest eigenvalue of $(ix-\MM^{[\le n]}(x;\h))E$
we can follow a strategy adapted from the proof of Lemma 2.
Let $v_\a(x;\h)=\pmatrix{u\cr v\cr}$ be an eigenvector of
$(ix-\MM^{[\le n]}(x;\h))E$ with eigenvalue $\a$ (here $u$ and $v$ are two
column vectors of dimension $N$):
$$ \eqalign{
& \pmatrix{-R^{[\le n]}(x;\h)&-ix+Q^{[\le n]}(x;\h)\cr
ix+Q^{[\le n]\dagger}(x;\h)& P^{[\le n]}(x;\h)}\pmatrix{u\cr v\cr}= \cr
& \qquad \qquad =
\pmatrix{-R^{[\le n]} u+(-ix +Q^{[\le n]})v\cr 
(ix+Q^{[\le n]\dagger}) u+P^{[\le n]} v}=
\a\pmatrix{u\cr v\cr} . \cr}
\Eqa(A7.1)$$
To estimate the eigenvalue $\a$ with smallest absolute value,
we restrict attention to the case $P^{[\le n]}-\a\ge P^{[\le n]}/2$, 
as in the opposite case $|\a|> {\m_1^{(n)}}/2$.
Then $(P^{[\le n]}-\a)$ is invertible, $(P^{[\le n]}-\a)^{-1}=
[P^{[\le n]}]^{-1}+O(\a)$ and we can rewrite \equ(A7.1) as
$$ \eqalign{
& v =-(P^{[\le n]}-\a)^{-1}(ix+Q^{[\le n]\dagger}) u , \cr
& -R^{[\le n]}-(-ix+
Q^{[\le n]})(P^{[\le n]}-\a)^{-1}(ix+Q^{[\le n]\dagger}) u=
\a u . \cr}
\Eqa(A7.2)$$
 From the second of \equ(A7.2) we see that $\a$ must solve an equation
of the form
$$\det\Big[N^{[\le n]}-\a\big(1+O(x)+O(\e)\big)\Big]=0 ,
\Eqa(A7.3)$$
where we defined $N^{[\le n]}\defin-R^{[\le n]}-(-ix+Q^{[\le n]})
[P^{[\le n]}]^{-1}(
ix+Q^{[\le n]\dagger})$ and we used
that $Q^{[\le n]}=O(\e)$. This means that $\a$ is an eigenvalue
of a Hermitian matrix of the form $N^{[\le n]}\big(1+O(x)+O(\e)\big)$,
so that, calling $\l_i^{[n]}$ the eigenvalues of $N^{[\le n]}$, we have 
(see \cita{Ka})
$$\a=\l_i^{[n]}\big(1+O(x)+O(\e)\big) ,
\Eqa(A7.5)$$
for some $i=1,\ldots,N$. The conclusion of the previous discussion is
that the problem of computing the smallest eigenvalue of
$(ix-\MM^{[n]}(x;\h))E$ is essentially equivalent to the
problem of computing the smallest eigenvalue of $N^{[\le n]}$.

An explicit computation of $N^{[\le n]}$
shows that it can be written in the form
$$\eqalign{N^{[\le n]}&=-x^2\HHH_0^{-1}+\big\{-
R^{[\le n]}+x^2\HHH_0^{-1}-(-ix+Q^{[\le n]})[P^{[\le n]}]^{-1}(ix+
Q^{[\le n]\dagger})\big\}=\cr
&=-x^2\HHH_0^{-1}+\pmatrix{0&i\e xD(\h)\cr
-i\e xD^\dagger(\h)&-\e\h^{k_0-1}\widetilde M_{\b\b}(\h)\cr}+O(\e x^2) ,
\cr} . \Eqa(A7.6)$$
where $\HHH_0\defin\pmatrix{\dpr^2_{\AA\AA}H_0 &\dpr^2_{\AA B}H_0\cr
\dpr^2_{B\AA}H_0&\dpr^2_{BB}H_0\cr}$,
$\widetilde M_{\b\b}(\h)=-ck_0\b_1^{k_0-1}(1+O(\h))$ and
$D(\h)=D_0+O(\h)$, with
$$\pmatrix{D_0\cr0\cr}\defin-\pmatrix{1&0\cr 0&0\cr}\HHH_0^{-1}\pmatrix{
\dpr^2_{\AA \b}f_\V0(\b_0)\cr\dpr^2_{B \b}f_\V0(\b_0)}\Eqa(A7.7)$$
The off--diagonal $O(\e x)$ elements can
be eliminated through a small rotation of $N^{[\le n]}$.
In fact, defining the Hermitian matrix $Z$ as:
$$Z=\pmatrix{0&[\widetilde M_{\b\b}(\h)]^{-1}D(\h)\cr
[\widetilde M_{\b\b}(\h)]^{-1}D(\h)&0\cr} ,
\Eqa(A7.8)$$
a computation shows that
$$\eqalign{e^{i {x\h^{-k_0+1}}Z}
N^{[\le n]}e^{-i x\h^{-k_0+1}Z}&\=-x^2\HHH_0^{-1}+\d N^{[\le n]}=\cr
&=-x^2\HHH_0^{-1}+\pmatrix{0&0\cr
0&-\e\h^{k_0-1}\widetilde M_{\b\b}(\h)\cr}+O(\h^{1/2} x^2) , \cr} 
\Eqa(A7.9)$$
where we used the property $x\h^{-k_0+1}\le (\r\h)^{1/2}$,
if $x^{2} \le \r \e \h^{k_{0}-1}$. Multiplying the r.h.s.
of \equ(A7.9) times $\HHH_0^{1/2}$ both from the left
and from the right (remark that $\HHH_0^{1/2}$ is a well--defined 
positive $O(1)$ operator) we see that the norm of $N^{[\le n]}$
can be bounded above and below by an $O(1)$ constant times the norm of
$$ -x^2+ \HHH_0^{1/2}\d N^{[\le n]}\HHH_0^{1/2} .
\Eqa(A7.10)$$
Finally, using the explicit structure of $\d N^{[\le n]}$, see
\equ(A7.9), $N-1$ of the eigenvalues of \equ(A7.10) are equal to
$-x^2\big(1+O(\h^{1/2})\big)$, while the last one is
$-x^2\big(1+O(\h^{1/2})\big)-\e\h^{k_0-1}
\widetilde M_{\b\b}(\e)\dpr^2_{B}H_0$.

 From this computation, the bound \equ(6.1) of Lemma 5 follows.  To
address the differentiability of $\l^{[n]}(x;\h)$ with respect to $\e$
and $x$, we note that $\HHH_0^{1/2}\d N^{[\le n]}\HHH_0^{1/2}$ is $C^\io$ in
the sense of Whitney (because such is $\MM^{[\le n]}$, see 
the corresponding discussion in Appendix A3 of \cita{GG2}) and its
derivatives admit the following dimensional bounds:
$$ \|\,\dpr_\e\big(\HHH_0^{1/2}\d N^{[\le n]}\HHH_0^{1/2}\big)\|\le 
C\h^{k_0-1} , \qquad
\|\,\dpr_x\big(\HHH_0^{1/2}\d N^{[\le n]}\HHH_0^{1/2}\big)\|\le C\h^{k_0} .
\Eqa(A7.11) $$
Given this and the fact that the
last eigenvalue of $\HHH_0^{1/2}\d N^{[\le n]}\HHH_0^{1/2}$ is isolated
(it is non degenerate and its distance from the others is $O(\e\h^{k_0-1})$),
we can represent $\l^{[n]}(x;\h)$ in the form:
$$\l^{[n]}(x;\h)=-\h^{2k_0-1}\Tr\,\Big( \fra1{2\p i}\oint_{\g}
\fra{z \, {\rm d}z} {z-\h^{1-2k_0}\HHH_0^{1/2}\d N^{[\le n]}
\HHH_0^{1/2}}\Big) ,
\Eqa(A7.12)$$
where $z$ is a complex variable {\it independent of} $\e,x$ and
$\g$ is a circle in the complex plane around
$\widetilde M_{\b\b}(\h)\dpr^2_{B}H_0$ not sorrounding the origin,
with $\e$--independent radius.
The derivatives of $\l^{[n]}(x;\h)$ with respect to $\e$ and $x$
can be computed differentiating the r.h.s. of \equ(A7.12) and,
using the dimensional bounds \equ(A7.11),
we find the same dimensional bounds for
the derivatives of $\l^{[n]}(x;\h)$:
$$|\dpr_\e\l^{[n]}(x;\h)|\le C\h^{k_0-1} ,\qquad
|\dpr_x\l^{[n]}(x;\h)|\le C\h^{k_0}.
\Eqa(A7.13)$$
The proof of Lemma 5 is concluded.

\*\*
\appendix(A8, Excluded values of the perturbation parameter)

\0In this section we want to check that, 
by imposing the first condition in \equ(6.8),
the measure of excluded values of $\e$ can be bounded by
the second of \equ(6.8). To estimate the
measure of the excluded $\e$'s we proceed as in Appendix A2 of \cita{GG2}.
We present a proof valid for any $n\ge n_0$.
The dimensional bounds \equ(6.2) and the property
$|\l^{[n]}(\h)-\l^{[n-1]}(\h)|\le Ce^{-\k_1 2^{n/\t_1}}$
(see Lemma 5) implies the following dimensional bound
for the derivative of $\Sqrt{{\l}^{[n]}(\h)}$:
$$|\dpr_\e \Sqrt{{\l}^{[n]}(\h)}|\le C'\h^{-1/2} .
\Eqa(A8.1)$$
Then the first condition in \equ(6.8)
excludes, for each $\nn$, a subinterval of $I(\lis\e)$ whose
measure is bounded by
$$ {C_0 2^{-m/2}\h^{1/2}\over C'|\nn|^{\t_1}}
\Eqa(A8.2)$$
The Diophantine condition on $\oo$ implies that if
the first condition in \equ(6.8) is invalid
then $|\nn|$ cannot be too small:
$$ {C_02^{-m/2}\over |\nn|^{\t_1}}+\Sqrt{\l^{[n]}(\h)}\ge |x|
\ge C_{0}|\nn|^{-\t_{0}}\;.\Eqa(A8.3)$$
Therefore ${C_0\over 2|\nn|^{\t_0}} \ge C''\h^{k_0-1/2}$, hence in
this case we only have to consider the values of $\nn$
such that $|\nn|^{\t_{0}} \ge (C_{0}/2C'')\h^{{1\over 2}-k_0}$.
Summing \equ(A8.2) over the $\nn$'s satisfying this constraint,
and using the definition $\t_1=\t_0(1+\d_1)+N$
we find that the total measure of excluded $\e$'s can be bounded by
$${C_0 2^{-m/2}\h^{1/2}\over C'}\Big({2C''\h^{k_0-1/2}\over C_0}\Big)^{
1+\d_1}\sum_{|\nn|\neq 0}{1\over |\nn|^{N}} \=
KC_02^{-m/2}|\e|\h^{\d_1(k_0-1/2)} ,
\Eqa(A8.4)$$
which is in fact the second inequality in \equ(6.8).


\rife{BT}{1}{
H. Broer, F. Takens,
{\it Unicity of KAM tori},
Preprint, Groningen, 2005.}
\rife{Ch1}{2}{
Ch.-Q. Cheng,
{\it Birkhoff-Kolmogorov-Arnold-Moser tori in convex
Hamiltonian systems},
Communications in Mathematical Physics
{\bf 177} (1996), no. 3, 529--559. }
\rife{Ch2}{3}{
Ch.-Q. Cheng,
{\it Lower-dimensional invariant tori in the regions of instability
for nearly integrable Hamiltonian systems},
Communication in Mathematical Physics
{\bf 203} (1999), no. 2, 385-419. }
\rife{CF}{4}{
L. Chierchia, C. Falcolini,
{\it A direct proof of a theorem by Kolmogorov in Hamiltonian systems},
Annali della Scuola Normale Superiore di Pisa Classe Scienze (4)
{\bf 21} (1994), no. 4, 541--593. }
\rife{E1}{5}{
L.H. Eliasson,
{\it Perturbations of stable invariant tori for Hamiltonian systems},
Annali della Scuola Normale Superiore di Pisa Classe Scienze (4)
{\bf 15} (1988), no. 1, 115--147. }
\rife{E2}{6}{
L.H. Eliasson,
{\it Absolutely Convergent Series Expansions
for Quasi Periodic Motions},
Mathematical Physics Electronic Journal
{\bf2} (1996), no. 4, 1--33. }
\rife{H}{7}{
M. Herman,
{\it Some open problems in dynamical systems},
Proceedings of the International Congr\-ess of Mathematicians,
Vol. II (Berlin, 1998), Documenta Mathematica 1998,
Extra Vol. II, 797-808. }
\rife{JLZ}{8}{
\`A. Jorba, R. de la Llave, M. Zou,
{\it Lindstedt series for lower-dimensional tori},
Hamiltonian systems with three or more degrees of freedom
(S'Agar\'o, 1995), 151--167,
NATO Advanced Science Institutes Series C: Mathematical
and Physical Sciences, 533, Kluwer Acad. Publ., Dordrecht, 1999. }
\rife{G1}{9}{
G. Gallavotti,
{\it Twistless KAM tori, quasi flat homoclinic intersections, and
other cancellations in the perturbation series of certain completely
integrable hamiltonian systems. A review},
Reviews in Mathematical Physics
{\bf 6} (1994), no. 3, 343--411. }
\rife{G2}{10}{
G. Gallavotti,
{\it Twistless KAM tori},
Communications in Mathematical Physics
{\bf 164} (1994), no. 1, 145--156. }
\rife{GBG}{11}{
G. Gallavotti, F. Bonetto, G. Gentile,
{\it Aspects of the ergodic, qualitative and statistical theory of motion},
Springer-Verlag, Berlin, 2004. }
\rife{GG1}{12}{
G. Gallavotti, G. Gentile,
{\it Hyperbolic low-dimensional invariant tori and summations
of divergent series},
Communications in Mathematical Physics
{\bf 227} (2002), no. 3, 421--460. }
\rife{GG2}{13}{
G. Gentile, G. Gallavotti,
{\it Degenerate elliptic resonances},
Communications in Mathematical Physics
{\bf 257} (2005), no. 2, 319--362. }
\rife{Ge1}{14}{
G. Gentile,
{\it Quasi-periodic solutions for two-level systems},
Communications in Mathematical Physics
{\bf 242} (2003), no. 1--2, 221--250. }
\rife{Ge2}{15}{
G. Gentile,
{\it Diagrammatic techniques in perturbations theory, and applications},
Symmetry and perturbation theory (Rome, 1998), 59--78,
World Sci. Publishing, River Edge, NJ, 1999. }
\rife{Ge3}{16}{
G. Gentile,
{\it Diagrammatic techniques in perturbation theory},
Encyclopaedia of Mathematical Physics,
Eds. J.-P. Francoise, G. Naber and T. Sh. Tsun,
Elsevier, Amsterdam, to appear. }
\rife{GBC}{17}{
G. Gentile, D.A. Cortez, J.C.A. Barata,
{\it Stability for quasi-periodically perturbed Hill's equations},
Communications in Mathematical Physics, to appear. }
\rife{GM}{18}{
G. Gentile, V. Mastropietro,
{\it Methods for the analysis of the Lindstedt series for KAM tori and
renormalizability in classical mechanics. A review with some applications},
Reviews in Mathematical Physics
{\bf 8} (1996), no. 3, 393--444. }
\rife{GP}{19}{
P. Goldreich, S. Peale,
{\it Spin-orbit coupling in the solar system},
Astronomy Journal
{\bf 71} (1966), no. 6, 425--438. }
\rife{Ka}{20}{
T. Kato,
{\it Perturbation theory for linear operators}, 
Springer-Verlag, Berlin-New York, 1976. }
\rife{P}{21}{
H. Poincar\'e,
{\it Les m\'ethodes nouvelles de la m\'ecanique c\'eleste},
Tome II, Gauthier-Villars, Paris, 1893. }
\rife{Po}{22}{
J. P\"oschel,
{\it On elliptic lower-dimensional tori in Hamiltonian systems},
Mathematische Zeit\-schrift 
{\bf 202} (1989), no. 4, 559--608. }
\rife{R1}{23}{
H. R\"ussmann,
{\it Invariant tori in non-degenerate nearly integrable Hamiltonian systems},
Regular and Chaotic Dynamics
{\bf 6} (2001), no. 2, 119--204. }
\rife{R2}{24}{
H. R\"ussmann,
{\it Stability of elliptic fixed points of analytic area-preserving
mappings under the Bruno condition},
Ergodic Theory and Dynamical Systems
{\bf 22} (2002),  no. 5, 1551-1573. }
%
%

\biblio

\ciao